\documentclass[10pt,journal,compsoc]{IEEEtran}

\usepackage{amsfonts}
\usepackage{microtype}
\usepackage{tabularx}
\usepackage{afterpage}
\usepackage{flushend}
\usepackage{enumitem}
\usepackage{hyperref}
\usepackage{framed}
\usepackage{multicol}
\usepackage{bm}
\usepackage{balance}

\setlist{itemsep=1pt, topsep=3pt, leftmargin=1em}

\makeatletter
\newcommand{\removelatexerror}{\let\@latex@error\@gobble}
\makeatother

\usepackage{url}
\usepackage{amsmath}
\usepackage{graphicx}
\usepackage{stfloats}
\usepackage{subfloat}
\usepackage{subfig}
\usepackage{makecell}
\usepackage{xspace}
\usepackage{color}
\usepackage{multirow}
\usepackage{pifont}
\definecolor{darkblue}{rgb}{0.0, 0.0, 0.55}

\usepackage{amsthm}

\newcommand\vfont[0]{cmtt}

\renewcommand\v[1]{\mbox{\fontfamily{\vfont}\selectfont{#1}}}

\newcommand\proposal[1]{$\mbox{\fontfamily{\vfont}\selectfont{proposal}}_{#1}$}
\newcommand\emp[1]{$\mbox{\fontfamily{\vfont}\selectfont{empty}}_{#1}$}

\usepackage[linesnumbered,vlined,ruled]{algorithm2e}
\makeatletter
\renewcommand{\@algocf@capt@plain}{above}% formerly {bottom}
\makeatother

\makeatletter
\renewcommand{\@algocf@capt@plain}{above}% formerly {bottom}
\makeatother

\SetCommentSty{mycommfont}
\SetKwProg{Fn}{function}{:}{}
\SetKw{In}{in}
\DontPrintSemicolon
\SetInd{0.3em}{0.5em}
\SetVlineSkip{0.05em}
\SetKwIF{If}{ElseIf}{Else}{if}{:}{elif}{else:}{}
\SetKwFor{For}{for}{:}{}

\begin{document}

\SetKwBlock{Upon}{upon}{end}

\SetKwFunction{Reply}{reply} 
\SetKwFunction{Append}{append} \SetKwFunction{NormalPropose}{normalPropose}
\SetKwFunction{CheckPropose}{checkPropose} \SetKwFunction{Broadcast}{bcast}
\SetKwFunction{CanFinalizeEmpty}{canFinalize}

\SetKw{In}{in}

\setlength{\abovedisplayskip}{3pt}
\setlength{\belowdisplayskip}{3pt}
\newcommand{\cmark}{\ding{5 1}}
\newcommand{\xmark}{\ding{55}}
\newcommand{\myfig}[1]{#1}
\def\f#1{{\fontfamily{cmtt}\selectfont #1}}

% What about Themis: https://en.wikipedia.org/wiki/Themis
\newcommand{\xxx}{\textsc{Eges}\xspace} 
\newcommand{\eg}[0]{e.g.}
\newcommand{\ie}[0]{i.e.}
\newcommand{\us}[0]{us}
\newcommand{\fig}[1]{\hyperref[#1]{\color{darkblue} Figure~\ref{#1}}}
\newcommand{\algo}[1]{\hyperref[#1]{\color{darkblue} Algorithm~\ref{#1}}}
\newcommand{\lref}[1]{\hyperref[#1]{\color{darkblue} Line~\ref{#1}}}
\newcommand{\citemoving}[0]{~\cite{cho2020toward, venkatesan2016moving}\xspace}

\newcommand{\tab}[1]{\hyperref[#1]{\color{darkblue} Table~\ref{#1}}}
\newcommand{\algoline}[2]{\hyperref[#2]{\color{darkblue} Algorithm~\ref{#1}\xspace Line~\ref{#2}}}
\newcommand{\algostartend}[3]{\hyperref[#2]{\color{darkblue} Algorithm~\ref{#1}\xspace Line~\ref{#2}$\sim$\ref{#3}}}

\newcommand{\chref}[1]{\hyperref[#1]{\color{darkblue}\S\ref{#1}}}
\newcommand{\TODO}[1]{{#1}}
\newcommand{\github}[0]{{\color{darkblue}\url{github.com/hku-systems/eges}}}
\newcommand{\bfheading}[1]{\smallskip\noindent\textbf{#1\ }}
\newcommand{\nth}[1]{$#1^{th}$}
\newcommand{\smart}[0]{BFT-SMaRt\xspace}

\newcommand{\algorand}[0]{Algorand\xspace}
\newcommand{\scifer}[0]{\textsc{Scifer}}

\newcommand{\tputmax}[0]{88.5X\xspace}
\newcommand{\tputmin}[0]{4.9X??\xspace}
\newcommand{\tputbft}[0]{3.4X??\xspace}

\title{Efficient and DoS-resistant Consensus for Permissioned Blockchains}
% \date{}
% \author{Submission \#8}

% \settopmatter{printfolios=true}

\author{Xusheng Chen, Shixiong Zhao, Ji Qi, Jianyu Jiang, Haoze Song, Cheng Wang, Tsz On Li,\\T.-H. Hubert Chan, Fengwei Zhang, Xiapu Luo, Sen Wang, Gong Zhang, and Heming Cui.}

\IEEEtitleabstractindextext{
\begin{abstract}
Existing permissioned blockchain systems designate a fixed and explicit group of
committee nodes to run a consensus protocol that confirms the same sequence of
blocks among all nodes. Unfortunately, when such a permissioned blockchain runs
in a large scale on the Internet, these explicit committee nodes can be easily
turned down by denial-of-service (DoS) or network partition attacks. Although
work proposes scalable BFT protocols that run on a larger number of committee
nodes, their efficiency drops dramatically when only a small number of nodes are
attacked.

In this paper, our \xxx protocol leverages Intel SGX to develop a new
abstraction called ``stealth committee'', which effectively hides the committee
nodes into a large pool of fake committee nodes. \xxx selects a distinct group of
stealth committee for each block and confirms the same sequence of blocks among
all nodes with overwhelming probability. Evaluation on typical geo-distributed settings shows that: (1) \xxx is
the first permissioned blockchain's consensus protocol that can tolerate tough
DoS and network partition attacks; and (2) \xxx achieves comparable throughput
and latency as existing permissioned blockchains' protocols.

\end{abstract}
}
\maketitle

\theoremstyle{plain}
\newtheorem{invariant}{Invariant}

\theoremstyle{plain}
\newtheorem{lemma}{Lemma}
% \IEEEdisplaynontitleabstractindextext

\section{Introduction} 

A blockchain is a distributed ledger recording transactions maintained by nodes
running on a peer-to-peer (P2P) network. These nodes run a consensus protocol to
ensure consistency: nodes confirm the same sequence of blocks (no forks). Each
block contains the hash of its previous block, forming an immutable hash chain.
A blockchain can be permissioned or permissionless. A typical permissionless
blockchain does not manage membership for nodes and is usually attached with a
cryptocurrency mechanism (e.g., Bitcoin~\cite{nakamoto2008bitcoin}) to incite
nodes to follow the blockchain's protocol. \looseness=-1

In contrast, a permissioned blockchain runs on a set of identified member nodes
and can leverage the mature Byzantine Fault-Tolerant (BFT)
protocols~\cite{bftsmart:dsn18,pbft:osdi99,SBFT,yin2019hotstuff} to achieve
better efficiency (i.e., throughput and latency on confirming blocks). This
paper focuses on permissioned blockchains because their decoupling from
cryptocurrencies 
% that are usually attached to
% computing power~\cite{nakamoto2008bitcoin,buterin2014ethereum} or nodes'
% wealth~\cite{gilad2017algorand,buchman2016tendermint}. Many rel
has facilitated the deployment of many real-world 
% is more suitable for deploying 
general data-sharing applications, including a
medical chain among UK hospitals~\cite{medicalchain}, IBM supply
chains~\cite{supplychain}, and the Libra payment system~\cite{libra:smr}. \looseness=-1

For performance and regulation reasons (e.g., meeting the honesty threshold of
BFT protocols~\cite{lamport1982byzantine}), a permissioned blockchain (e.g.,
Hyperledger Fabric~\cite{androulaki2018hyperledger}) typically runs its
consensus protocol on a static and explicit committee. This static committee
approach is already robust for a permissioned blockchain among a small
scale of enterprises~\cite{supplychain}. 

Unfortunately, as permissioned blockchains become popular and are deployed in
large scales on the Internet, this static committee approach is vulnerable to
targeted Denial-of-Service (DoS) and network partition~\cite{gilad2017algorand,
natoli2016balance,heilman2015eclipse} attacks. For instance,
Libra~\cite{libra:smr} aims to build a global payment system (there are more
than 5,000 banks in the US~\cite{usbanks}) and identifies DoS attacks as a
significant threat, but it only provides partial mitigation
(\chref{sec:background:permissioned}). \looseness=-1

Indeed, great progress has been made in designing scalable BFT protocols (\eg,
SBFT~\cite{SBFT}) running on a larger group of committee nodes and tolerating
more nodes being attacked. However, these protocols designate a small number of
committee nodes to finish critical tasks (e.g., combing ACKs), making these
protocols' efficiency drop dramatically if these nodes are under DoS attacks
(\chref{sec:background:permissioned}). \looseness=-1

% In particular, \smart~\cite{bftsmart:dsn18} is an optimized BFT protocol for
% permissioned blockchains. \smart evaluated at most ten committee nodes; if
% four of these nodes are unreachable due to DoS attacks, the system cannot
% proceed or admit new committee nodes, but can only wait for these four nodes
% to recover.
With recent DoS attacks lasting for days~\cite{ddos:report, ddos:297hours},
tolerating such attacks is crucial, yet challenging, for applications deployed
on permissioned blockchains. 
\looseness=-1

To address such vulnerabilities of static committees, a promising direction is
to adopt the \textit{dynamic committee} merit from permissionless blockchain
systems~\cite{gilad2017algorand,buchman2016tendermint}. These systems select a
distinct committee for each block to handle the frequent leaving of nodes and to
provide fairness and can ensure liveness even if a committee fails to confirm a
block. \looseness=-1

% However, all existing blockchains~\cite{gilad2017algorand,
% buchman2016tendermint, kokoris2016enhancing, decker2016bitcoin} that adopt the
% dynamic committee merit, except for Algorand~\cite{gilad2017algorand}, cannot
% tolerate targeted DoS attacks.
Simply applying the dynamic committee approach, however, cannot ensure DoS-resistance.
% However, the dynamic committee approach does not brings immediate
% DoS-resistance. 
Another crucial requirement is unpredictability: the identifies of nodes in a
committee must be unpredictable to the attacker before the committee tries to
achieve consensus on a block. Otherwise, the attacker can adaptively attack the
ready-to-be committee and cause the system stuck.
% For instance, Tendermint~\cite{buchman2016tendermint} rotates committees
% deterministically, so a DoS attacker can adaptively attack those nodes that are
% going to be the next committee, causing the system to be stuck forever.
For instance, ByzCoin~\cite{kokoris2016enhancing} lets the proof-of-work winners
of recent blocks be the committee, but these explicit nodes are easily targeted
by a DoS attacker, causing ByzCoin to lose liveness
permanently~\cite{bycoin:stuck}. 
\looseness=-1

% After all, these blockchains use dynamic
% committees only g nodes, not to tackle
% DoS.

\looseness=-1
To the best of our knowledge, Algorand is the only work that can tolerate
targeted DoS attacks. 
% Algorand selects committees in an unpredictable way: each node uses verifiable
% random functions~\cite{micali1999verifiable} to independently determine
% whether it is in the committee for a given block.
However, as Algorand is designed for permissionless blockchains,
%  cannot control the exact number of nodes in a committee, 
it confirms a block with up to 15 rounds and minute-level latency
(discussed in \chref{sec:background:permissioned}), making it unsuitable for general
data-sharing applications on permissioned blockchains (e.g., Libra).

This paper aims to explore the new design point of building a permissioned
blockchain's consensus protocol that adopts the unpredictable dynamic committee
merit to defend against targeted DoS or partition attacks, and at the same time,
achieves comparable efficiency as existing BFT protocols (e.g., SBFT~\cite{SBFT}
has second-level latency). 

% Indeed, recent work (e.g., Tendermint~\cite{buchman2016tendermint},
% Tenderbake~\cite{tenderbake}, Tenderand~\cite{tenderand}) proposes BFT protocols
% using dynamic committees. 

% However, these protocols select committees in a
% deterministic way (e.g., round-robin~\cite{tenderbake}),
% % ensure consistency, 
% making them still vulnerable to targeted DoS or partition attacks: an attacker
% can adaptively attacks those to-be-committee nodes, causing the system stuck. 

A main obstacle 
% restricting the adoption of dynamic committees in permissioned blockchains 
is to ensure that any selected committee meets the honesty requirement for
byzantine problems: for consistency, each committee must have at most one-third
of nodes being malicious~\cite{lamport1982byzantine}. Permissionless blockchains
meet this requirement by selecting committees based on nodes' wealth in the
built-in cryptocurrency mechanism (i.e., proof-of-stake), but cryptocurrencies
are usually unavailable in permissioned blockchains. Consequently, 
to meet such a requirement, 
% 1that any selected committee has at most one-third
% malicious nodes, 
one has to unrealistically assume that almost all member nodes ($>$90\%) are
honest (calculated in \chref{sec:background}). 

% Particularly, fewer than one-third of \textit{all
% nodes} being malicious does not mean that fewer than one-third of any selected
% committee nodes are malicious ), and no BFT protocol can
% ensure consistency if more than one third of committee nodes are
% malicious~\cite{lamport1982byzantine}.
% % Therefore, existing permissionless protocols usually adopts their built-in 
% cryptocurrency mechanism to solve this problem, but 

Fortunately, recent work (e.g., Microsoft CCF~\cite{coco:ccf},
REM\cite{zhang2017rem}) shows that the code integrity feature of
SGX~\cite{sgx-explain} can regulate the misbehavior of blockchain nodes. For
instance, a recent implementation~\cite{minbft:sgx} of MinBFT leverages SGX to
ensure that a node cannot send conflicting messages to different nodes
and is incorporated into Hyperledger~\cite{androulaki2018hyperledger}.
\looseness=-1

We present \xxx\footnote{\xxx stands for Efficient, GEneral, and Scalable
consensus.}, the first efficient consensus protocol that can tackle targeted DoS
or partition attacks for a permissioned blockchain. \xxx adopts the dynamic
committee merit to select a distinct committee for confirming each block. To
defend against DoS or partition attacks targeting the committees, we leverage
the integrity and confidentiality features of SGX to present a new abstraction
called \textit{stealth committee}. \looseness=-1

\xxx's stealth committee has two new features. First, \xxx selects a
stealth committee in SGX: the selection progress has no communication among
committee nodes, and the selection result cannot be predicted from outside SGX.
This ensures that a committee node stays stealth (cannot be targeted by the
attacker) before sending out its protocol messages. 
Second, when nodes in a committee are trying to confirm a block, \xxx hides
these committee nodes into a large pool of fake committee nodes that behave
identically as the real ones observed from outside SGX, so that an attacker
cannot identify the real committees. 

%  Second, non-committee
% nodes randomly send covering (dummy) messages to conceal the true ones. 

% Third, to avoid attackers
% discovering the committee in the long run, \xxx selects a distinct committee for
% each block (i.e., moving the attack targets).

% The Second challenge
% Even with stealth committee, it is still 
% One challenge 
However, even equipped with SGX and stealth committee, it is still challenging
to efficiently ensure both consistency (i.e., no two member nodes confirm
conflicting blocks) and reasonable liveness (i.e., allow non-empty blocks to get
confirmed) in the asynchronous Internet due to the FLP
impossibility~\cite{fischer1985impossibility}. Specifically, suppose a
committee node $x$ for the \nth{n} block fails to receive the \nth{(n-1)} block
after a timeout, $x$ cannot distinguish whether it is because the committee for
the \nth{(n-1)} block failed to confirm the \nth{(n-1)} block, or because $x$
itself does not receive the confirmed \nth{(n-1)} block due to network problems.
As the committee nodes for the \nth{(n-1)} block may be under DoS attacks and be
unreachable, $x$ must have a mechanism to distinguish these two scenarios in
order to maintain both consistency and reasonable liveness in \xxx.

\xxx tackles this challenge using simple probability theory. \xxx's committee
for each block contains one proposer and $n_A$ (e.g., 300) acceptors, randomly
and uniformly selected from all nodes. The proposer broadcasts its block
proposal to all nodes by P2P broadcasts and seeks quorum \v{ACK}s from
the acceptors. \xxx models the randomly selected acceptors as a sampling of the
\textit{delivery rate} of the proposal in the P2P overlay
network~\cite{gilad2017algorand}. In the previous example, \xxx confirms the
proposal for the \nth{(n-1)} block only if the proposal is delivered to a large
portion of member nodes; if multiple rounds of the sampling show that very few
nodes have received that proposal for the \nth{(n-1)} block, nodes in \xxx
consistently confirm the \nth{(n-1)} block as an empty block (with an overwhelming
probability). 

% if only very few nodes (e.g., less
% than 10\%) received the proposal for the \nth{(n-1)} block, the probability that
% this proposal has been confirmed is overwhelmingly low, and thus an empty block
% can be confirmed as the \nth{(n-1)} block.  

% \xxx realizes this idea with a new consensus protocol. Specifically, \xxx lets
% proposers for subsequent blocks ask their own acceptors whether they received
% the proposal of the $(n-1)^{th}$ block along with broadcasting their own block
% proposals. If quorums (\eg, 59\%) of acceptors for the next $D$ (\eg, a system
% parameter, 4 by default) blocks replies, and they have never received that
% proposal, then we have overwhelming confidence (probability) to say that only
% very few nodes received that proposal and thus that proposal is never confirmed.
% We prove that our protocol enforces consistency with overwhelming probability

In sum, \xxx efficiently enforces consistency and can defend against targeted
DoS or partition attacks. Specifically, \xxx defends against such attacks by (1)
letting committee nodes stay stealth \textit{before} they start achieving
consensus for a block, (2) using fake committee nodes to conceal real committee
nodes \textit{while} they are achieving consensus for a block, and
(3) switching to a different committee and consistently confirming a block
even if the attacker luckily guesses most real committee nodes for this block. 
% and DoS them. 
% for
% the attacked committee. the real ones when a committee is achieving consensus on
% a block; even if an attacker happens to guess and to target most of the real
% committee nodes, \xxx switches

In essence, \xxx's stealth committee is a moving target defense
approach~\citemoving. \xxx unpredictably replaces the critical components (i.e.,
the committee) in a large system so that a DoS attacker cannot launch effective
attacks targeting these components.  
\xxx's consensus protocol also achieves good efficiency because confirming a
block in a gracious run (e.g., the proposer can reach most acceptors) only
involves two P2P broadcasts and a half UDP round trip (\chref{sec:protocol}). We
provide a rigorous analysis of \xxx's DoS-resistance and proof of \xxx's
consistency guarantee in \chref{sec:analysis}.

	% \small
% \usepackage{hhline}

\begin{table}[tb]
	\centering
	\footnotesize
	\centering
	\setlength\tabcolsep{1pt} % default value: 6pt
	\begin{tabular}{l|c|c|c|c|c|c} 
	\Xhline{2\arrayrulewidth}
	\multicolumn{1}{c|}{\begin{tabular}[c]{@{}c@{}}\textbf{protocol}\\\textbf{ name} \end{tabular}} & \multicolumn{1}{c|}{\begin{tabular}[c]{@{}c@{}}\textbf{DoS}\\\textbf{resistanc}\end{tabular}} & \begin{tabular}[c]{@{}c@{}}\textbf{with}\\\textbf{SGX}?\end{tabular} & \begin{tabular}[c]{@{}c@{}}\textbf{consensus}\\\textbf{ model} \end{tabular} & \begin{tabular}[c]{@{}c@{}}\textbf{number }\\\textbf{ of nodes} \end{tabular} & \begin{tabular}[c]{@{}c@{}}\textbf{tput}\\\textbf{ (txn/s)} \end{tabular} & \begin{tabular}[c]{@{}c@{}}\textbf{confirm}\\\textbf{ latency (s)} \end{tabular}  \\ 
	\Xhline{2\arrayrulewidth}
	\multicolumn{1}{l|}{\multirow{2}{*}{\xxx}} & \multicolumn{1}{c|}{\multirow{2}{*}{high}} & \multicolumn{1}{c|}{\multirow{2}{*}{Yes}} & \multicolumn{1}{c|}{\multirow{2}{*}{hybrid}} & \multicolumn{1}{c|}{300} & \multicolumn{1}{c|}{3226} & \multicolumn{1}{c}{0.91} \\ \cline{5-7} 
	\multicolumn{1}{l|}{}                      & \multicolumn{1}{c|}{}                      & \multicolumn{1}{c|}{}                     & \multicolumn{1}{c|}{}                        & \multicolumn{1}{c|}{10k} & \multicolumn{1}{c|}{2654} & \multicolumn{1}{c}{1.13} \\ \hline
	Algorand                                                                                        & high                                                                                          & No                                                                   & BFT                                                                                                                                               & 10K                                                                           & $\sim$727                                                                 & $\sim$22s                                                                         \\\hline
	PoET                                                                                            & high                                                                                          & Yes                                                                  & Hybrid                                                                                                                                              & 100                                                                           & 149                                                                       & 45.2                                                                              \\\hline
	Ethereum                                                                                        & high                                                                                          & No                                                                   & BFT                                                                                                                                                 & 100                                                                           & 178                                                                       & 82.3                                                                              \\\hline
	SBFT                                                                                            & medium                                                                                        & No                                                                   & BFT                                                                                                                                            & 62                                                                            & 1523                                                                      & 1.13                                                                              \\\hline
	MinBFT                                                                                          & low                                                                                           & Yes                                                                  & Hybrid                                                                                                                                              & 64                                                                            & 2478                                                                      & 0.80                                                                              \\\hline
	\smart																						    & low                                                                                           & No                                                                   & BFT                                                                                                                                              & 10                                                                            & 4512                                                                      & 0.67                                                                              \\\hline
	Tendermint                                                                                      & low                                                                                           & No                                                                   & BFT                                                                                                                                             & 64                                                                            & 2462                                                                      & 1.31                                                                              \\\hline
	HotStuff                                                                                        & low                                                                                           & No                                                                   & BFT                                                                                                                                                & 64                                                                            & 2686                                                                      & 2.63                                                                           \\\hline
	HoneyBadger                                                                                     & low                                                                                           & No                                                                   & BFT                                                                                                                                                 & 32                                                                            & 1078                                                                      & 9.39                                                                           \\\hline
	\Xhline{2\arrayrulewidth}
	\end{tabular}

	% \multicolumn{1}{c}{\textbf{\begin{tabular}[c]{@{}c@{}}protocol\\ name\end{tabular}}} & \multicolumn{1}{c}{\textbf{\begin{tabular}[c]{@{}c@{}}with\\ SGX?\end{tabular}}} & \multicolumn{1}{c}{\textbf{\begin{tabular}[c]{@{}c@{}}consensus\\ model\end{tabular}}} & \multicolumn{1}{c}{\textbf{\begin{tabular}[c]{@{}c@{}}DoS\\ resistance\end{tabular}}} & \multicolumn{1}{c}{\textbf{\begin{tabular}[c]{@{}c@{}}number \\ of nodes\end{tabular}}} & \multicolumn{1}{c}{\textbf{\begin{tabular}[c]{@{}c@{}}tput\\ (txn/s)\end{tabular}}} & \multicolumn{1}{c}{\textbf{\begin{tabular}[c]{@{}c@{}}confirm\\ latency (s)\end{tabular}}} \\
	\caption{Comparison of \xxx to baseline protocols. Analysis of DoS
	resistance is in \chref{sec:background}; evaluation setup is covered in
	\chref{sec:eval}.}\label{tab:performance}
	\vspace{-1em}
	\end{table}

\looseness=-1
We implemented \xxx using the codebase from Ethereum~\cite{buterin2014ethereum}
and compared \xxx with nine consensus protocols for blockchain systems,
including five state-of-the-art efficient BFT protocols for permissioned
blockchains (\smart~\cite{sousa2017byzantine}, SBFT~\cite{SBFT},
HoneyBadger~\cite{miller2016honey}, and HotStuff~\cite{yin2019hotstuff}), two
SGX-powered consensus protocols for permissioned blockchains
(Intel-PoET~\cite{poet} and MinBFT~\cite{veronese2013efficient}), the default
consensus protocol in our codebase (Ethereum-PoW~\cite{buterin2014ethereum}),
and two permissionless blockchains' protocols that run on dynamic committees
(Algorand~\cite{gilad2017algorand} and Tendermint~\cite{buchman2016tendermint}).
We ran \xxx on both our cluster and AWS. Evaluation shows that:

\begin{itemize}
    \setlength\itemsep{-0em}
    \item \xxx is robust. Among all consensus protocols for permissioned
    blockchains, \xxx is the only protocol that can defend against targeted DoS and
    network partition attacks, by both a theoretical analysis
    (\chref{sec:analysis}) and evaluation (\chref{sec:eval:robust}). 
    
    \item \xxx is efficient. \xxx confirms a block with 3000 transactions in
    less than two seconds in typical geo-distributed settings, comparable to
    evaluated consensus protocols that cannot tolerate targeted DoS attacks.
    % suitable for existing applications running on permissioned blockchains
    % (e.g., the Libra payment system). 

    \item \xxx's throughput and latency are scalable to the number of nodes. When
    running 10k nodes, \xxx showed 2.3X higher throughput and 16.8X lower
    latency than Algorand with 10k nodes.
    % \item \xxx greatly improves the reliability of both \xxx-DB and \xxx-ToR.
\end{itemize}
\pagebreak

Compared to existing BFT protocols~\cite{sousa2017byzantine, SBFT,
miller2016honey,yin2019hotstuff} and SGX-powered consensus protocols~\cite{poet,
veronese2013efficient} for permissioned blockchains, \xxx is the only protocol
that can tolerate targeted DoS attacks, and \xxx's efficiency is comparable to
the fastest of these protocols. 
% achieves comparable efficiency 
Compared to Algorand, the only known DoS-resistant
consensus protocol for permissionless blockchains, \xxx has much higher
throughput and lower latency. \looseness=-1

Our contribution is two-fold. First, \xxx leverages SGX to explore the new
design point of tackling DoS attacks while enforcing both consistency
and reasonable liveness (including efficiency) for a permissioned blockchain in
the asynchronous Internet. Second, we designed the new stealth committee
abstraction and implemented \xxx's consensus protocol. \xxx's source code is
available on \github. \xxx can facilitate the deployments of various
mission-critical, DoS-resistant permissioned blockchain applications on the
Internet (e.g., e-voting~\cite{riemann2017distributed} and payment~\cite{libra:smr}). 
% we leverage the integrity and
% confidentiality features of SGX to build a new stealth committee abstraction:
% \xxx selects a fixed number of committee nodes in SGX, which is unpredictable
% from outside SGX and has no communication among committee nodes. This
% abstraction also uses fake committee nodes to conceal the real ones, enabling
% \xxx to defend against DoS and partition attacks targeting committee nodes.
% Second, with this new abstraction, 
% Second, 
% we design and implement a new 
% consensus protocol that can ensure efficiently ensure consistency in a permissioned blockchain.

% \xxx has the potential to deploy
% existing centralized SGX-protected applications (e.g.,
% SGX-ToR~\cite{sgxtor:nsdi17} with a centralized directory service) in a
% decentralized manner (discussed in \chref{sec:eval:discussion}). 

\looseness=-1 In the rest of the paper, \chref{sec:background} introduces \xxx's
background and motivation; \chref{sec:model} defines the model of \xxx;
\chref{sec:high} gives a high-level overview of \xxx; \chref{sec:protocol}
introduces \xxx's consensus protocol; \chref{sec:analysis} analyzes the safety
and liveness of \xxx; \chref{sec:eval} shows our evaluation, and \chref{sec:con}
concludes.

\section{Background and Related Work}\label{sec:background}

We discuss targeted DoS and network partition attacks together in this paper
because these two attacks cannot be effectively distinguished in an asynchronous
network. Particularly, when a node cannot reach a remote node, the node cannot
determine whether it is because the remote node is under DoS attacks or because
these two nodes are partitioned in the network. Therefore, \xxx maintains consistency by
handling both cases together. 

When discussing targeted DoS or partition attacks in this subsection, we assume
that the attacker has an attack budget $B$ (e.g., $B = 300$): the attacker can
adaptively target $B$ nodes at a time. This model is the same as Algorand's,
which we will formally define in \chref{sec:model}. \looseness=-1

\subsection{Intel SGX}\label{sec:background:sgx}
Intel Software Guard eXtension (SGX)~\cite{sgx-explain} is a hardware feature on
commodity CPUs. SGX provides a secure execution environment called enclave,
where data and code execution cannot be seen or tampered with from outside. Code
outside enclaves can enter an enclave by ECalls, and SGX uses remote
attestations~\cite{sgx-explain} to prove that a particular piece of code is
running in an enclave on a genuine SGX-enabled CPU. SGX provides a trustworthy
random source (\v{sgx\_read\_rand}), which calls the hardware pseudo-random
generator through the RDRAND CPU instruction seeded by on-chip entropy
sources~\cite{sgx-explain}. Previous studies show that this random source
complies with security and cryptographic standards and cannot be seen or
tampered with from outside enclaves~\cite{sgxrand:blackhat,
hamburg2012analysis}. \looseness=-1

Recent work shows that SGX can be leveraged to improve diverse aspects of
blockchain systems. Intel's PoET~\cite{poet} replaces the PoW puzzles
with a trusted timer in SGX; \xxx is more efficient than PoET
(\chref{sec:eval:perf}). REM~\cite{zhang2017rem} uses SGX to replace the useless
PoW puzzles with useful computation (e.g., big data), orthogonal to \xxx.
Microsoft CCF~\cite{coco:ccf} (originally named CoCo) is a permissioned
blockchain platform using SGX to achieve transaction privacy, but it does not
include a DoS-resistance approach. Scifer~\cite{ahmed2018identity} uses SGX's
attestation to establish reliable identities of nodes and maintains their
identities on the blockchain, which is adopted in \xxx
(\chref{sec:impl:membership}).

Ekiden~\cite{cheng2018ekiden} and ShadowEth~\cite{yuan2018shadoweth} offload the
execution of smart contracts to SGX-powered nodes to avoid redundant execution
and to preserve privacy; TEEChain~\cite{teechain} uses SGX to build an efficient
and secure off-chain payment channel; Town Crier~\cite{zhang2016town} uses SGX
to build a trustworthy data source for smart contracts;
Tesseract~\cite{bentov2017tesseract} uses SGX to build a cross-chain coin
exchange framework; Obscuro~\cite{tran2017obscuro} uses SGX to improve bitcoin's
privacy; these systems do not focus on consensus protocols and are orthogonal to
\xxx. 

% cite more SGX blockchain work here. 

\subsection{Consensus for Permissioned Blockchains}\label{sec:background:permissioned}

We briefly introduce recent notable consensus protocols for permissioned
blockchains, which are also \xxx's evaluation baselines. Overall, \textit{all}
these protocols run on a static committee. To ensure liveness under a DoS
attacker with an attack budget of $B$, these protocols must scale to $3\times B
+ 1$ nodes (for BFT protocols) or $2\times B + 1$ nodes (for SGX-powered
protocols). However, to our best knowledge, no existing protocol can achieve
such scalability. 

\smart~\cite{bftsmart:dsn18} is an optimized implementation of
PBFT~\cite{pbft:osdi99}. As each node broadcasts consensus messages to all other
nodes, \smart has $O(n^2)$ message complexity to the number of
committee nodes, resulting in poor scalability. Its paper~\cite{bftsmart:dsn18}
only evaluated up to 10 nodes. 
SBFT~\cite{SBFT} is a scalable BFT protocol that uses a new type of committee
nodes called collectors. A node sends its consensus messages to only $c$
(usually $c<8$) explicit collectors who will then broadcast a combined message
using the threshold signature. SBFT's fast path can commit a block if fewer than $c$
nodes have failed; however, SBFT's performance drops dramatically if an attacker
targets the $c$ collectors (\chref{sec:eval:sbft}).

% if there are 
% nodes attacked.
% . This makes SBFT's 

%
HotStuff~\cite{yin2019hotstuff} is a BFT protocol optimized for frequent leader
changes, and Libra~\cite{libra:smr} leverages Hotstuff to tolerate targeted DoS
attacks \textit{on leaders}. However, since Hotstuff reports a near-linear
increment of latency with an increasing number of nodes, it only evaluated up to
128 nodes, where an attacker can DoS attack or partition one-third of all nodes
rather than finding the leader.
HoneyBadger~\cite{miller2016honey} uses randomization to remove the partial
synchrony assumption of PBFT. However, both its paper and our evaluation shows
that HoneyBadger achieves high latency due to multiple rounds of its
asynchronous byzantine agreements. 

MinBFT~\cite{veronese2013efficient} is an SGX-powered BFT protocol that has the
same fault model as \xxx. MinBFT reduces the number of rounds in PBFT and can
tolerate more node failures (one-half instead of one-third), but MinBFT still
has $O(n^2)$ message complexities, so its performance is not scalable to the
number of nodes. 
\looseness=-1

% However, to maintain
% liveness in the face of an attacker with an attack budget $B$, MinBFT must scale
% to $2\times B + 1$ committee nodes. 

% MinBFT's SGX
% implementation~\cite{minbft:sgx} is in the mainstream of the Hyperledger
% project, supporting the \xxx's choice of using SGX to regulate nodes' behavior
% in a permissioned blockchain. 

\subsection{Consensus for Permissionless Blockchains}\label{sec:background:public}

Existing permissionless blockchains can be divided into two categories based on
how they confirm blocks. The first category confirms block with variants of the
longest-chain rule (i.e., Nakamoto consensus~\cite{nakamoto2008bitcoin}),
including BitCoin~\cite{nakamoto2008bitcoin},
Ethereum~\cite{buterin2014ethereum}, BitCoin-NG~\cite{eyal2016bitcoinng},
% Fruitchain~\cite{pass2016fruitchains}, 
Snow-White~\cite{bentov2016snow}, Ouroboros~\cite{kiayias2016ouroboros},
Paros~\cite{david2017ouroboros}, Genesis~\cite{badertscher2018ouroboros}, and
GHOST~\cite{sompolinsky2013accelerating}. Specifically, each node asynchronously
selects the longest chain it received and confirms a block when there are $k$
blocks succeeding it. However, waiting for $k$ more blocks leads to a long
confirm latency, and previous work~\cite{gervais2016security} shows that this
$k$ must be large enough to ensure consistency. Moreover, the longest-chain rule
cannot ensure consistency under partition attacks~\cite{apostolaki2017hijacking,
heilman2015eclipse,gilad2017algorand}. Intuitively, during a network partition,
each partition will independently grow a chain; if these chains diverge for more
than $k$ blocks, nodes in different partitions will confirm conflicting blocks.

The second category of permissionless blockchains confirms blocks using the
committee-based BFT approach, which can confirm a block as soon as the BFT
consensus is achieved. This category includes Algorand~\cite{gilad2017algorand},
ByzCoin~\cite{kokoris2016enhancing}, Tendermint~\cite{buchman2016tendermint},
and PeerCensus~\cite{decker2016bitcoin}. These systems select distinct
(dynamic) committees for different blocks based on the content (e.g., nodes'
wealth) on the blockchain for fairness and for handling nodes joining or
leaving. Similar to \xxx, these systems run a tailored consensus protocol (BA* in
Algorand~\cite{gilad2017algorand}, Tendermint~\cite{buchman2016tendermint},
Tenderbake~\cite{tenderbake}, and Tenderand~\cite{tenderand}) on dynamic
committees to confirm blocks.

\looseness=-1
However, these protocols cannot be ported to a permissioned
blockchain because of the tight coupling with cryptocurrency. For instance,
although Tendermint~\cite{tendermint:opodis} and Tenderbake~\cite{tenderbake}
are described as stand-alone BFT protocols, they assume that in any
committee, fewer than one-third of nodes are malicious. 
In a permissioned blockchain without cryptocurrency, if we want to ensure that
any randomly selected committee (say 100 nodes) from a large number of (say 10k)
nodes meets this requirement with overwhelming probability ($>1-10^{-10}$), we
need to assume over 91\% of all nodes being honest (by the hypergeometric
distribution), which is an overly-strong assumption for a practical large-scale
blockchain system (e.g., a global payment system~\cite{libra:smr}) on the
Internet. 

\looseness=-1
Moreover, these systems (except Algorand) cannot ensure liveness under targeted
DoS attacks because they select committees in a predictable way so that all
nodes can verify the identities of committees. For instance,
ByzCoin~\cite{kokoris2016enhancing} lets the proof-of-work winners of recent
blocks be the committee. However, these nodes with explicit identities are
easily targeted by a DoS attacker, and ByzCoin may lose liveness
permanently~\cite{bycoin:stuck} if more than one-third of these nodes are
attacked. Algorand defends against 
% is the only known protocol that can ensure practical liveness
targeted DoS attacks by letting each node use verifiable random functions to
determine its committee membership. We provide a detailed comparison
showing why \xxx is more efficient than Algorand in
\chref{sec:high}.

% sharding
% Elastico~\cite{luu2016secure},
% OmniLedger~\cite{kokoris2017omniledger}
\section{System Model}\label{sec:model}

\looseness=-1 \xxx is a consensus protocol for a permissioned blockchain running
on $M$ member nodes (\textit{nodes} for short) connected with an
asynchronous network. Each node is equipped with an attested SGX enclave running the \xxx
protocol (\chref{sec:protocol}).

\looseness=-1 \xxx adopts the hybrid fault model used in existing SGX-powered
consensus
protocols~\cite{chun2007attested,veronese2013efficient,zhang2018research}, where
each node has a trusted module (i.e., the SGX enclave) that will only fail by
crashing, and all other components can behave arbitrarily.

\xxx has the following design goals:

\begin{itemize}
    \item \textbf{Safety (consistency)}. \xxx ensures safety
    in an asynchronous network. Formally, if a node confirms a block $b$ as the
    $n^{th}$ block on the blockchain, the probability that another node confirms
    $b'\neq b$ as the $n^{th}$ block is overwhelmingly low ($<10^{-10}$).
    
    \item \textbf{DoS-resistance (liveness)}. In addition to safety, \xxx can
    make progress (i.e., allow non-empty blocks to be confirmed) with two
    additional assumptions about the DoS attacker's capability as described
    below.
\end{itemize}

\smallskip\noindent\textbf{SGX's threat model.} \xxx has the same threat model
for SGX as typical SGX-based systems~\cite{bigmatrix:ccs17,enclavedb:sp18,glamdring:atc17,minbft:sgx,uranus:asiaccs20}. We trust the
hardware and firmware of Intel SGX, which ensures that code and data in an
enclave cannot be seen or tampered with from outside. We trust that the remote
attestation service can identify genuine SGX devices from fake ones (e.g.,
emulated with QEMU). Side-channel and access pattern attacks on SGX are out of
the scope of this paper. Moreover, the adversary cannot break standard
cryptographic primitives, including public-key based signatures and
collision-resistant hash functions.

\smallskip\noindent\textbf{Communication model.} \xxx maintains safety in an
asynchronous network, where network packets can be dropped, delayed, or
reordered arbitrarily. Nodes may be nonresponsive, either due to going offline
or due to targeted DoS attacks (e.g., botnet DDoS attacks~\cite{ddos:report}) by
a DoS attacker. When a node cannot reach a remote node, the node cannot
determine whether the remote node is under DoS attack (or is offline) or the
network packets are delayed. 

% , the adversary can arbitrarily drop or delay
% network packets; the adversary can cause arbitrary nodes to be nonresponsive
% (i.e., targeted DoS attacks) by botnet DDoS attacks~\cite{ddos:report}, ISP
% insider attacks~\cite{duncan2012insider}, or simply shutting down those nodes'
% machines;

\looseness=-1
To achieve liveness, \xxx has the "strong synchrony" assumption, same as
Algorand~\cite{gilad2017algorand}. Specifically, \xxx assumes that messages
between two nodes not under DoS attacks can be delivered within a known time
bound, and the assumptions about the attacker's capability are described below.

\looseness=-1 
Nodes are connected with a P2P overlay network, same as existing large-scale
blockchain systems~\cite{nakamoto2008bitcoin,gilad2017algorand}. Specifically,
each node has a P2P module connecting to a random set of other nodes and relays
messages using the gossip protocol~\cite{kermarrec2007gossiping}. 

A node's P2P module is outside SGX and can be controlled by the attackers: the
attacker can partition some nodes from other
nodes~\cite{natoli2016balance,heilman2015eclipse}) or selectively pass consensus
messages to nodes' SGX enclaves. However, such manipulations are already
included in \xxx's asynchronous network assumption. For safety, \xxx leverages
the sampling merit to estimate the delivery rate of a specific block proposal
and derives overwhelming probability, regardless of how nodes are connected. For
liveness, the adversary can control the P2P modules of a number of nodes with
the restriction of the adversary's attack budget described below. 

\smallskip\noindent\textbf{Capability of DoS attackers.} 
% Each \xxx node has three modules, including an
% \textit{\xxx enclave module} that runs \xxx's consensus protocol and
% consistently maintains \xxx's member list, a \textit{P2P module} that connects
% to a random set of nodes' P2P modules and relays message using the Gossip
% protocol~\cite{kermarrec2007gossiping}, and a \textit{blockchain storage} module
% that stores the node's local blockchain. 
% To maintain liveness, 
\xxx has three assumptions on the capability of a DoS attacker, same as
Algorand~\cite{gilad2017algorand} and existing move target defense (MTD)
systems~\citemoving. First, the adversary has a targeted attack budget $B$
(e.g., $B$ = 300 or 10\% of the total number of nodes): the adversary cannot
constantly cause more than $B$ targeted nodes in \xxx to be nonresponsive. This budget is
adaptive: the adversary can attack different nodes at different times, but the
number of attacked nodes at a time cannot be constantly larger than $B$. 

Second, the attacker can conduct ubiquitous DoS attacks (without targeting
specific nodes) or partition a number of nodes from other nodes (e.g., by
manipulating nodes' P2P modules~\cite{heilman2015eclipse,natoli2016balance}).
However, the P2P overlay network should have a large enough portion (e.g., 65\%)
of nodes connected. We provide a quantitative analysis of how \xxx
can preserve liveness under such attacks in
\chref{sec:analysis:parameter}.

Third, the adversary cannot constantly succeed in mounting an attack 
targeting a node within the time window for the node to send out an \xxx protocol
message. Specifically, \xxx protocol messages are larger than the network
maximum packet size and are fragmented into multiple packets; an \xxx committee
node's identity is unknown to the adversary before sending out the first packet,
and we assume that the adversary cannot mount targeted DoS attacks until the
node sends out all packets belonging to this message (at most hundreds of kB and
can be sent within one second).

% Among these three modules, only the \xxx enclave module runs in SGX. 

% The P2P module and blockchain storage module is outside SGX and can be
% manipulated by the adversary arbitrarily. 

% These two assumptions are already weak enough,
\looseness=-1
\xxx already assumes a strong enough attacker for practical distributed systems on the
Internet. As pointed out by Algorand~\cite{gilad2017algorand}, a more powerful
adversary than our model usually controls the internet service provider and can
prevent all \xxx nodes from communicating at all: no practical system can ensure
liveness under such a strong adversary, and such attacks can be easily detected.
We will provide a rigorous analysis of \xxx's DoS resistance in
\chref{sec:analysis:liveness}.

% \xxx is oblivious to how blocks and transactions are stored in each node.
% However, as confirmed transactions form a hash chain, the attacker cannot
% manipulated the confirmed blocks. 

% \xxx assumes a powerful
% adversary that can control the code outside SGX enclaves of arbitrary \xxx
% nodes, including their local blockchain storage modules, P2P modules, and
% operating systems. 

% \xxx also achieves practical \textbf{liveness} (

\section{\xxx's High-level Idea}\label{sec:high}

% \noindent\textbf{Key management.} 

% Such a design has two benefits in \xxx. First, each consensus messages must be 

% a node does Although an attacker can selectively drop Second, a committee member
% does not need to prove its membership when sending protocol messages as a
% non-committee node cannot forge a valid consensus messages. 

% By doing so, \xxx
% fault tolerant (BFT) problem for a permissioned blockchain into a \textbf{crash
% fault tolerant (CFT)} problem. 

\xxx has three important features to achieve DoS resistance. First, \xxx
randomly selects a distinct group of committee for each block. The selection is
done inside the SGX enclaves of a previous committee, and the selection result is
encrypted on the confirmed blockchain. By doing so, a committee node can
determine its committee membership without interactions with other nodes, making
it \textit{stay stealth} before trying to achieve consensus on its block. 

Second, when a committee is achieving consensus for a given block, \xxx uses
fake committee nodes to conceal the real ones by sending dummy messages. Since
whether a node is a real committee node is only known within the node's SGX
enclave, and the dummy messages are of the same format as real ones, a DoS
attacker cannot distinguish the real committee nodes from the fake ones.
Therefore, the attacker must have an unrealistic large attack budget to attack
all the real and fake committees; otherwise, he has to randomly guess who are the real ones. 
% ones and attack them, or he needs an unrealistic large 

Third, even if the attacker luckily guesses the real ones (and he may
eventually succeed if tried persistently), \xxx can ensure safety with
overwhelming probability leveraging the delivery rate on all nodes of the unique
proposal for each block. Specifically, even if a committee cannot confirm its
own block, committees for subsequent blocks can help to consistently confirm
this block by repetitive querying. This feature is in contrast to most existing
consensus protocols (i.e., all except Algorand~\cite{gilad2017algorand}), where
the system must wait statically until a quorum of nodes become reachable. 
% derive a new consensus protocol that can ensure safety with overwhelming probability 
% even 

% inside SGX and encrypt the committee identities in the confirmed
% blockchain, predictable way, avoiding the DoS attacker targeting the committee
% of a specific node. Second, \xxx uses fake 

% First, 
% \xxx uses fake committee nodes to conceive the 

% \xxx select
% a distinct committee for each block, avoids the attacker 
%  Second, 

\xxx's consensus protocol has three parameters, $n_A$ (default 300), $\tau$
(default 59\%), and $D$ (default 4), where $n_A$ is the number of acceptors,
$\tau$ is the quorum ratio, and $D$ is the finalize depth for an empty block. We
will show how to select these parameters in \chref{sec:analysis:parameter} and how 
these parameters affect \xxx's performance in \chref{sec:eavl:sensitiviy}.

For each block index $n$, \xxx selects (\chref{sec:protocol:select}) only one
committee from all nodes. The committee contains two types of nodes: \textbf{one
proposer} ($P_n$) and \textbf{a group of stealth acceptors} ($\mathbb{A}_n$)
with the count of $n_A$. \xxx ensures the following invariant (see
\chref{sec:analysis:safety} for proof).

\begin{invariant}\label{invariant:proposal} 
    
    For any block index $n$, at most one unique block proposal (\proposal{n}) is
    generated; a node can only confirm \proposal{n} or a default empty block
    (\emp{n}) as its \nth{n} block.
    
\end{invariant}

% \begin{table}[t]
%     \small
%     \begin{tabularx}{\columnwidth}{lX}
%     \hline 
%     % \multicolumn{2}{c} \\\hline
%     \v{chain} & the node's local blockchain \\
%     $\mathbb{U}$ & indices of unknown blocks on \v{chain}\\
%     \v{MC} & index of the last confirmed block on \v{chain} \\
%     learntProposals & proposals learnt for blocks in $\mathbb{U}$ \\
%     \v{cache} & a map saving received propose requests \\\hline
%     \end{tabularx}
%     
%     \caption{Each node's local state.}\label{tab:localstate}
%     
% \end{table}
\myfig{
\begin{figure}[t]
    \centering
    \includegraphics[width=0.95\columnwidth]{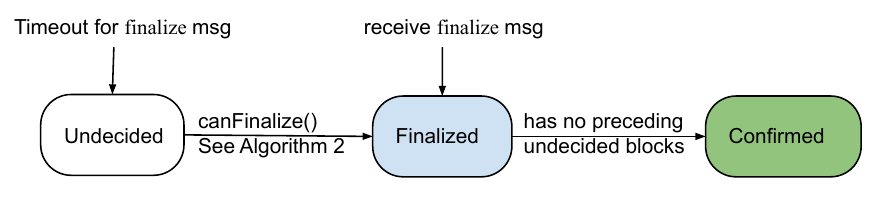}
    \caption{\xxx's block status diagram.}\label{fig:block}
\end{figure}
}

\xxx uses different steps to confirm \proposal{n} or \emp{n}. We make the steps
of confirming \proposal{n} as lightweight as possible for high efficiency in
gracious runs.

\smallskip\noindent\textbf{Confirming $\bm{\mbox{proposal}_n}$.} In a gracious
run, \xxx confirms \proposal{n} in three steps: $P_n$ broadcasts \proposal{n}
with a \v{propose} request through the P2P network; acceptors send \v{ACK}s to
$P_n$ after receiving \proposal{n}; $P_n$ broadcasts a \v{finalize} message
confirming \proposal{n} once receiving \v{ACK}s from a quorum ($\tau\times n_A$)
of acceptors.

\smallskip\noindent\textbf{Confirming $\bm{\mbox{empty}_n}$}. For consistency, when
confirming \emp{n}, \xxx must ensure that no node has confirmed \proposal{n}.
Existing consensus protocols on static nodes use view-change
protocols~\cite{raft:usenix14,
pbft:osdi99,buchman2016tendermint,yin2019hotstuff} that count how many nodes
have sent out \v{ACK}s and leverage the quorum intersection
property~\cite{howard2016flexible} to determine whether a proposal may have been
confirmed. However, in \xxx, this method is not viable because \xxx must ensure
liveness even if most nodes in $\mathbb{A}_n$ are DoS attacked after sending out
their \v{ACK}s.

\looseness=-1 
\xxx's protocol for confirming \emp{n} leverages the idea of
repeated sampling to predicate that the probability of \proposal{n} having been
confirmed is overwhelmingly low. If \proposal{n} is confirmed on some nodes,
\proposal{n} should have been delivered to a large-enough portion of nodes in the
P2P network; this is because confirming if \proposal{n} needs quorum
\v{ACK}s from nodes in $\mathbb{A}_n$, and $\mathbb{A}_n$ is uniformly
selected from all nodes. Therefore, if one repetitively (>$D$) samples many
nodes (>$\tau\times n_A$) from all nodes, and no node has received \proposal{n},
she can predicate only a small portion of (or no) nodes have received
\proposal{n}, and thus the probability of \proposal{n} having been confirmed is
overwhelmingly low. 

To be DoS-resistant, these multiple rounds of checking must be initiated by
different nodes, so \xxx lets the proposers for subsequent blocks (i.e.,
$P_{n+1}$, $P_{n+2}$, etc.) do such samplings at the same time of seeking
\v{ACK}s for their own proposals. Suppose $P_{n}$ failed before sending out
\proposal{n}, the next 4 proposers will all report that no node has received
\proposal{n}, so \emp{n} can be confirmed as empty together with
\proposal{n+1}$\sim$\proposal{n+4}.

\bfheading{Comparison with Algorand.} \xxx and Algorand both select a distinct
committee for each block in an unpredictable way, and both use the delivery rate
of a block proposal to confirm a block. Algorand leverages its built-in
cryptocurrency to incite committee nodes to follow its protocol (i.e., proof of
stake). However, even if one runs Algorand within SGX in a permissioned
blockchain, there are still two major design differences making \xxx more
efficient than Algorand. 

First, Algorand uses verifiable random functions (VRF) to determine committees,
so it can only control the expected count of proposers for each block without an
exact number ($1\sim70$ in their experiment~\cite{gilad2017algorand}). This
design makes Algorand's consensus protocol not responsive~\cite{yin2019hotstuff,
libra:smr}: informally, a responsive protocol lets nodes wait for a number of
messages rather than a large amount of time in each protocol step, which ensures
a good performance when the network is in good condition. For each
block, Algorand selects $1\sim70$ proposers, and each proposer broadcasts a
block proposal with a distinct priority level. Then, Algorand selects one of
these proposals by letting nodes vote for the received proposal with the highest
priority. Since the total number of proposers is unknown, each node must wait
for a conservatively long time (e.g., 10s) before voting to ensure it received
most proposals. In contrast, \xxx selects one proposer for each block, without
the necessity for the selection progress, and \xxx's protocol is responsive
(\chref{sec:protocol}).

Second, \xxx adopts an optimistic design while Algorand adopts a pessimistic
design. Specifically, Algorand uses a heavy step for both confirming a
non-empty block and confirming an empty block. In contrast, \xxx optimistically
makes its gracious runs (i.e., confirming \proposal{n}) fast and shifts the
burden of maintaining consistency to the rare failure cases (i.e., confirming
\emp{n}). 

% Moreover, \xxx lets the confirmation of \emp{n} run in
% parallel with the consensus of subsequent blocks to minimize the disturbance on
% the performance. 

\bfheading{Why does \xxx use SGX?} \xxx chooses to use SGX for three main
reasons. First, \xxx uses SGX to regulate the behaviors of randomly selected
committee nodes. As discussed in \chref{sec:background:public}, using randomly
selected committees in a permissioned blockchain may result in an unsolvable
byzantine problem where more than one-third of committee nodes are
malicious~\cite{lamport1982byzantine}. 

\looseness=-1
In \xxx, each node have a private key that is
\textit{only visible} within the node's SGX enclave, and the corresponding
public key works as the node account saved in all other nodes' SGX enclaves
(\chref{sec:impl:membership}). A valid protocol message must carry a valid
signature using the sender node's private key, proving that the message is
generated in the sender node's enclave with code integrity. By doing so, a node
cannot equivocate (i.e., sending conflicting messages to different other nodes)
or forge protocol messages (e.g., a proposer sending out a \v{finalize} message
without receiving quorum \v{ACK}s). 

Second, \xxx leverages SGX to make its committee's identities stealth: only a
node's enclave knows whether itself is a committee member for the current block.
This not only enables \xxx to maintain practical liveness under DoS
attacks but, more importantly, makes \xxx's consistency model resistant to
targeted attacks. Specifically, \xxx leverages probability theory to model
randomly selected acceptors as a uniform sampling of the delivery rate of a
block proposal in the P2P network (same as Algorand~\cite{gilad2017algorand}).
If acceptors' identities are public, an attacker can selectively transfer or
drop packets towards them, breaking \xxx's safety. 

Third, the usage of SGX enables \xxx to select a stealth committee with a known
count. As discussed above, this makes \xxx more efficient than Algorand's. 
%  in contrast to only an expected count if selected with VRF, which makes \xxx
% more efficient than Algorand, as discuss above. 

% Specifically, 
% As will be
% explained in \chref{sec:background:public}, selecting one single proposer has
% the potential to build more efficient DoS-resistant consensus protocols than
% selecting multiple proposer; we will also discuss in
% each block has better liveness than only controlling the expected number of
% acceptors.  

\section{\xxx Consensus Protocol}\label{sec:protocol}

\subsection{Protocol Preliminaries}\label{sec:protocol:pre}

\textbf{Block structure.} 
\xxx adds one data field to the block structure of common blockchain
systems~\cite{buterin2014ethereum,nakamoto2008bitcoin}: the encrypted committee
identities for a future block (\chref{sec:protocol:select}). \xxx is oblivious
to how transactions are stored or executed.

\bfheading{Each node's local states.}
Each \xxx node maintains three major local states: a local blockchain (the
\v{chain}), a proposal cache (the \v{cache}), and a set of \v{learntProposals},
The \v{cache} is maintained in the node's \xxx enclave. When the node receives
\proposal{n}, it puts the proposal into the \v{cache}, in case the committees of
future blocks query the delivery rate of \proposal{n}.\looseness=-1

\bfheading{Block status.}
Each block in a node's \v{chain} has three states: \textit{undecided},
\textit{finalized}, and \textit{confirmed}. An undecided \nth{n} block can only
be \emp{n}. A node appends \emp{n} to its \v{chain} when the node triggers a
timeout waiting for the \v{finalize} message for the \nth{n} block; the
block is in the undecided state because the node cannot determine (for now)
whether it should confirm \proposal{n} or \emp{n}.

A finalized \nth{n} block can be either \proposal{n} or \emp{n}. \xxx ensures
that, if a node's \nth{n} is finalized as one choice, no other nodes will
finalize the \nth{n} block as the other choice. A node appends finalized
\proposal{n} if it receives the \v{finalize} message for \proposal{n}; a node
changes the \emp{n} from the undecided state to finalized if the node can
predicate that no node has finalized \proposal{n} (\chref{sec:protocol:confirm}).

A node confirms its finalized \nth{n} block if all blocks with indices smaller
than $n$ in its \v{chain} are finalized.
% , and transactions in a confirmed block can be executed. 
Note that although \xxx may finalize blocks out of order, \xxx confirms blocks
sequentially, same as typical blockchains~\cite{buterin2014ethereum,
androulaki2018hyperledger}.  \looseness=-1

\looseness=-1
The \v{chain} on each node is divided into two parts: the confirmed part and the
unconfirmed part. We use \v{MC} to represent the maximum confirmed index and
$\mathbb{U}$ to represent the indices of undecided blocks in \v{chain}. The
confirmed part of \v{chain} (i.e., indices$\leq$\v{MC}) are
cryptographically-chained by hash values and can be saved out of the enclave and
get executed, while unconfirmed parts are saved in the \xxx enclave. 

The \v{learntProposals} is the set of known proposals for undecided blocks on this
node and is saved in \xxx enclave. \looseness=-1

\smallskip
\bfheading{Membership and key managements.} In \xxx, each node $i$ has a key
pair $\langle pk_i, sk_i \rangle$, with the public key $pk_i$ as its account,
and its secret key $sk_i$ is \textit{only visible within its enclave}: even this
node's administrator cannot see the plain-text of its secret key. We use the
notations from PBFT~\cite{pbft:osdi99}: we denote a message $m_1$ sent to node
$i$ encrypted by $i$'s public key $pk_i$ as $\{m_1\}_{pk_i}$; we denote a message
$m_2$ generated by node $i$'s enclave and signed by $sk_i$ as $\langle
m_2\rangle_{\sigma_i}$. For efficiency, \xxx only signs on message digests.

% In \xxx, each node's enclave uses its secret key to sign every protocol message
% generated from it. This design ensures that only a true committee node for a
% block can generate a valid protocol message for this block (e.g.,
% \v{ACK}s), where a valid message is a message associated with a valid
% signature using a node's private key. This is because \xxx's consensus protocol
% runs within each node's enclave for integrity, and a node's enclave invokes the
% functions for generating protocol messages for the \nth{n} block only after
% confirming its committee membership. This simple design eliminates the necessity
% of an \xxx committee node to prove its committee membership when sending a
% consensus message. We will cover \xxx's implementation in
% \chref{sec:impl:enclave}.

To ease understanding, in this section, we assume that \xxx runs on a fixed
membership, where all nodes' accounts (public keys) are loaded
to nodes' \xxx enclaves a priori, and all nodes' \xxx enclaves are
attested. We will show how \xxx supports dynamic membership and attestations in
\chref{sec:impl:membership}.

\myfig{
\begin{figure*}
\fbox{
\begin{minipage}{0.97\textwidth}
\begin{multicols*}{2}
\footnotesize
% \small
\begin{tabularx}{\columnwidth}{lX}
\hline
\multicolumn{2}{c}{\bf Propose} \\
\Xhline{2\arrayrulewidth}
n & The index of the proposal \\
blk & the content of the block to be proposed \\
$\v{MC}_p$ & the \v{MC} value of the proposer \\
{\color{darkblue} $\mathbb{U}_p$} & The $\mathbb{U}$ list of the proposer \\
{\color{darkblue} \proposal{u_m}} & $u_m$ = max($\mathbb{U}_p$). The proposer tries to finalize
this proposal together with its proposed \nth{n} block \\
\hline
\multicolumn{2}{c}{\bf ACK} \\
\Xhline{2\arrayrulewidth}
n & The index of corresponding proposal \\
sender & sender's public key (account) to identify it \\
{\color{darkblue} notifications} & proposals that the sender received and want to notify the
proposer  \\
isReal & for identifying real ACK from cover messages \\
nonce & random padding to make cipher text unpredictable \\
\hline
\multicolumn{2}{c}{\bf Finalize} \\
\Xhline{2\arrayrulewidth}
n & The index of the block finalized \\
{\color{darkblue} $u_m$} & The index of the block that is finalized together \\
{\color{darkblue} learnt} & Proposals learnt from acceptors notifications \\
% status & confirm or pending \\
\hline
\end{tabularx}
% \end{table}
\captionof{table}{\xxx's consensus messages' fields. Blue fields are only used
in the checking mode and are left as \v{nil} in the normal mode.}\label{tab:messages}

\vfill
\input{proposer.tex}
\newpage
\input{allnodes.tex}
\vfill
\input{acceptor.tex}
\end{multicols*}
\end{minipage}}
\caption{\xxx's consensus protocol.}
\end{figure*}
}

\subsection{Selecting a Stealth Committee}\label{sec:protocol:select} For each
block, \xxx selects a committee, including \textbf{one proposer} and
\textbf{$\bm{n_A}$ acceptors}, in an unpredictable way without communication
among nodes.

The committee members for the \nth{n} block is selected in the \xxx enclave of
$P_{n-lb}$, and these committee nodes' identities are encrypted in the
\nth{(n-lb)} block. $lb$ (look-back) is a system parameter and needs to be large
enough (e.g., the number of blocks confirmed in days) to ensure that even when
the network condition is poor, and new blocks cannot be confirmed in time, \xxx
can still derive committees for future blocks. We assume that the first $lb$
committees' identities are encrypted in the genesis (\nth{0}) block by a trusted
party, or that the blockchain is bootstrapped in a controlled domain for at
least $lb$ blocks. Note that the value of $lb$ does not affect \xxx's safety. 

Occasionally, a node may be selected as the committee for a future block and
then leave the system, which \xxx already tolerates as a failed node. If the
\nth{(n-lb)} block happens to be an empty block, \xxx uses the committee
identities encrypted in the \nth{(n-2lb)} block (and identities in the
\nth{(n-3lb)} block if the \nth{(n-2lb)} block is also empty, recursively).
Although this proposer's identity is already explicit when confirming the
\nth{(n-lb)} block and may be targeted, \xxx can tolerate it as a failed
proposer and uses subsequent committees to confirm \emp{n}.

$P_{(n-lb)}$ selects the committee for the \nth{n} block with two steps, which
are done in $P_{(n-lb)}$'s \xxx enclave to ensure both integrity (i.e., an
attacker cannot control the selection) and confidentiality (i.e., an attacker
cannot know the selection result). In the first step, $P_{(n-lb)}$ randomly
selects the committee members from all member nodes following the uniform
distribution. Recall that the member list is loaded in the \xxx enclave on each
node (\chref{sec:protocol:pre}), so $P_{(n-lb)}$ simply selects $n_A + 1$ nodes
from the list using the SGX's trustworthy pseudo-random number generator as the
random source, which has been shown to be cryptographically-secure and cannot be
seen or tampered with from outside enclave (\chref{sec:background:sgx}).

\looseness=-1
In the second step, for each selected committee node, $P_{(n-lb)}$ generates one
certificate, which is the cipher-text of the concatenation of a predefined byte
string and a random nonce (for making the cipher-text unpredictable), encrypted
with that committee node's public key. Then, $P_{(n-lb)}$ includes these
$(n_A+1)$ certificates in the \nth{(n-lb)} block's proposal. The first
certificate is for the proposer, and the other $n_A$ certificates are for
acceptors. When a node confirms this block, it tries to decrypt one of these
certificates using its own secret key in its enclave; if the node can get the
predefined string, it predicates that it is a committee node for the \nth{n}
block. 

Despite using asymmetric cryptography, this mechanism is efficient in \xxx
because both encryption and decryption are done asynchronously off the consensus's
critical path. For encryption, since $P_{(n-lb)}$'s enclave knows it is the
proposer for the \nth{(n-lb)} block after confirming the \nth{(n-2lb)} block,
$P_{(n-lb)}$ starts selecting the committees and encrypting the certificates as
soon as confirming the \nth{(n-2lb)} block. Similarly, the decryption is also off the
critical path as the decryption result is used $lb$ blocks later.

\looseness=-1 \xxx's committee selection mechanism is unpredictable and
non-interactive because: (1) the random source cannot be seen or tampered with
from outside the enclave of $P_{n-lb}$, and the certificates can only be
verified within a selected committee node's enclave; and (2) the selection
process is solely done within the \xxx enclave of $P_{n-lb}$. These two features
ensure the committee nodes' identities are not exposed during the selection, so
the committee nodes cannot be targeted before sending out protocol messages for
the \nth{n} block.

\bfheading{Discussions.} 
\xxx selects only one proposer for each block to achieve good efficiency: \xxx
only needs to achieve a binary consensus on whether to confirm the unique
proposal by this proposer or a default empty block. For acceptors, however, an
alternative design is to let each node independently determine whether it is an
acceptor for the current block with a probability, and \xxx only knows an
\textit{expected} total count. However, this alternative design will lead to a
much larger quorum ratio (i.e., $\tau$) to ensure safety and thus worse
liveness (quantitative analysis in \chref{sec:eavl:sensitiviy}).

\subsection{Confirming a Block}\label{sec:protocol:confirm}

A proposer $P_n$ has two operation modes: \textit{normal} mode and
\textit{checking} mode. $P_n$ is in the normal mode if all blocks in its
\v{chain} before $n$ are confirmed (i.e., $\mathbb{U} = \emptyset$), and $P_n$
tries to confirm \proposal{n} quickly. Otherwise, $P_n$ is in the checking mode:
while proposing \proposal{n}, it also checks the status of the undecided blocks in
its \v{chain}.
\looseness=-1

\looseness=-1
\bfheading{Normal mode.}
\algostartend{algo:proposer}{proposer:normal:start}{proposer:normal:end} shows
how a normal mode $P_n$ tries to confirm \proposal{n} in a gracious run.
% the workflow of a normal mode proposer with three steps. 
First, $P_n$ broadcasts a \v{propose} request through the P2P network carrying
\proposal{n} and its \v{MC}. The \v{MC} value helps nodes align confirmed parts
of their \v{chain}: if a node's \v{MC} is smaller than the proposer's, the node
asks for the missing confirmed blocks from its peers
(\algoline{algo:allnodes}{allnode:askfor}). Upon receiving this \v{propose}
request, an acceptor replies an \v{ACK} using UDP directly to $P_n$
(\algoline{algo:acceptor}{acceptor:reply}). Second, $P_n$ waits for quorum
($\tau\times n_A$) \v{ACK}s from $\mathbb{A}_n$. $P_n$ does not know which nodes
are acceptors, but \xxx's ensures that a non-acceptor cannot send valid \v{ACK}s
(\chref{sec:model}). Third, $P_n$ broadcasts a \v{finalize} message; on
receiving this message, a node finalizes \proposal{n}.

\bfheading{Checking mode.} 
$P_n$ is in the checking mode if it has undecided blocks (i.e., $\mathbb{U}$ is
non-empty), and its workflow is shown in
\algostartend{algo:proposer}{proposer:checking:start}{proposer:checking:end}.
$P_n$ checks the status of its undecided blocks and tries to finalize
them (if possible) by adding additional fields to the \v{propose} message.
% seeking \v{ACK}s for \proposal{n}.

Each $u\in\mathbb{U}$ in $P_n$'s \v{chain} is categorized into one of the two
types: (1) if $P_n$ has learnt the unique \proposal{u}, either from the
\v{propose} messages from $P_u$ from the notifications of other nodes, we call
$u$ a ``known undecided'' block; (2) otherwise we call $u$ ``unknown
undecided".
% proposal i.e., undecided undecided blocks), and (2) those blocks that $P_n$ has
% learnt their corresponding unique proposals but these blocks are not finalized
% (known undecided block). 
For each unknown undecided block $u$, $P_n$ tries to learn \proposal{u} from
$\mathbb{A}_n$. If $P_n$ learns the proposal, $P_n$ carries it in the
\v{finalize} message in order to let subsequent proposers finalize it;
otherwise, $P_n$ carries a message stating that most acceptors in $\mathbb{A}_n$
never received \proposal{n}, and a node receiving this message finalize
\emp{n} if the node received such messages form more than $D$ consecutive
proposers (\algoline{algo:allnodes}{allnode:canfinalize:start}). 

For known undecided blocks, $P_n$ helps to finalize only \proposal{u_m} where
$u_m = max(\mathbb{U})$ and leaves other blocks for subsequent proposers. We
will explain later that only finalizing $u_m$ is essential for \xxx's safety
(\fig{fig:corner}).

\fig{fig:help} shows how a proposer $P_{102}$ helps to finalize an undecided
\proposal{100}. Suppose $P_{100}$ failed just before broadcasting its
\v{finalize} message. Therefore, the \nth{100} block's state is undecided
among all nodes. Then, $P_{101}$ learns \proposal{100} and carries it in its
\v{finalize} message, and $P_{102}$ learns it. Then $P_{102}$ finalizes
\proposal{100} together with \proposal{102}. Moreover, since all blocks
before 102 are finalized, the chain is confirmed up to 102. 

\fig{fig:empty} shows how an undecided block is finalized as empty. Suppose
$P_{200}$ failed before broadcasting its proposal, and $D=4$. When $P_{201}\sim
P_{204}$ asks whether their acceptors ($\mathbb{A}_{201}\sim\mathbb{A}_{204}$)
receive \proposal{200}, they get no positive answers. Therefore, these four
blocks are all finalized, carrying a message stating that four samplings have
been conducted on the delivery rate of \proposal{200}, but \textit{no replied
node has received} \proposal{200}. This indicates that the probability that
\proposal{200} is finalized at some nodes is overwhelmingly low. Therefore, a
node can \textit{independently} finalize \emp{200}, after which the
\v{chain} is confirmed to 204. 

\myfig{
\begin{figure*}[t]
    \centering
    \begin{minipage}[t]{.38\textwidth}
        \centering
    \includegraphics[width=\textwidth]{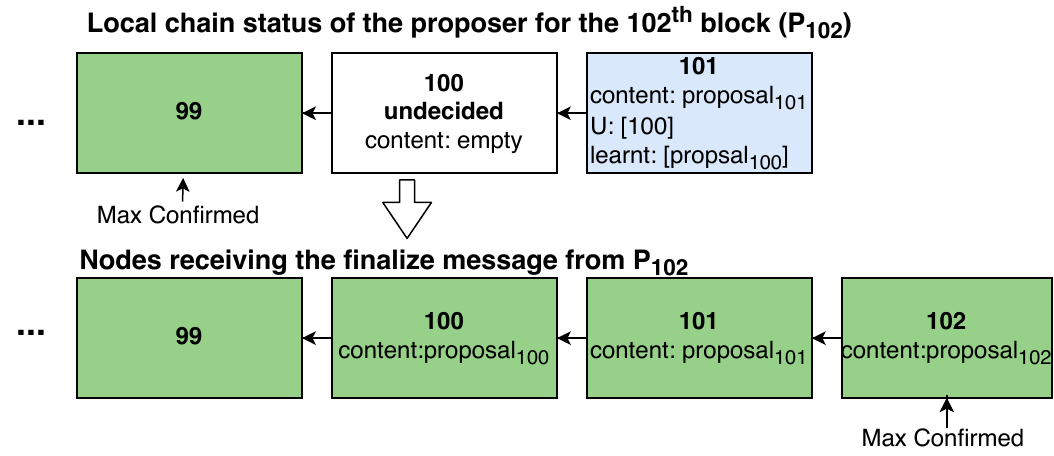}
    \caption{An example where $p_{102}$ helps finalizing \proposal{100}  
        while proposing its (\nth{102}) block. }
    \label{fig:help}
    \end{minipage}
    \hfill
    \begin{minipage}[t]{.55\textwidth}
        \centering
            \includegraphics[width=\textwidth]{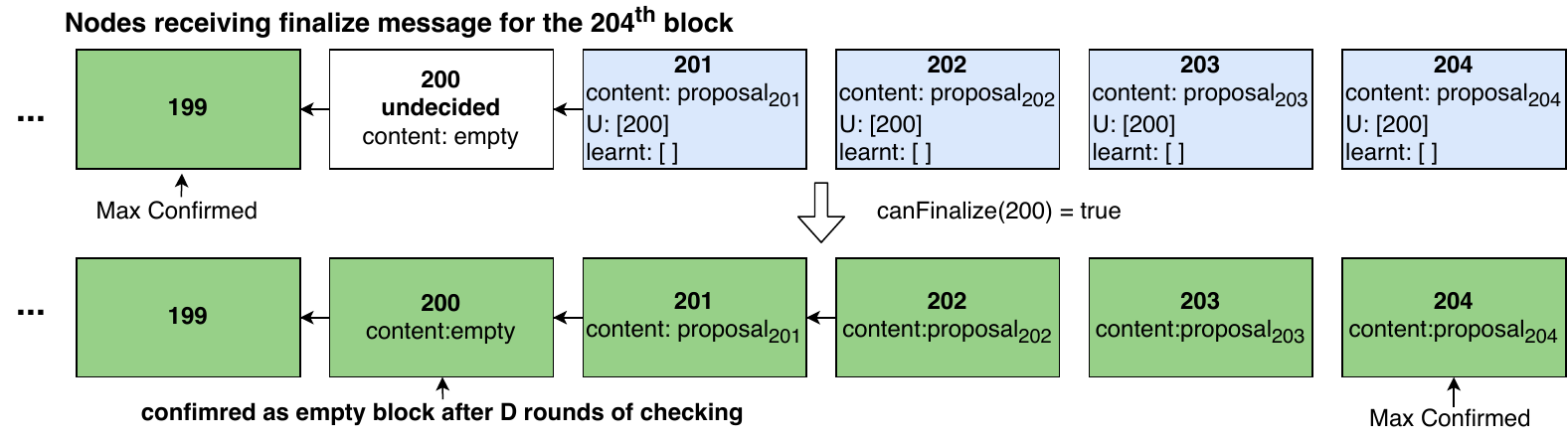}
            \caption{\small An example for confirming an empty block (\nth{200})
            after $D=4$ succeeding blocks containing 200 in $\mathbb{U}$ are
            finalized.}
            \label{fig:empty}
    \end{minipage}
    \begin{minipage}[t]{1\textwidth}
        \centering
        \includegraphics[width=\textwidth]{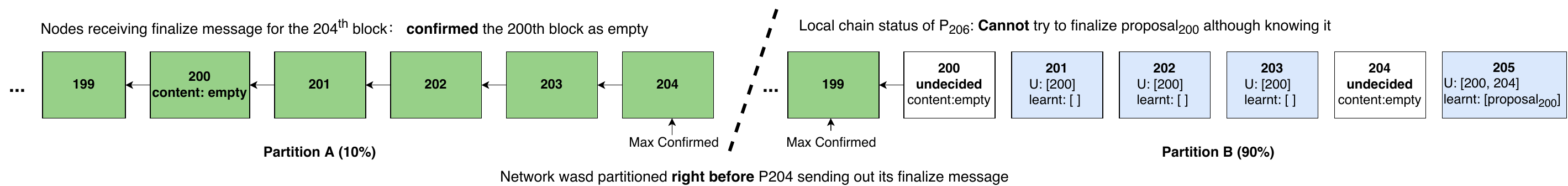}
        \caption{An example showing why a checking mode proposer can finalize only the block proposal with index = $max(\mathbb{U})$. $D=4$ in this example.}       \label{fig:corner}
    \end{minipage}
\end{figure*}
}

% Recall we mentioned that one caveat is a . 
\fig{fig:corner} shows why it is essential that a checking mode proposer can
only finalize \proposal{u_m} where $u_m = max(\mathbb{U})$. Suppose we remove
this restriction, and we consider the following scenario. (1) $P_{200}$ fails
after broadcasting \proposal{200}, and only very few nodes received it: none
of $P_{201}\sim P_{204}$ learns \proposal{200}. (2) The network is divided
into two partitions A\&B just before $P_{204}$ broadcasting its \v{finalize}
message; $P_{204}$ is in partition A, so nodes in partition A confirm
\proposal{204} and confirm \emp{200} (3) $P_{205}$ and $P_{206}$ are in
partition B, so they timeout waiting for the \v{finalize} message for
\proposal{204} and mark the \nth{204} block as undecided. (4) $P_{205}$ learns
\proposal{200} from one node from $\mathbb{A}_{205}$ (who happens to be in the
very few nodes) and carries \proposal{200} with its \v{finalize} message,
which is learnt by $P_{206}$. (5) $P_{206}$ finalizes \proposal{200} and
causes inconsistency: nodes in partition A confirm \emp{200}, while nodes in
partition B confirm \proposal{200}.

% which is then learnt by . If we remove
% this restriction, nodes in partition A finalizes and confirms the \nth{200}
% block as empty, but $P_{206}$ may succeeds finalizing \proposal{200} in
% partition B, causing inconsistency. 
% 
% The rationale behind this restriction is that,

The inconsistency happens because, without this restriction, when a node
finalizes \emp{n}, it only ensures proposers for blocks with index $\leq n+ D$
has not finalized \proposal{n}; adding this restriction helps to ensure that
proposers for blocks with index $> n +D$ cannot finalize \proposal{n}.
Specifically, if this restriction presents in the previous example, nodes in
partition B need to first finalize the \emp{204} finalizing
\proposal{200}. However, since \proposal{204} has already been delivered to a
large portion of nodes, this inconsistency cannot happen. We show the proof in
\chref{sec:analysis:safety}.

\subsection{Handling DoS Attacks Targeting
Proposers}\label{sec:protocol:arbiter} In \xxx, a proposer stays stealth before
proposing its block, but its identity becomes explicit after broadcasting its
proposal. If the proposer is DoS attacked at this time, this block cannot be
finalized in time, impairing \xxx's liveness. 

% This attack window is roughly the
% time for one P2P broadcast and half UDP round-trip (less than 2 seconds in our
% evaluation). Despite small, this attack window may still be exploited by
% powerful attackers. 

To mitigate this problem, we propose a new role of nodes called \v{arbiter}. An
arbiter for a block index $n$ does \textit{not} generate new block proposals but
only helps the proposer to finalize its proposal. For each block index $n$, \xxx
has many arbiters (larger than the attack budget $B$) doing the same tasks to
tolerate DoS attacks on them. 

Since we do not need to know the exact number of arbiters count and their
identities, \xxx simply let each node's \xxx enclave independently determine
whether it is an arbiter for the \nth{n} block with a probability of $p_a$ after
receiving \proposal{n}. An arbiter for the \nth{n} block broadcasts an
\v{arbit} request  
following the same protocol as the proposer (\algo{algo:proposer}), and an
acceptor responds to an \v{arbit} message with the same logic responding to
a \v{propose} message.

% However, arbiters of the same block may receive
% \v{ACK}s from different quorums of acceptors and make different decisions.
% This does not affect consistency for finalizing a proposal but may cause
% inconsistency for finalizing an empty block. In the example in \fig{fig:empty},
% if an arbiter for the \nth{204} block happens to receive \proposal{200}, it
% may incur inconsistency. Therefore, one more alteration is that when a node
% tries finalizing an empty block by counting succeeding blocks
% (\algostartend{algo:allnodes}{allnode:canfinalize:start}{allnode:canfinalize:end}),
% it skips blocks finalized by an arbiter.

\bfheading{Discussions.} With the help of arbiters, a proposer's critical task
is only to send out its \v{propose} request, and the arbiters can help to
finalize it. However, responding to both proposer and multiple arbiters makes
acceptors targets of DoS attacks. Therefore, \xxx lets normal nodes also
randomly send out fake (dummy) \v{ACK}s to cover the real acceptors.
(\algoline{algo:allnodes}{allnode:fake}). Since real or fake ACKs are all
encrypted with $P_n$'s public key, only $P_n$'s \xxx enclave can decrypt them
within its enclave and distinguish the real ones, so an attacker cannot know 
who are the real acceptors. 
\section{Security Analysis}\label{sec:analysis}
\subsection{Safety with Overwhelming Probability}\label{sec:analysis:safety}
\xxx ensures safety with overwhelming probability (i.e., $>1- 10^{-10}$).
Formally, if a node confirms a block $b$ as the $i^{th}$ block on the
blockchain, the probability that another member node confirms $b'\neq b$ as the
$i^{th}$ block is $< 10^{-10}$. 

We prove the safety guarantee of \xxx by induction: suppose \xxx guarantees
safety from the $0^{th}$ block to the $(n-1)^{th}$ block ({\color{darkblue}
hypothesis 1\phantomsection\label{hypo:first}}), and we prove that there is only
one unique block that can be confirmed as the $n^{th}$ block among nodes in the
blockchain. The base case is trivial because all nodes start from the same
\nth{0} block. 

\begin{lemma}
% ({\color{darkblue} Invariant~\ref{invariant:membership}}), 
if two nodes have the same maximum confirmed block in their \v{chain} (i.e.,
\v{MC} = $n-1$ due to {\color{darkblue} hypothesis 1}), then during
consensus for the \nth{(n+i)} block where $i\geq0$, as long as \v{MC} is not
changed, these two nodes see the same member list. 
\end{lemma}

\begin{proof}
Proving this lemma is trivial if \xxx works on a
fixed member list, and we will show in \chref{sec:impl:membership} that
\xxx's protocol for dynamic memberships also ensures this lemma.
\end{proof}

\begin{lemma}
({\color{darkblue} Invariant~\ref{invariant:proposal}} in
\chref{sec:high}): at most one proposal can be generated for the
\nth{n} block. 
\end{lemma}

\begin{proof}
This lemma is proved by two steps. First, as the proposer for
the \nth{n} block is encrypted in the \nth{(n-lb)} block, and the \nth{(n-lb)}
block is the same among nodes because of \hyperref[hypo:first]{hypothesis 1}.
Therefore, there is only one proposer (may have failed) for the \nth{n} block.
Second, this proposer generates at most one proposal, and non-proposer nodes cannot
generate valid proposals for the \nth{n} block because \xxx's consensus module
runs in SGX. 
\end{proof}

\bfheading{Proof of the induction step.} 
% On each node, the \nth{n} block can only be finalized as either the unique
% \proposal{n} or an empty block (Lemma 2), and 
In \xxx, each block has only two choices ({\color{darkblue}Lemma 2}), and a
confirmed block must be first finalized (\chref{sec:protocol:pre}). Therefore,
it is sufficient to prove the following
{\color{darkblue} proposition 1\phantomsection\label{proposition}}: the
probability that one node finalizes \emp{n} (event A), and another node
finalizes \proposal{n} (event B) is overwhelmingly small.

\looseness=-1
For event A, suppose node $X$ finalizes \emp{n}. We use $f_{m_x}$ to denote
the maximum finalized block index on node $X$. Consider blocks with indices in
$[n+1,f_{m_x}]$. Since \xxx finalizes \textit{empty} blocks in descending order
(\algostartend{algo:allnodes}{allnode:descending:start}{allnode:descending:end}),
% we can predicate that 
there are no undecided blocks 
% (i.e., either finalized as
% proposal or finalized as empty) 
in $[n+1,f_{m_x}]$, and we can have another
level of induction by supposing blocks finalized as empty in $(n+1, f_{m_x}]$
are finalized consistently (name it {\color{darkblue} hypothesis
2\phantomsection\label{hypo:backfinalize}}). For event B, \proposal{n} can be
finalized either by $P_n$ (call it event B1) or by subsequent proposers that
have learnt this proposal (event B2). 

\looseness=-1
First, we prove that the probability that event A and event B1 happen together
is overwhelmingly low. Suppose that a portion $p$ of all $M$ \xxx nodes
received and cached the \proposal{n}, and we calculate the probability for
event B1. We use \texttt{Re} to denote the number of acceptors for the \nth{n}
block that REceived \proposal{n}. Since \proposal{n} is broadcasted in \xxx's
P2P network and the stealth acceptors are selected uniformly, \texttt{Re}
follows hypergeometric distribution $\texttt{Re}\sim H(M, n_A, p\times M)$.
Thus, the probability that $P_n$ finalizes \proposal{n} is 
$$Prob (B1) = Prob(\texttt{Re}>\tau\times n_A)$$ 
We then calculate the probability
of event A. Event A infers that after the 
\nth{n} block,
there are at least $D$ non-empty blocks that are finalized and carrying $n$ in the undecided list. This
means each proposer of these $D$ blocks received $(\tau \times n_A)$ ACKs from
their acceptor group, and none of these acceptors sending those ACKs has
received \proposal{n}. For each of the $D$ blocks, the number of acceptors
$\texttt{NR}$ not receiving \proposal{n} follows hypergeometric distribution
$\texttt{NR}\sim H(M, n_A, (1-p)\times M))$.

\smallskip
\noindent
Therefore, the probability of event A is $$Prob(A) = (Prob(\texttt{NR} > \tau
\times n_A))^D$$ The calculation shows that the probability of event A and event B1
happening together $Prob(A) \times Prob(B1) $ is overwhelmingly low for any
delivery rate $p$ by setting $\tau$ and $n_A$
(\chref{sec:eavl:sensitiviy}). For instance, our evaluation chose $\tau$ as
59\%, $D$ as 4, $n_A$ as 300, $M$ as 10K, and the probability of \xxx enforcing
safety is $1-10^{-9}$. In real deployments, $M$ may change due to 
membership changes; however, when $M$ is much bigger (\eg, 20X) than $n_A$, this
probability is not sensitive to $M$ because hypergeometric distribution is
approximate to binomial.

\looseness=-1
For the second step, we prove that event A and event B2 cannot happen together.
For event B2, we suppose that proposer $P_i$, where $i>n$, learns and finalizes
\proposal{n}. We discuss by comparing $i$ and $f_{m_x}$ and derive
contradictions. If $i\leq f_{m_x}$, \hyperref[hypo:backfinalize]{hypothesis 2}
infers that $X$ did not finalize \emp{i}, so \proposal{n} is finalized
together with \proposal{i} at node X, contradicting to event A. Else if $i>
f_{m_x}$, since a proposer can only finalize the maximum index in its local
$\mathbb{U}$ list (\algoline{algo:proposer}{proposer:maxu}), for node $P_i$ we
can predicate that blocks with index in $[n+1, i)$ are finalized. Due to
\hyperref[hypo:backfinalize]{hypothesis 2}, blocks within $[n+1, f_{m_x}$] are
finalized the same as node X, and therefore $P_i$ should also finalize the
\nth{n} block as empty, causing contradiction.

Putting the two steps together, we proved \hyperref[proposition]{proposition 1}
and thus proved the induction step that the \nth{n} block must be confirmed consistently
among nodes with overwhelmingly high probability. Therefore, \xxx ensures safety
with overwhelmingly high probability.

\subsection{Liveness under Targeted DoS Attacks}\label{sec:analysis:liveness}
\xxx can defend against DoS attacks and partition attacks targeting committee
nodes under the threat model in \chref{sec:model}. Since from a single
node's point of view, when it cannot reach a remote node, it cannot distinguish
whether a remote node is under DoS attack or is partitioned, \xxx handles these
two attacks altogether.
%  in the
% same way. 

\xxx has two types of committee nodes for each block, a proposer and $n_A$
acceptors, randomly selected from all nodes in the system. Because of \xxx's
stealth committee abstraction (\chref{sec:protocol:select}), the identity of
each committee node is unknown to attackers outside SGX enclaves before the node
sending out its first protocol message. 

\looseness=-1
We first discuss acceptors. An acceptor sends \v{ACK} messages to both the
proposer and arbiters, but its identity becomes explicit after sending out its
first \v{ACK} message, so \xxx uses fake acceptors to conceal the real
ones. If observed from outside enclaves, the fake acceptors behave identically
as real ones so that an attacker cannot differentiate the real acceptors and
attack them. \xxx achieves this with three design points. 

First, fake acceptors are randomly selected from all nodes for each block so
that an attacker cannot determine the fake acceptors by monitoring network
packets (\chref{sec:protocol:select}). Second, real and fake acceptors' \xxx
enclaves respond to protocol messages (\v{propose} and \v{arbit}) in
the same way if observed outside their enclaves (\chref{sec:protocol:confirm}),
so that an attacker cannot distinguish real or fake acceptors by watching their
behaviors. Third, all messages from real or fake acceptors are of the same
format, encrypted with the receiver's public key (\ref{sec:protocol:confirm})
and can only be decrypted in the receiver's enclave, so that an attacker cannot
differentiate real acceptors from fake ones by watching the packet content. 

\looseness=-1
Therefore, if an attacker targeting acceptors in \xxx, it can only randomly
select $B$ nodes from both real acceptors and fake acceptors. For instance, if
\xxx has $n_A = 300$, $\tau = 59\%$ and has (expected) 600 fake acceptors; if
the attacker's attack budget is 300, the probability that the attacker can
luckily attack more than $(1-\tau)\times n_A$ real acceptors for one block is
0.3\%. Moreover, even if the attacker is so lucky that it successfully guessed
more than $(1-\tau)\times n_A$ acceptors for some block (say \nth{n}), \xxx can
still consistently determine whether to confirm \proposal{n} or \emp{n} using
the committees for subsequent blocks (\chref{sec:protocol:confirm}); in other
words, the attacker must constantly be so lucky to make \xxx lose liveness. 
%  because future committees can safely determine
% whether to confirm this blocks as their corresponding unique proposals or
% default empty blocks, making \xxx stay live. 
% k
% \emp{n}, as long as the attacker cannot \textit{continuously} guess the real
% acceptors, \xxx can stay live.

% A proposer (say $P_n$) stays stealth before broadcasting its block proposal
% (\chref{sec:protocol:select}), so it cannot be targeted by an attacker. 
Then we discuss proposers. The identity of a proposer (say $P_n$) becomes
explicit after broadcasting its \proposal{n} and may be targeted attacked when
it is waiting for quorum \v{ACK}s. However, since \xxx has many (larger
than $B$) arbiters that can help to finalize the \proposal{n}, attacking $P_n$
will not affect \xxx's liveness. \xxx lets each node independently determine
whether it is an acceptor for each block (\chref{sec:protocol:arbiter}) so that
an attacker cannot keep attacking them. Note that one single proposer or arbiter
is enough for \xxx to stay alive because any of them can finalize the current
\nth{n} block. 

% In contrast, for existing permissioned blockchain's consensus protocols
% running on static committees, they must scale to $3\times B + 1$ committee nodes
% in order to tolerate an attacker with an attack budget of $B$. As far as we
% know, no existing consensus protocols for permissioned blockchains can scale to
% these many nodes. 

\looseness=-1
Note that \xxx's targeted attack model (\chref{sec:model}) handles only
DoS or partition attacks targeting specific \xxx nodes. A more powerful attacker may
also target \xxx's major communication links. From the protocol aspect, \xxx
avoids such vulnerabilities in the Network layer (in the OSI model~\cite{osi})
by using distinct committees for different blocks: \xxx's protocol traffic is
spread among the whole P2P network rather than centralized among a few dedicated
nodes. However, when \xxx is deployed, \xxx's communication messages may be
aggregated in the Link or Physical layer. For instance, if a large number of
\xxx nodes are hosted in the same data center (DC), the links connecting this DC
and the Internet may be susceptible to attacks. However, such attacks are not
adaptive, and as long as a great majority of nodes are connected, \xxx can achieve
practical liveness. \chref{sec:eavl:sensitiviy} shows the relation between
\xxx's liveness and the maximum connected component size in the P2P network. 
Nevertheless, \xxx cannot ensure liveness under arbitrary partitions, and previous
work shows that it is impossible to ensure both consistency and liveness under
partitions~\cite{fischer1985impossibility,gilbert2002brewer}.

\subsection{Parameter Selection}\label{sec:analysis:parameter}
\begin{figure}[tb]
	\centering
    \begin{minipage}[t]{0.47\columnwidth}
	    \centering
        \includegraphics[width=\textwidth]{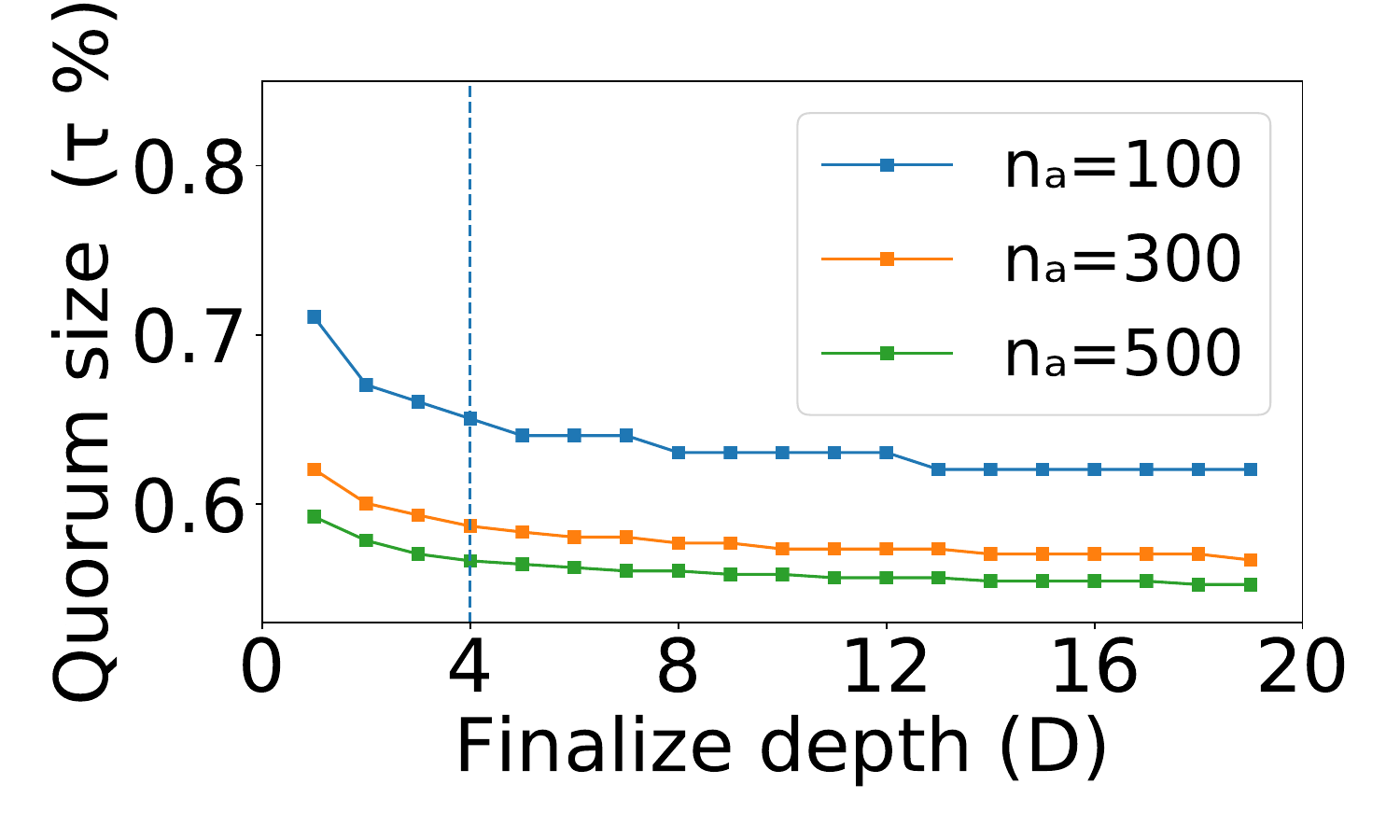}
		\caption{\small Parameter selection for $\tau$ and $D$ 
		% to ensure safety
		% with overwhelmingly probability ($>$1$-$$10^{-10}$) 
		on 10K nodes for
        different $n_A$ values.}\label{fig:safety}
    \end{minipage}
    \hfill
    \begin{minipage}[t]{0.48\columnwidth}
	    \centering
        \includegraphics[width=\textwidth]{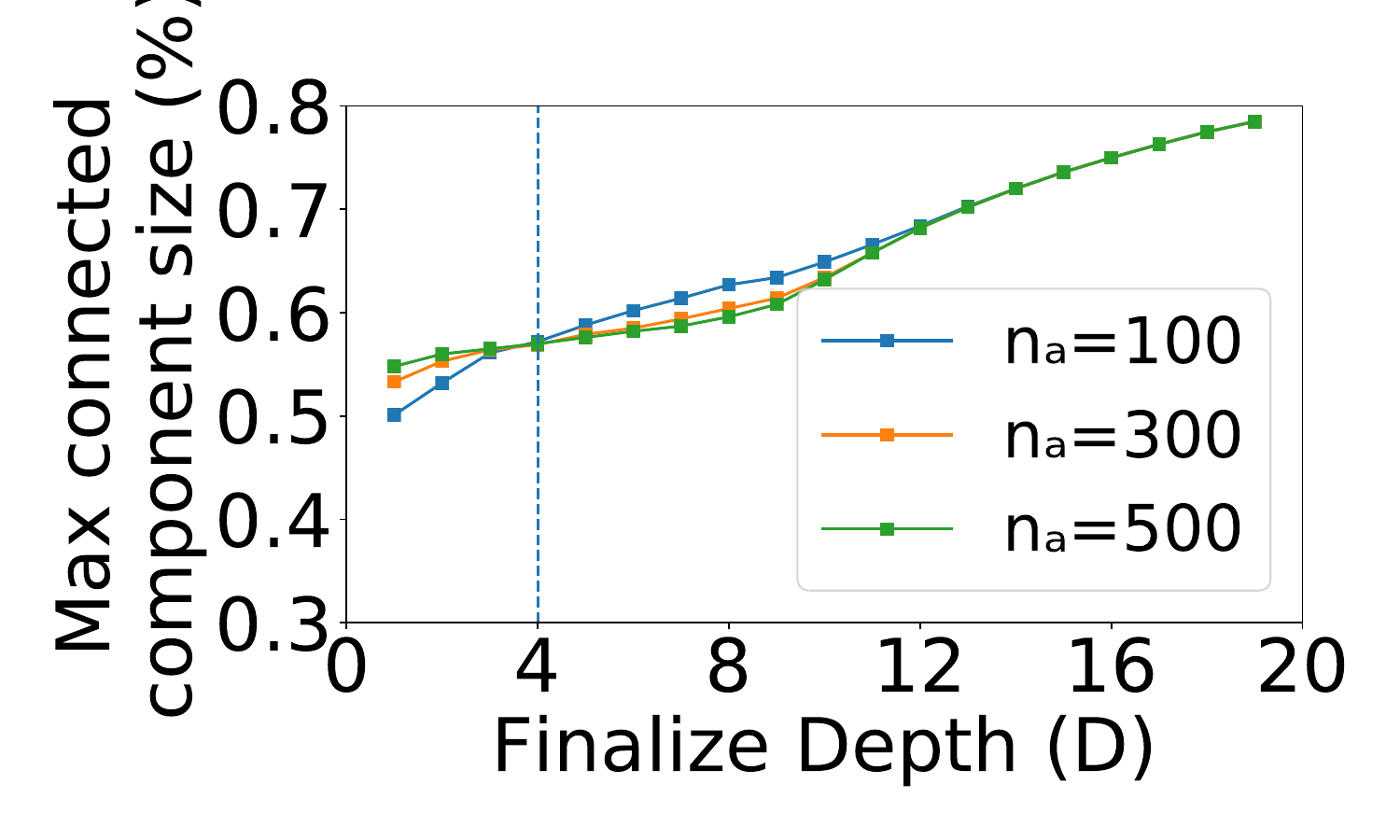}
        \caption{\small Connected component size required to ensure
		liveness with different $D$ values.
		%  (with corresponding $\tau$ values in \fig{fig:safety}).
		}\label{fig:liveness}
    \end{minipage}
    
\end{figure}

\fig{fig:safety} shows the relation between $\tau$ and $D$ in order to ensure safety.
With a smaller $\tau$, a proposer $P_n$ can finalize \proposal{n} after
collecting fewer ACKs from acceptors,
% (i.e., with a lower dissemination rate
% threshold), 
so subsequent proposers need more rounds of checking (larger $D$) when trying to
finalize the \emp{n} (\chref{sec:protocol:confirm}).

The $\tau$ and $D$ values also affect \xxx's ability to achieve liveness
(confirm non-empty blocks) on network partitions (or ubiquitous DoS attacks). We
quantify this ability to the ``minimum largest connected component size''
($cc\%$) required in the P2P graph, provided that nodes in a connected component
can reach each other before a timeout. A smaller $cc$ means that \xxx is more
robust to partition and ubiquitous DoS attacks. From a mathematical aspect, as
long as the probability of finalizing a proposal is non-zero, the probability
$p_D$ that $D$ consecutive proposals are successfully finalized is always larger
than zero, inferring that \textit{eventually} \xxx can achieve liveness.
However, we conservatively calculate the required $cc$ to make the $p_D$ larger
than 5\% for practical liveness, as shown in \fig{fig:liveness}. In our
evaluation, we chose $\tau$ as 59\%, $D$ as 4, $n_A$ as 300, which ensures both
safety and achieves good liveness on partition attacks
(\chref{sec:eval:robust}).

\fig{fig:binomial} shows the parameter selections if \xxx does not use its
stealth committee mechanism, but lets each node \textit{independently} determine
whether it is an acceptor with the probability of $M / n_A$, with $n_A$ being
the \textit{expected} number of acceptors for each block. If \xxx makes such a
design choice, the $\texttt{Re}$ (\chref{sec:analysis:safety}) becomes a
binomial distribution with the probability of $p\times M / n_A$, and other
distribution changes similarly. As shown in \fig{fig:binomial}, \xxx would need
a larger quorum size $\tau\times n_A$, and worse liveness on network partition.

\myfig{
\begin{figure}[h]
	\centering
    \begin{minipage}[t]{0.47\columnwidth}
	    \centering
        \includegraphics[width=\textwidth]{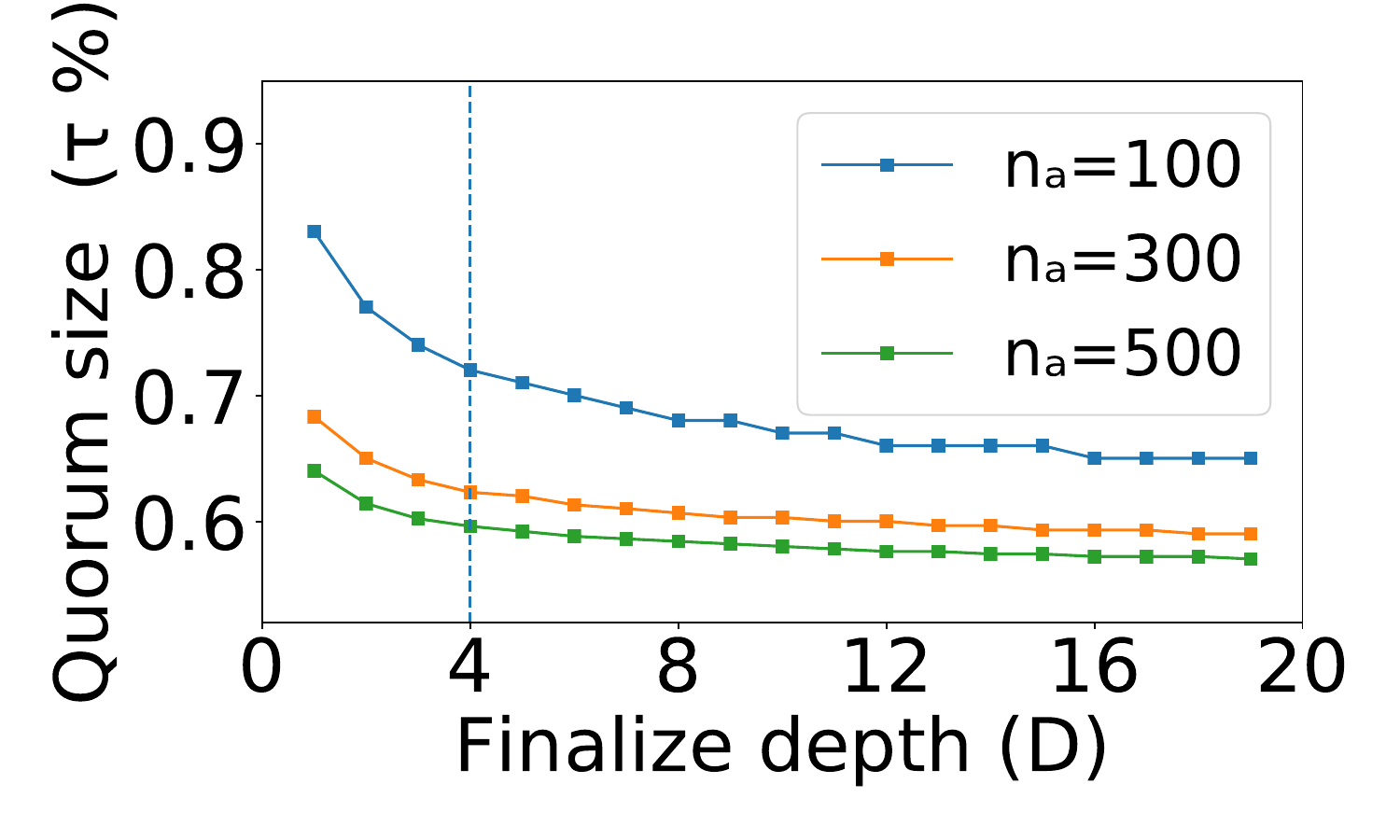}
    \end{minipage}
    \hfill
    \begin{minipage}[t]{0.47\columnwidth}
	    \centering
        \includegraphics[width=\textwidth]{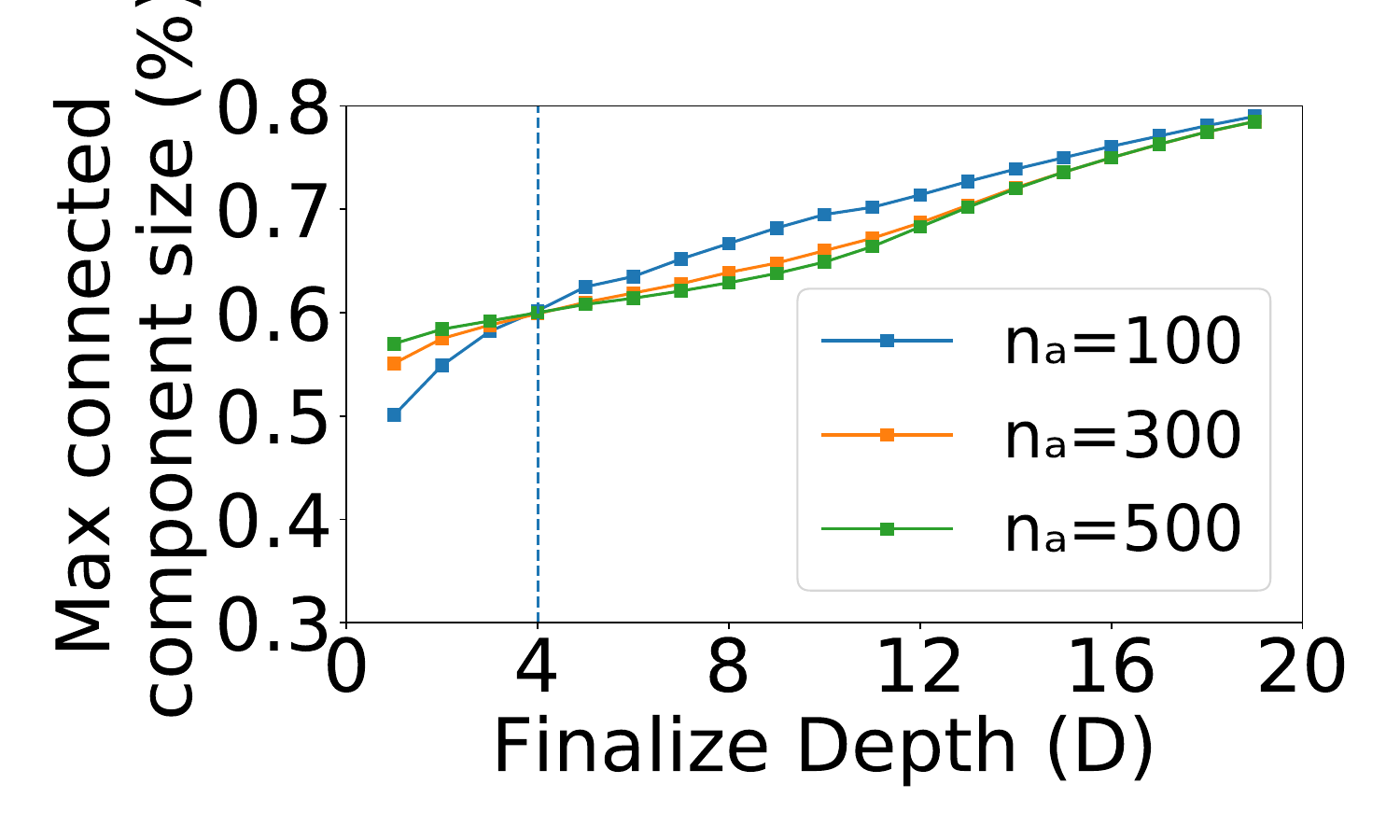}
    \end{minipage}
    
    \caption{\small Parameter selection and liveness requirements if \xxx lets each node to independently decide its committee membership.}\label{fig:binomial}
    
\end{figure}
}

\section{Implementation}\label{sec:impl} 

We selected the Golang implementation of Ethereum (i.e., geth) as our codebase
because geth is heavily tested on the Internet. We leveraged the P2P libraries
from geth and rewrote the functions for generating, verifying, and handling new
blocks. Since SGX only provides SDKs in C/C++, we used CGo to invoke ECalls. We
modified 2073 lines of Golang code and implemented the consensus protocol for
1943 lines of C code. For asymmetric key based encryption, we used ECC-256 from
the API provided by the SGX SDK. 

Each \xxx node has three modules: a consensus module running the \xxx consensus
protocol and storing nodes' member list (\chref{sec:protocol}), a P2P module
connecting to a random set of peers and relaying messages using the
Gossip~\cite{kermarrec2007gossiping} protocol, and a blockchain core module
storing confirmed parts of \v{chain} and transactions submitted by clients.

\looseness=-1
As stated in \chref{sec:model}, only the consensus module runs in the node's SGX
enclave. We put the P2P module outside SGX because this module is only
responsible for relaying messages and putting it inside SGX cannot effectively
bring more benefits: a malicious node can still selectively drop packets at the
network layer. We put the blockchain core module outside enclave because
immutable properties of the hash-connected blockchain make this module's
actions easily verifiable, and putting this module in enclave will consume much
memory.

In \xxx, a node may finalize a block before knowing its preceding blocks.
Therefore, when an \xxx proposer proposes a block or a node finalizes a block,
it lefts the block's field of ``hash value for the previous block'' empty, and
\xxx's enclave computes this field when confirming this block. In other words,
\xxx actually achieves consensus on a totally ordered sequence of transactions,
same as Hyperledger Fabric~\cite{androulaki2018hyperledger}, and encapsulate
these baches into a hash-chain of blocks while confirming them. 

\subsection{Membership and attestation}\label{sec:impl:membership}

To support dynamic memberships, \xxx leverages the idea from
\textsc{Scifer}~\cite{ahmed2018identity} to record the joining of new nodes as
transactions on the blockchain. This mechanism ensures Lemma 1 because all
updates to the member list are only determined by confirmed blocks. 

When a node $i$ wants to join the system, it needs to find a member node $j$ to
do attention (\chref{sec:background}) through an out-of-band peer discovery
service. We assume that node $i$ knows the genesis (\nth{0}) block, so node
$i$ can inductively verify the blockchain, without relying on whether peer $j$ is malicious.

A node $i$ joins \xxx with three steps. First, $i$ launches its \xxx enclave,
which generates its account $(pk_i, sk_i)$ and creates the hardware monotonic
counter $c$ for defending forking attacks (\chref{sec:impl:attacks}).
The node's account is securely saved to permanent storage using SGX's seal
mechanism~\cite{sgx-explain} for recoveries from machine failures (e.g., power
off). Then, $i$ sends a \textit{join} request to $j$. Second, $j$ does a
standard SGX remote attestation~\cite{sgx-explain}, which succeeds with a
signed quote $Q_i$ from Intel's attestation service, and $i$'s enclave transfers
its public key $pk_i$ and counter value $c$ to $j$'s enclave through the secure
communication channel between two enclaves created during attestation. Third,
node $j$'s enclave creates a signed registration transaction including $pk_i, c,
addr_i, Q_i$ and $i$'s ip address. 
% broadcasts it to all nodes. 
Node $i$ joins \xxx when the
transaction is included in a confirmed block. 

% \bfheading{Removing inactive nodes.} As a performance optimization, \xxx
% carries a mechanism to track active nodes and removes inactive nodes using a
% lease-based protocol. After registration, an \xxx node's membership expires
% after $T$ blocks and needs to redo the joining. When a proposer broadcasting its
% \v{finalize} message, it includes the accounts of those acceptors responding
% ACKs as a proof of participation, where these nodes' membership is renewed to
% $T$. $T$ is a heuristic value without affecting safety. In our evaluation, we
% set $T = 4 * M/ n_A$ ($M$ is the current number of nodes), which efficiently
% removed offline nodes (\chref{sec:eval:robust}). Note that the triggering of
% inactive-check only happens when a block is \textit{confirmed} to maintain Lemma
% 1.
\begin{figure}
    \centering
    \subfloat[][On confirming a block.]{
        \makebox[0.5\columnwidth][c]{
            \includegraphics[width=0.395\columnwidth]{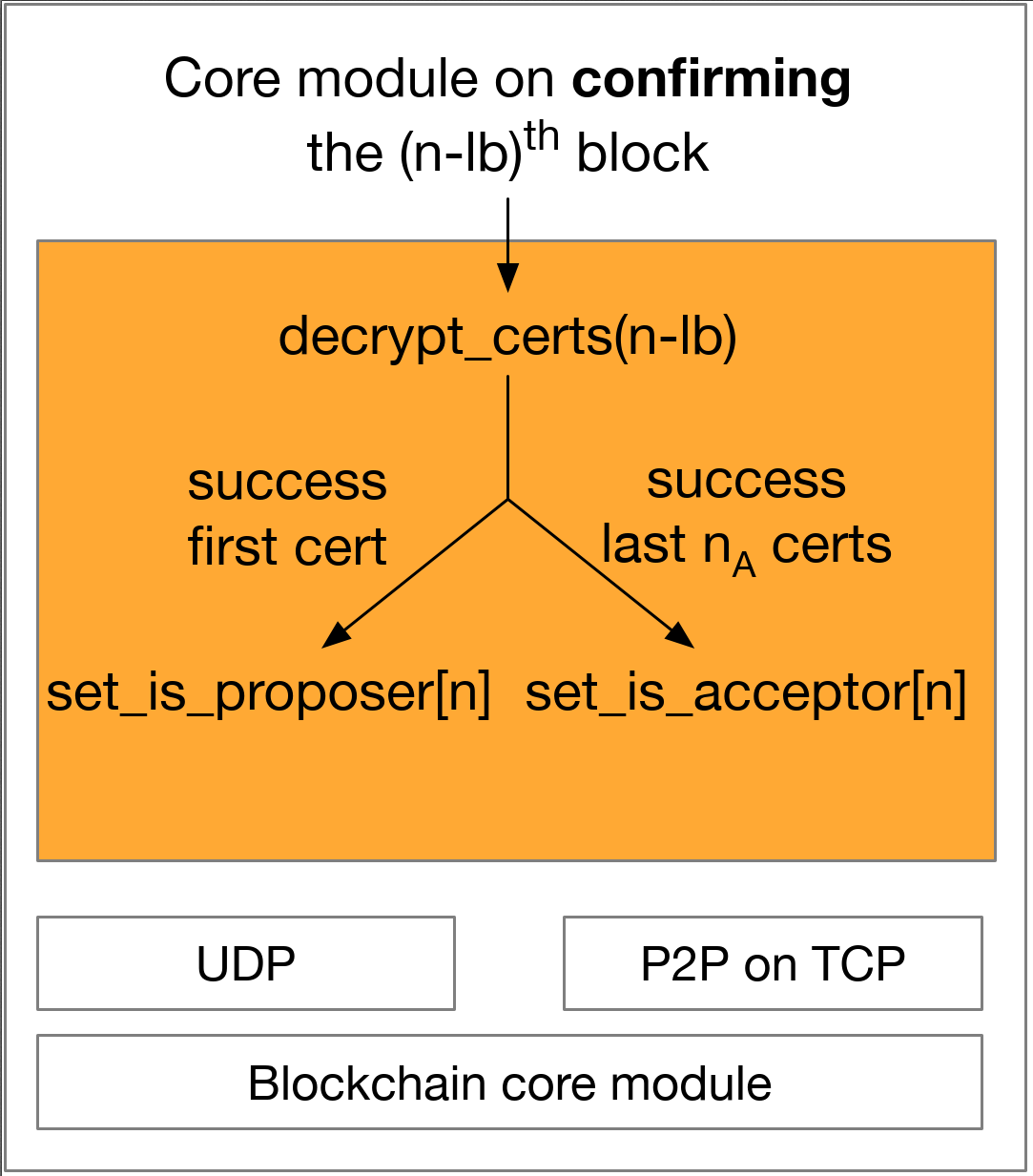}
            \label{fig:ecall:one}
        }
    }
    \subfloat[0.5\columnwidth][On appending a block.]{
        \makebox[0.5\columnwidth][c]{
            \includegraphics[width=0.345\columnwidth]{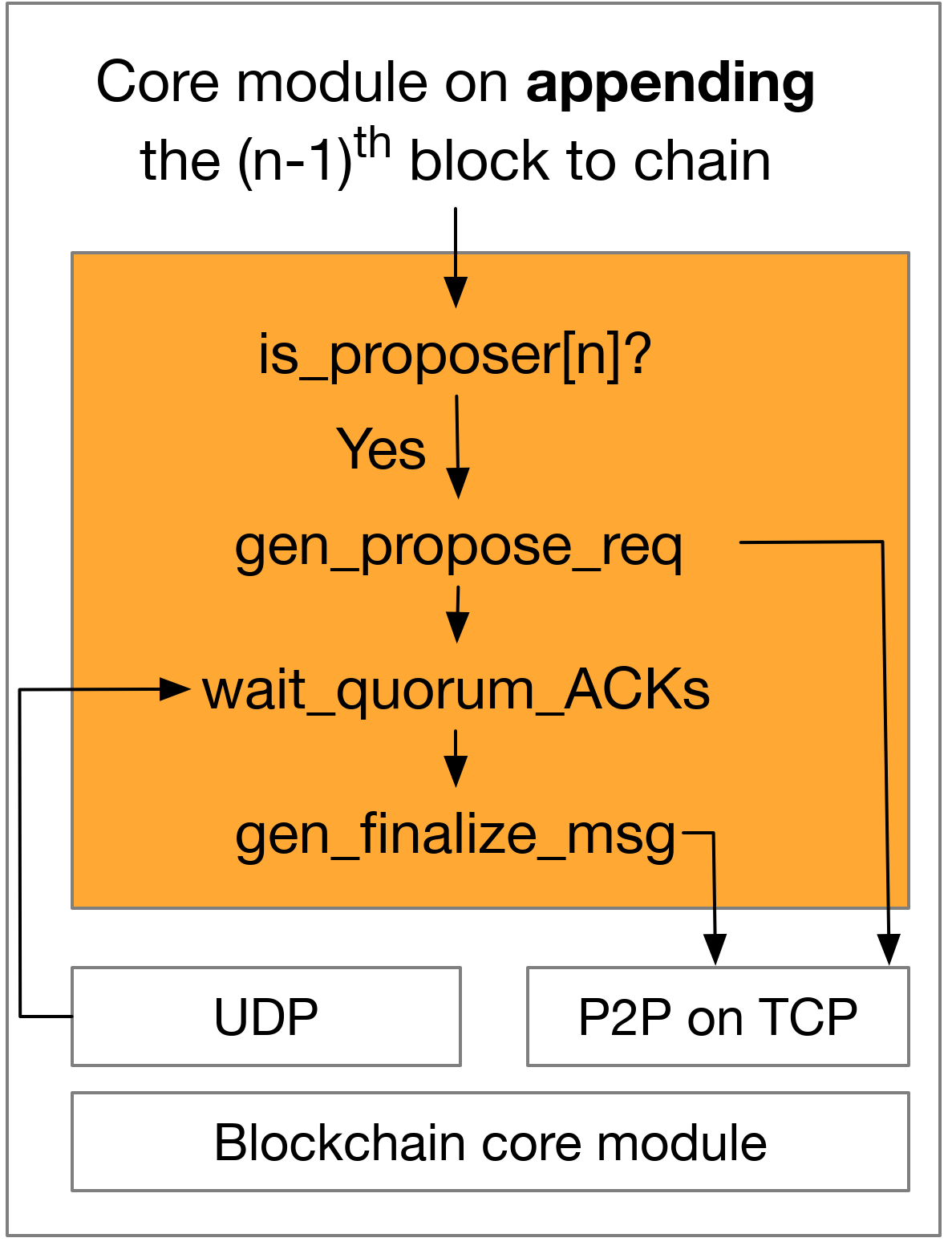}
            \label{fig:ecall:two}
        }
    }
    \\
    \subfloat[\columnwidth][On receiving a \v{propose} request]{
        \makebox[\columnwidth][c]{
            \includegraphics[width=0.6\columnwidth]{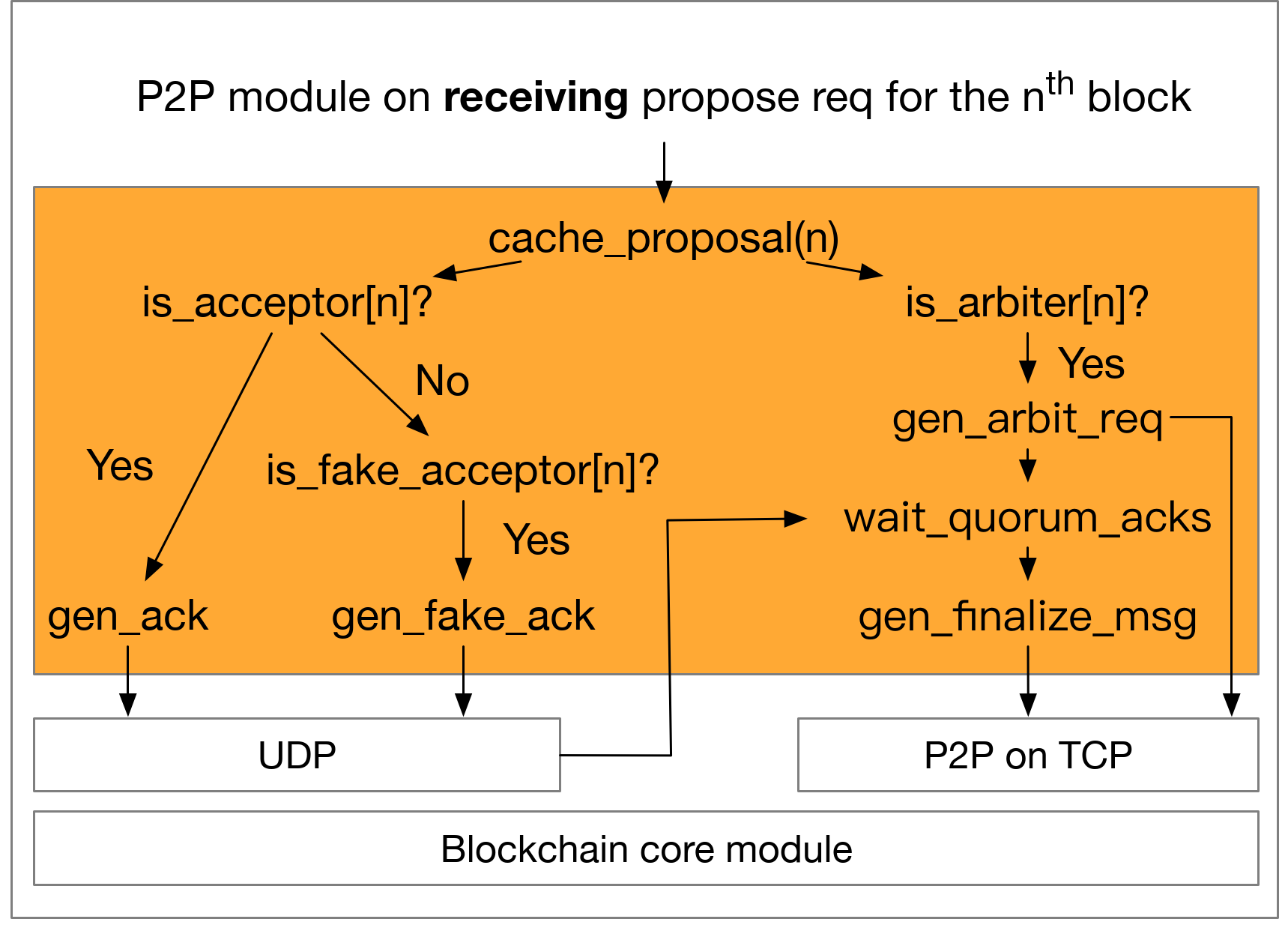}
            \label{fig:ecall:three}
        }
    } 
    \caption{\xxx's enclave interactions (ECalls and OCalls). The enclave
    module is shaded in orange.}\label{fig:enclave}
\end{figure}

\subsection{Enclave Interactions}\label{sec:impl:enclave} 

\fig{fig:enclave}
shows the implementation of \xxx enclave. An \xxx enclave holds three data
structures shared among ECalls: \v{cache}, \v{is\_proposer}, and
\v{is\_acceptor}, each as a hash map. As explained in \chref{sec:protocol:pre},
the \v{cache} saves received block proposals. In our implementation, the
\v{cache} only keeps the hash values of block headers instead of whole blocks to
save enclave memory. \v{is\_proposer} and \v{is\_acceptor} are hash maps with
block indices as keys and boolean as values, saving whether this node will be in
the committee for a future block. \looseness=-1 

In \fig{fig:ecall:one}, when a node's blockchain core module confirms the \nth{(n-lb)}
block, it asynchronously invokes an ECall letting the enclave check whether it
will be the proposer/acceptor for the \nth{n} block
(\chref{sec:protocol:select}). In \fig{fig:ecall:two}, when a node's core module
appends the \nth{(n-1)} block, it invokes an ECall with a batch of transactions,
and the enclave will follow the protocol in \algo{algo:proposer} if it is the
proposer for the \nth{n} block. In \fig{fig:ecall:three}, when a node's P2P module
received a \v{propose} request, it invokes an ECall passing this request to the
enclave, and the enclave will generate an ACK that will be sent through UDP
using an OCall if it is an acceptor (\algo{algo:acceptor}) or fake acceptor
(\algoline{algo:allnodes}{allnode:fake}). If it is an arbiter for the \nth{n}
block, it will also start working as an arbiter (\chref{sec:protocol:arbiter}).
\looseness=-1
% after a timeout.

\subsection{Handling Attacks on Enclaves}\label{sec:impl:attacks}

\bfheading{Forking attacks.} In the P2P scenario, one challenge is
enclave forking attacks~\cite{rote}. \xxx must permit a node to reuse its sealed
account (\chref{sec:impl:membership}) in case the node restarts its machine.
However, a malicious node can create multiple copies of \xxx enclaves with the
same account $pk$, directs different messages to them, and lets them generate
conflicting messages (e.g., block proposals). Existing defending
techniques~\cite{rote, gu2017secure} work in the client-server manner, where
clients attest and communicate to only a single server. 
% to avoid the
% server computer forking multiple enclave copies. 
These techniques are not
suitable for P2P settings because they will need every two \xxx nodes to connect
and attest each other. 

\xxx defends such attacks using SGX's platform counter~\cite{sgx-explain}, which
is monotonic among all enclaves on the same machine. When a node launches its \xxx
enclave, the enclave increments and read this counter value $c$ and enclose this
$c$ to its registration transaction: the node's membership is
bound to the enclave with counter value $c$ but not the account $pk$. 
% and subsequent forks will have larger counter values. 
When the enclave sees a registration with the same account but a higher 
counter value, it quits automatically.

\bfheading{Premature timeout.} \xxx uses the timeout mechanism to suspect node
failures and to achieve liveness. A premature timeout does not affect 
\xxx's safety because \xxx's safety argument does not rely on timing constraints (\chref{sec:analysis:safety}).

However, a premature timeout may affect \xxx's performance, and we implemented a
simple mechanism to mitigate this problem by leveraging the trusted timer API
\textit{sgx\_get\_trusted\_time} provided by the SGX platform. This API provides a
verifiable timer with the granularity of seconds. To avoid the heavy time cost
(tens of milliseconds) caused by frequently checking this timer, \xxx maintains
an untrusted timer outside. When the untrusted timeout is triggered, the
untrusted code invokes the timeout logic in \xxx's consensus module in the enclave,
and the consensus module checks the trusted timer before invoking \xxx's
consensus logic.  
\section{Evaluation}\label{sec:eval}

\bfheading{Evaluation setup.}
Our evaluation was done on both our own cluster with 30 machines and the AWS
cloud, with parameters shown in \tab{tab:parameters}. In our
cluster, each machine has 40Gbps NIC, 2.60GHz Intel E3-1280 V6 CPU
with SGX, 64GB memory, and 1TB SSD. On AWS, we started up to 100 c5.18xlarge
instances (VMs) running in the same city, each of which has 72 cores, 128GB
memory, and up to 25 Gbps NIC. We ran up to 100 \xxx nodes on each VM 
(10k nodes in total), with each \xxx node running in a docker container. 
\looseness=-1

To evaluate \xxx and baseline protocols in a typical geo-replicated setting, while
running \xxx on both our cluster and AWS, we emulated the world scale Internet
by using the Linux traffic control (TC) command to limit the RTT between every
two nodes to a random value between 150ms and 300ms. These settings are
comparable to Algorand's setting on AWS. As AWS does
not provide SGX hardware, we ran \xxx in the SGX simulation mode on AWS and in
the SGX hardware mode on our cluster; we show that \xxx's performance in
simulation mode is roughly the same as hardware mode because \xxx's performance
is bound to network latency in WAN (\chref{sec:eval:perf}). The scalability
(\fig{fig:eval_scale}) and robustness (\fig{fig:eval_down}) experiments were done
on AWS, and the rest were in our cluster. 

\myfig{
\begin{table}[t]
	\centering
	\footnotesize
	\begin{tabular}{lll}
    \Xhline{2\arrayrulewidth}
		\textbf{Config} & \textbf{Cluster} & \textbf{AWS Cloud} \\
    \Xhline{2\arrayrulewidth}
		\# Nodes        & 300              & up to 10K      \\
		% Proposer candidate size  & 30               & 30             \\
		Acceptor group size ($n_A$)   & 100              & 300            \\
		D & 4 & 4 \\
		$\tau$ & 65\% & 58\% \\
		$\v{LB}$ & 5000 & 10000\\
		$timeout$ & 2s & 3s\\
		% Membership Lease ($T$) & 30& 150 \\
		% \xxx code & full implementation & full implementation \\
		SGX mode & hardware mode & simulation mode \\
    \Xhline{2\arrayrulewidth}
	\end{tabular}
	\caption{\xxx's evaluation parameters.}\label{tab:parameters}
	
\end{table}
}

We evaluated \xxx{} with nine consensus protocols for blockchain systems,
including five state-of-the-art efficient BFT protocols for permissioned
blockchains (\smart~\cite{sousa2017byzantine}, SBFT~\cite{SBFT},
HoneyBadger~\cite{miller2016honey}, and HotStuff~\cite{yin2019hotstuff}), two
SGX-powered consensus protocols for permissioned blockchains
(Intel-PoET~\cite{poet} and MinBFT~\cite{veronese2013efficient}), the default
consensus protocol in our codebase (Ethereum-PoW~\cite{buterin2014ethereum}),
and two permissionless blockchains' protocol that runs on dynamic committees
(Algorand~\cite{gilad2017algorand} and Tendermint~\cite{buchman2016tendermint}).
A detailed description of these protocols is in
\chref{sec:background:permissioned}.

Since Algorand's open-source code is still under development, and we were unable
to deploy its latest release~\cite{algorand:release} to the same scale as \xxx
and Algorand's own paper (i.e., 10k nodes), we took Algorand's performance from
Figure 5 in its paper~\cite{gilad2017algorand}. To make the comparison fair, we
make \xxx's network setting more rigorous than Algorand's: Algorand divided
nodes into multiple cities where intra-city packets have negligible latency,
while \xxx lets the RTT among any two nodes be at least 150ms. 

\looseness=-1
For all evaluated protocols, we measured their performance when each of them
reached peak throughput. For an apple-to-apple comparison of latency, we adopted
Algorand's method to measure transactions' server-side confirmation time: from
the time a transaction is first proposed by a committee node to the time the
transaction is confirmed at this node, excluding the time for clients'
transaction submissions. We measured the server-side instead of the client-side
latency because this method precludes the disturbance of client behaviors as
these protocols run on different blockchain frameworks. For instance, in
Ethereum, PoET (running on Hyperledger Sawtooth~\cite{poet}), and
\xxx, a client submits a transaction to a random node, and the transaction is
disseminated via P2P networks; in \smart, a client submits transactions to a
fixed node (i.e., the leader); in Algorand, a consensus node directly packs a
block with a fixed amount of data (e.g., 1MB) instead of using separate
transactions. 

\looseness=-1
We set \xxx's transaction size to 250 bytes, a typical
transaction size for general data-sharing
applications~\cite{eth:transactionsize,tendermint:opodis}. Since Algorand
reported throughput on block size, we convert it to txn/s by assuming the same
size of transactions as \xxx's. The transaction sizes for the other eight
baseline protocols are either equal to or smaller than that of \xxx. Our
evaluation focuses on these questions:

\begin{itemize}[leftmargin=0.5em, align=left, labelsep=0em]
	\item[\chref{sec:eval:perf}]: Is \xxx efficient and scalable?
	\item[\chref{sec:eval:robust}]: What is \xxx's performance under targeted DoS attacks? 
	% robust is \xxx{} on
	% network anomalies?
	\item[\chref{sec:eavl:sensitiviy}]: How sensitive is \xxx{}
	to its parameters?
	\item[\chref{sec:eval:sbft}]: How do \xxx performance and fault
	tolerance compare with notable BFT protocols?
	\item[\chref{sec:eval:discussion}]: What are the limitations and potential future
	works of \xxx? 
\end{itemize}

\subsection{Efficiency and Scalability}\label{sec:eval:perf}

\tab{tab:performance} shows the performance comparison of \xxx and eight
baseline protocols. As Alogrand's paper~\cite{gilad2017algorand} evaluated at
least 2K nodes, we postpone the comparison between \xxx and Algorand to when we
evaluated \xxx's scalability.

\looseness=-1
Overall, in the geo-replicated setting, \xxx achieved comparable performance to
MinBFT, Tendermint, HotStuff, and SBFT. We ran \smart in its default setting
(ten nodes), and it showed higher throughput and lower latency than \xxx. \smart
is more suitable for small scale permissioned blockchains where a few companies
run nodes in a controlled environment, so it lets nodes send messages to each
other directly. In contrast, \xxx is designed for tolerating targeted
DoS attacks on committee nodes, so it has two P2P broadcasts to confirm a block.
\chref{sec:eval:sbft} shows that \xxx's scalability and fault tolerance are
better than \smart.

%  these many rounds of round trip limit its performance. 
SBFT and HotStuff had a lower throughput and a higher latency than \xxx. They
rely on designated nodes to collect the consensus messages that were originally
all-to-all broadcasted and to distribute a combined message to all nodes.
Although this approach improves scalability, it also incurs two more RTTs,
limiting their performance in a geo-distributed deployment. Moreover, an
attacker targeting these designated nodes will causes a dramatic performance drop
to the system, which is evaluated in \chref{sec:eval:sbft}.

% MinBFT uses SGX to reduces the number of rounds of PBFT and can tolerate more
% failures. However, it still needs $O(n^2)$ messages complexity in its last step,
% limiting its scalability. \xxx shows comparable performance to MinBFT, but \xxx
% can tolerate targeted DoS attacks while MinBFT cannot. 

% As \smart and SBFT
% outperform MinBFT in small-scale or large-scale settings, we focus on comparing
% with \smart and SBFT in the following subsections.

% Algorand is the only protocol we know that tolerates targeted DoS attacks. 
\looseness=-1
HoneyBadger showed a lower throughput and a higher latency than \xxx because
HoneyBadger uses multiple rounds of broadcasts for a single block,
which incurred a long latency in a geo-distributed setting. \xxx showed orders
of magnitude better performance than PoET and Ethereum, two PoW protocols. Their
performance is limited by the time for solving PoW puzzles (or sleep time) and
the number of blocks to wait for before confirming a block
(\chref{sec:background:public}). Our evaluation result for PoET is similar to a
recent study~\cite{deloitteReport}.
%  However, its throughput is still much lower
% than \xxx's.
% % +% Although EOS whitepaper~\cite{eos} estimates a throughput of 100K
% % +% transaction/s, we found its throughput 420 transaction/s in our cluster. This
% % +% gap is because EOS's latest open-source implementation was under development and
% % +% it lacked several crucial features (\eg, parallel chains and signature
% % +% verification) reported in their whitepaper. Moreover, EOS relies on
% % +% pre-determined super nodes~\cite{eos}, vulnerable to DoS attacks.
% +PoET achieved
% +a low throughput due to its long wait time (\chref{background}). This
% +result is similar to a recent study's~\cite{deloitteReport}.

% This is because Algorand
% have multiple proposers with unknown count for each block, and thus it wait for
% a long time 

%  and achieved at least one order of magnitude better performance
% than Algorand. 

\begin{figure}[tb]
	\centering
	\includegraphics[width=.6\columnwidth]{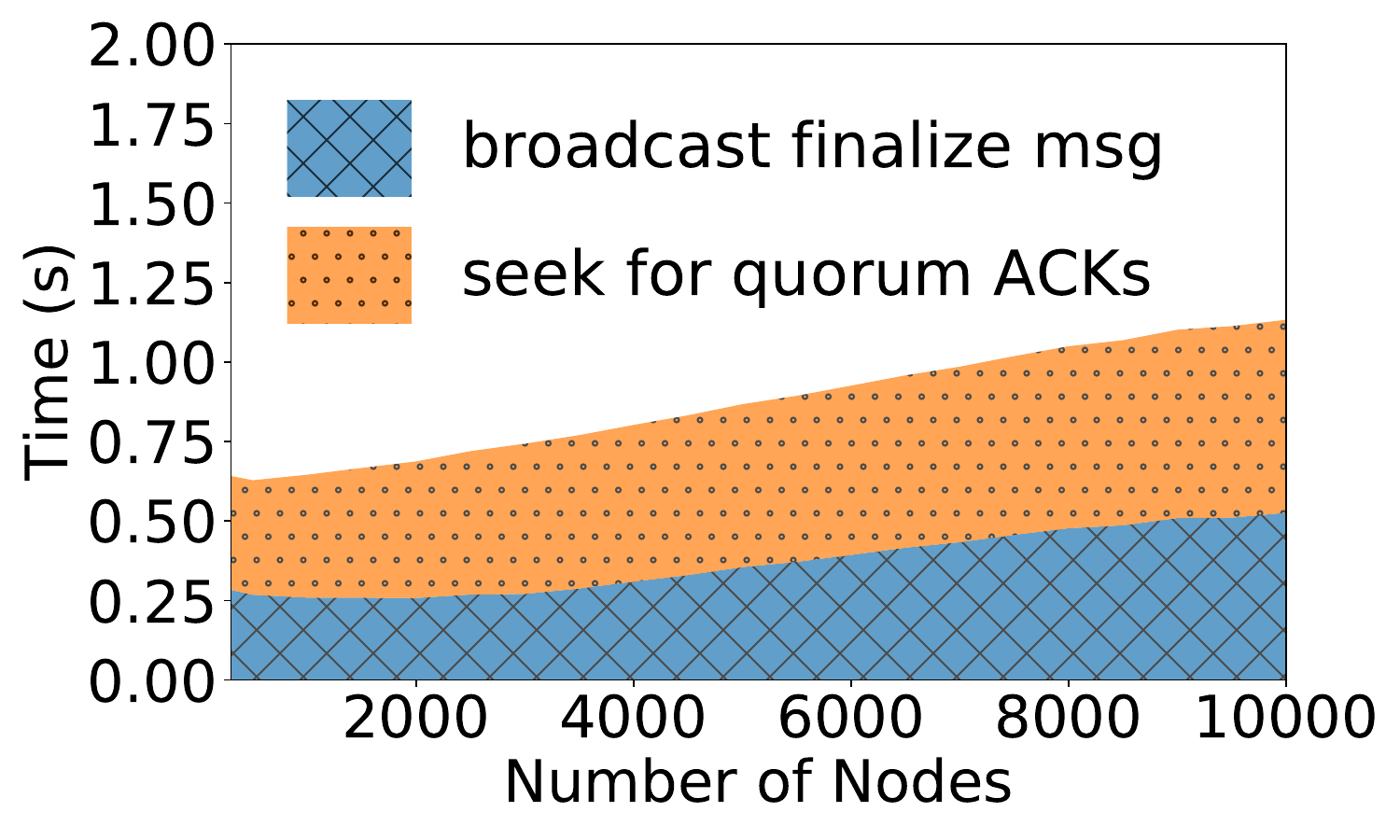}
	
	\caption{Scalability to the number of nodes on Internet.}
	\label{fig:eval_scale}
	
\end{figure}

\bfheading{Breakdown and micro-events.}
To understand \xxx's latency, we recorded the time taken for the two steps of
\xxx's protocol (\chref{sec:protocol:confirm}): seeking for quorum \v{ACK}s took
576ms; broadcasting \v{finalize} messages took 329ms. The first step took a longer
time because \xxx broadcasts the proposed block in its P2P network in this step.
This P2P broadcast time is essential in any blockchain system because new blocks
need to be broadcasted to all nodes.

\bfheading{SGX's overhead.}
\tab{tbl:micro} shows the micro-events of \xxx{}. The ECall column shows
the number of times that \xxx's proposer node entered its SGX enclaves on
finalizing a block. Since each ECall only takes around 3us~\cite{sgx-explain},
and \xxx's proposer only did 97 ECalls on average for each block, running in SGX
hardware mode and simulation mode makes little difference for \xxx's
performance.

\begin{table}[h]
	
	\footnotesize
	\centering
	\begin{tabular}{c|c|c|c|c}
		\Xhline{2\arrayrulewidth}
		\bf blk size & \bf txns/blk & \bf \# ECalls & \bf CPU usage   & \bf network usage  \\
        \Xhline{2\arrayrulewidth}
		750 KB   & 3000   & 97       & 12.4\% & 15.53 Mbps \\
	    \Xhline{2\arrayrulewidth}
	\end{tabular}
	\caption{Proposer's micro-events for finalizing a block.}\label{tbl:micro}
	
\end{table}

\looseness=-1
\bfheading{Scalability.} To evaluate \xxx's scalability, we ran 100-10000 nodes
on AWS and evaluated its confirm latency with the same block size as in the
cluster evaluation. \fig{fig:eval_scale} shows the result. The latency is
divided into two parts. The figure shows that the seeking for quorum \v{ACK}s
phase of \xxx (\chref{sec:protocol:confirm}) is the dominant factor because it
broadcasts the proposed 750KB block on the P2P network. Fortunately, a P2P
broadcast latency is proportional to approximately the \textit{log} of the
number of nodes~\cite{chord}, indicating \xxx's reasonable scalability. The
increase rate was slightly greater than the log scale because 100 nodes were run in
one VM with CPU and NIC contentions. While running on AWS, \xxx's latency was
slightly faster than on our cluster, because AWS CPUs are faster.

Compared to Algorand's performance in \tab{tab:performance}, \xxx showed 2.3X
higher throughput and 16.8X faster latency than Algorand. This is due to two
reasons. First, Algorand's VRF-based method selects multiple proposers for each
block, and Algorand uses a reduction step to select one proposal by these
proposers. Moreover, as the VRF-based approach cannot control the exact number
of proposers, nodes must wait for a conservatively long time in the reduction
step (\chref{sec:high}). In contrast, \xxx's stealth committee
abstraction selects one proposer for each block, without the need for such a
reduction step. Second, 
% \xxx leverages the code integrity
% feature of SGX to 
% effectively reduce the BFT problem to a CFT problem: 
\xxx's consensus protocol has only two rounds in gracious runs to confirm a
block (\chref{sec:protocol:confirm}). 

% This is
% because Algorand is designed for cryptocurrency without a known member list, so
% it relies on PoS for proposer sortition, which has two major drawbacks. First,
% Algorand has , so it needs to invoke a binary
% consensus phase trying to agree on one proposal, while \xxx ensures one unique
% proposal by design
%  ({\color{darkblue} Invariant~\ref{invariant:proposal}}).
% Second, as Algorand does not know the number of proposers for each block, each
% node must wait for a sufficiently long time instead of a fixed number of messages,
% leading to suboptimal performance (\chref{sec:background:public}).
% In
% contrast, in \xxx, a node only waits for a given number of messages in each step
% and can move forward as soon as the messages are received, ensuring good
% performance when the network is in good condition. 

\subsection{Performance on DoS Attacks}\label{sec:eval:robust}
\looseness=-1 

To evaluate \xxx's robustness under DoS attacks, we ran \xxx with 1000 nodes on
AWS with $n_A = 100$, and conducted targeted DoS attacks that are compliant
with our attack model (\chref{sec:model}): we assumed the attacker's
budget $B = 10\%$ of total nodes, and we set the expected count for fake acceptors and 
arbiters to be 200 and 50 correspondingly. 
% our cluster and did five times of targeted DoS attacks;
Each time we targeted the current proposer and 99 arbiters or real/fake acceptors 
(because we cannot distinguish real acceptors). For each attack, we
blocked all communication from the attacked nodes for 20 seconds. 

We deem such attacks to be powerful enough, as no existing protocols for
permissioned blockchain can maintain liveness under such powerful attacks. As
shown in \chref{sec:eavl:sensitiviy}, existing consensus protocols,
which ran on static committee nodes, lost liveness \textit{until} the DoS attack
ended. In contrast, each time after we attacked 100 nodes, \xxx's
throughput had a temporary drop and recovered \textit{before} the DoS attack
ended, which shows that \xxx can ensure practical liveness under such powerful
attacks.

% Each time when we conducted the attack, \xxx's throughput incurred a temporary
% drop due to an \xxx timeout (\chref{sec:protocol:confirm}) and recovered. 
After the first attack, the line started to go up after 11.3s, much slower than
the other attacks (about 3.1s). We inspected the execution log and found that
the slow recovery was because the proposer for the next block happened to be
attacked, and \xxx waited until $D$ more blocks to confirm that block as empty.
After the second attack, the line took about 7.2s to go up. This is because most
real acceptors happened to be attacked together with the proposer, which makes
the arbiters failed to finalize the block for the proposer
(\chref{sec:protocol:arbiter}). For the other three attacks, the arbiters
successfully helped corresponding proposers to finalize their blocks quickly.  

% Note that our evaluated scenario is very adversarial: for each block, 101 nodes
% (1 proposer and 100 real acceptors) are selected as committee nodes among total
% 300 nodes, so an attacked node has a high probability to be in the next
% committee. In a larger scale deployment (thousands of nodes), this probability
% will be much smaller, and \xxx can recover much faster (as the third and fifth
% attacks).

\begin{figure}[tb]
	\centering
	\subfloat[Targed DoS attacks]{
		\includegraphics[width=.48\columnwidth]{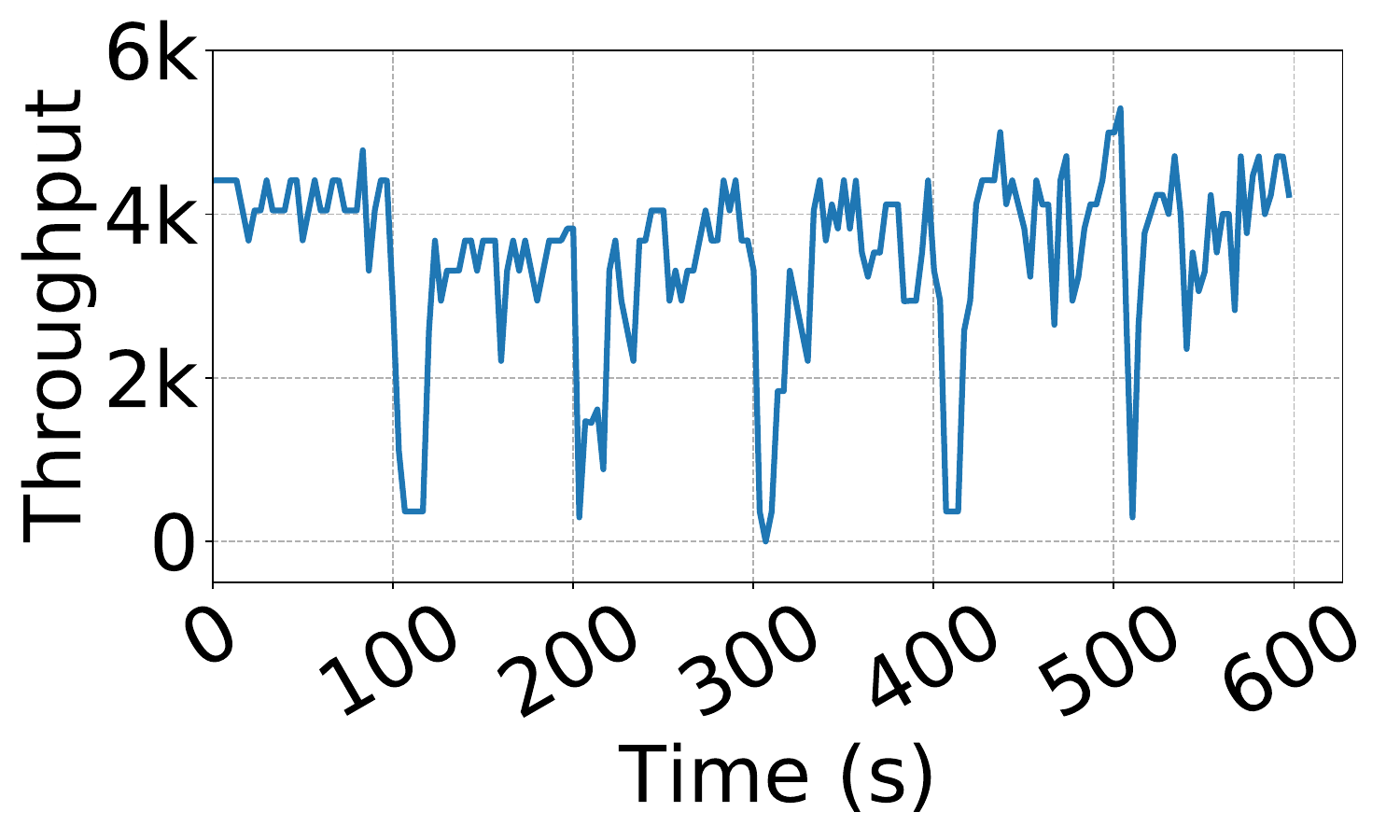}
		\label{fig:eval_down:1}
	}
	\subfloat[80\%-20\% partition]{
		\includegraphics[width=.48\columnwidth]{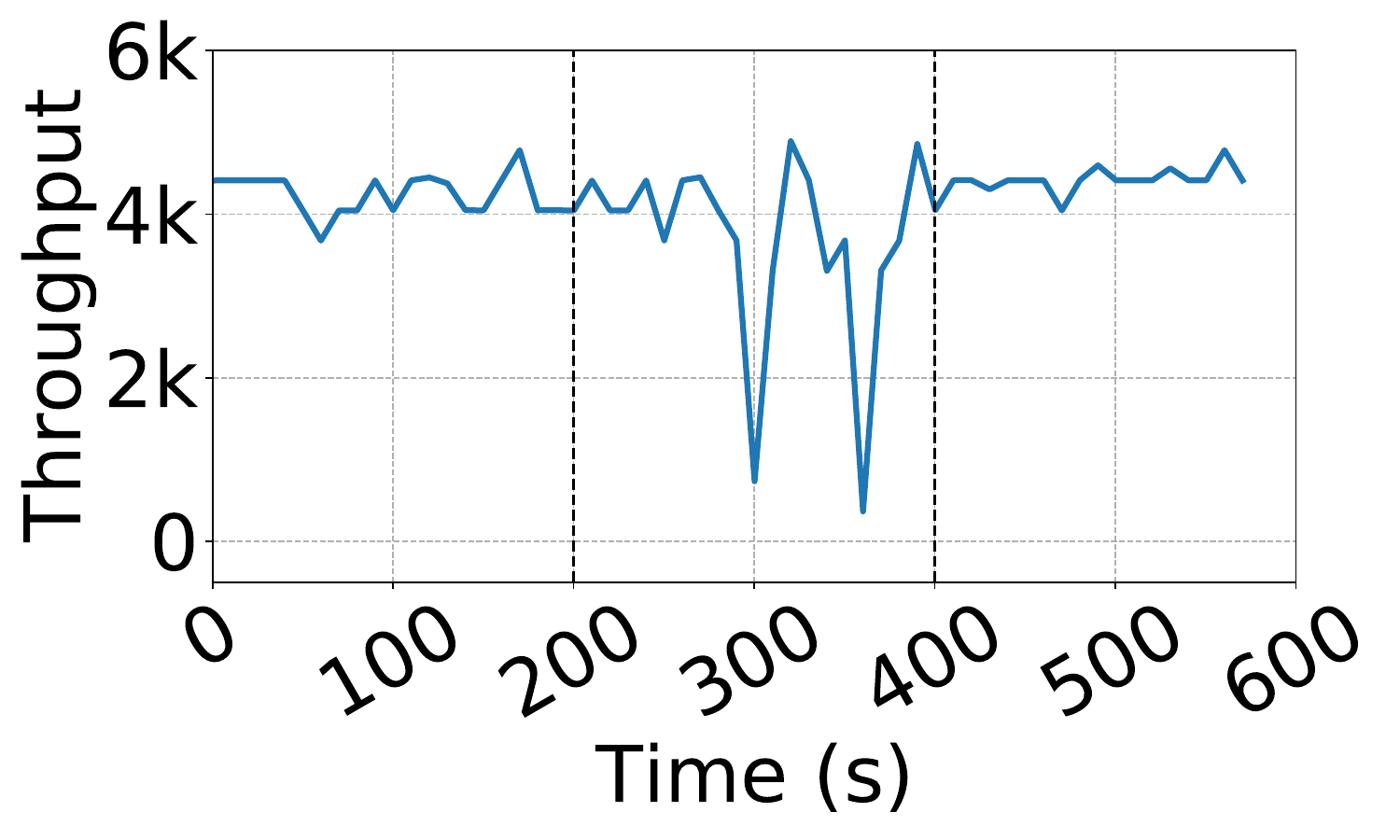}
		\label{fig:eval_partition}}
	
	\caption{\xxx's throughput on DoS and network partition attacks. There were 1000 nodes on AWS at 0s.}
	
	\label{fig:eval_down}
\end{figure}

\looseness=-1
To evaluate \xxx performance on network partitions, we manually divided the
network into two partitions at 200s and reconnected them at 400s, with one
partition containing 80\% nodes and the other containing 20\% nodes.
\fig{fig:eval_partition} shows the throughput measured in the large partition.
Overall, the large partition maintained liveness during the partition. The small
partition did not succeed in confirming any block during the partition and caught
up after the network reconnected, preserving safety. There are two obvious
throughput drops in the figure, which are caused by the pre-designated proposers
being in the small partition, and \xxx confirmed empty blocks for them. Note
that \xxx may temporarily lose liveness in catastrophic partitions (e.g., 50-50
or 40-30-30 partitions) but can preserve safety.
\chref{sec:eavl:sensitiviy} shows a quantitative analysis of how \xxx
can preserve liveness under network partitions. 

\subsection{Sensitivity}\label{sec:eavl:sensitiviy}
\xxx's throughput and confirm latency depend on three important protocol
parameters, the number of acceptors, the number of fake acceptors, and block
size. \fig{fig:eval_committee} shows the sensitivity results. \xxx's
performance turns out to be insensitive to the first two parameters because the
latency is dominated by the time for broadcasting the new block on the P2P
network.

\begin{figure}[tb]
	\centering
	
	\subfloat[Throughput]{
		\includegraphics[width=.45\columnwidth]{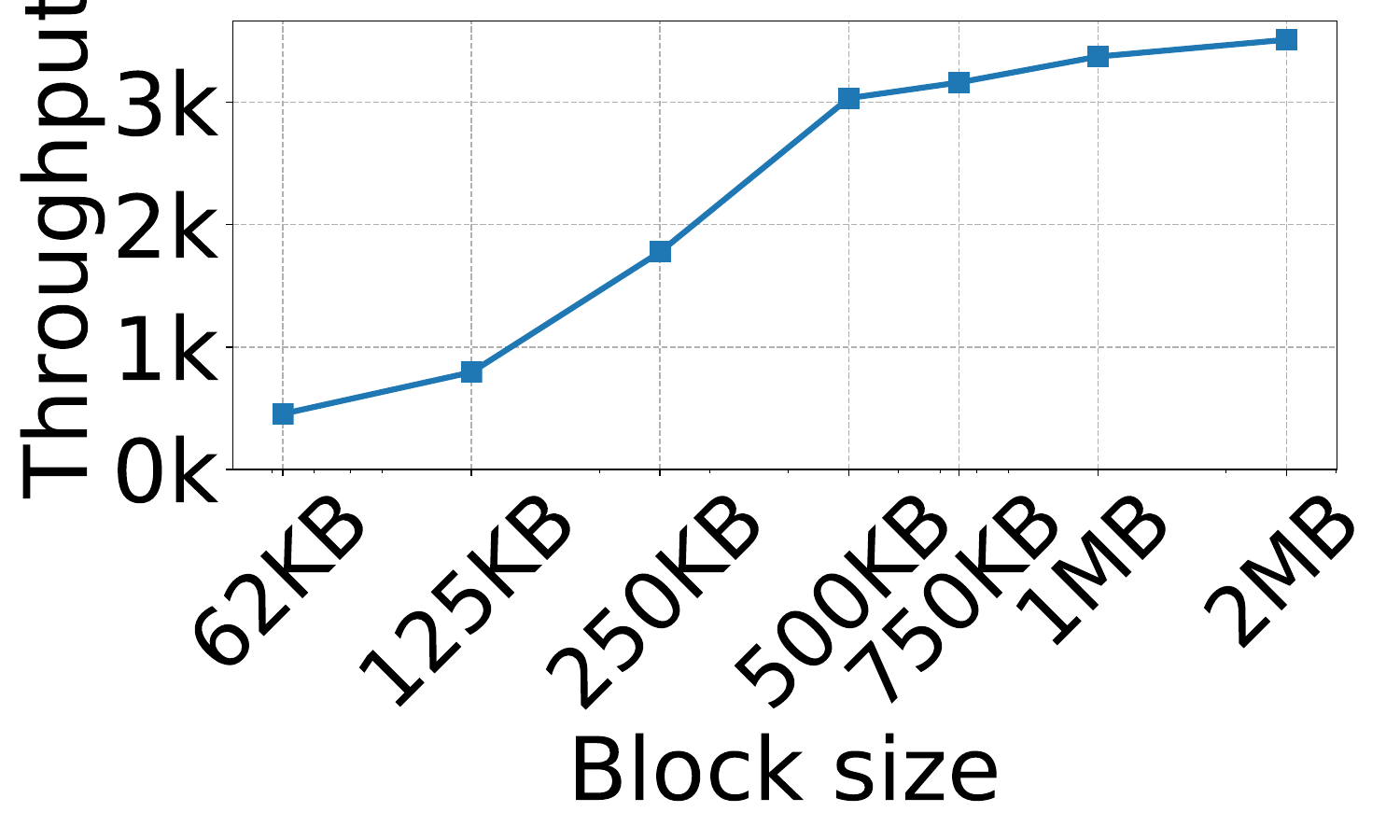}
		\label{fig:eval_blk_tput}}
	\xspace
	\subfloat[Confirm Latency]{
		\includegraphics[width=.45\columnwidth]{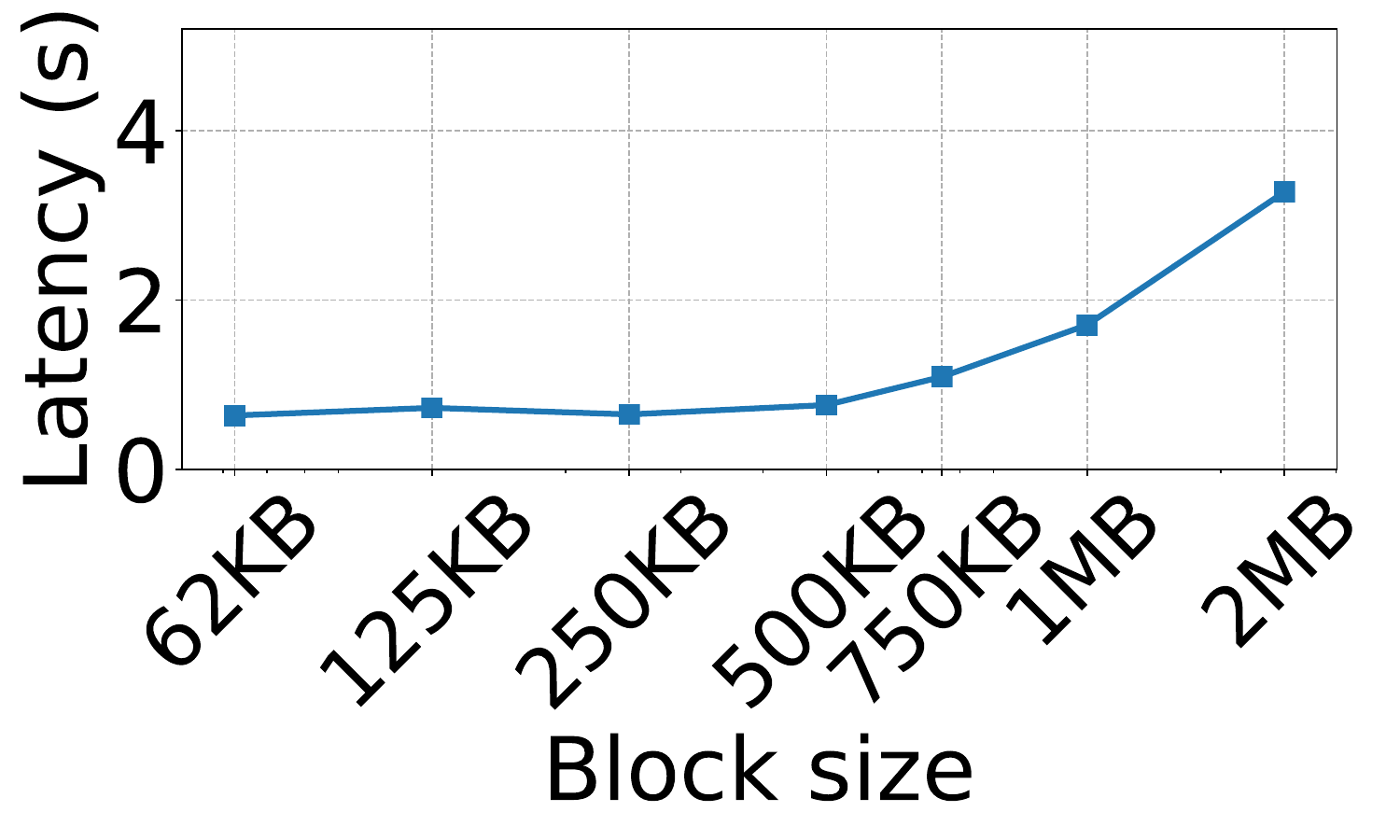}
		\label{fig:eval_blk_latency}}
	
	\caption{\xxx's sensitivity on block sizes (cluster setting).}\label{fig:eval_blk_size}
	
\end{figure}

\begin{figure}[tb]
	\centering
	\subfloat{
		\includegraphics[width=.45\columnwidth]{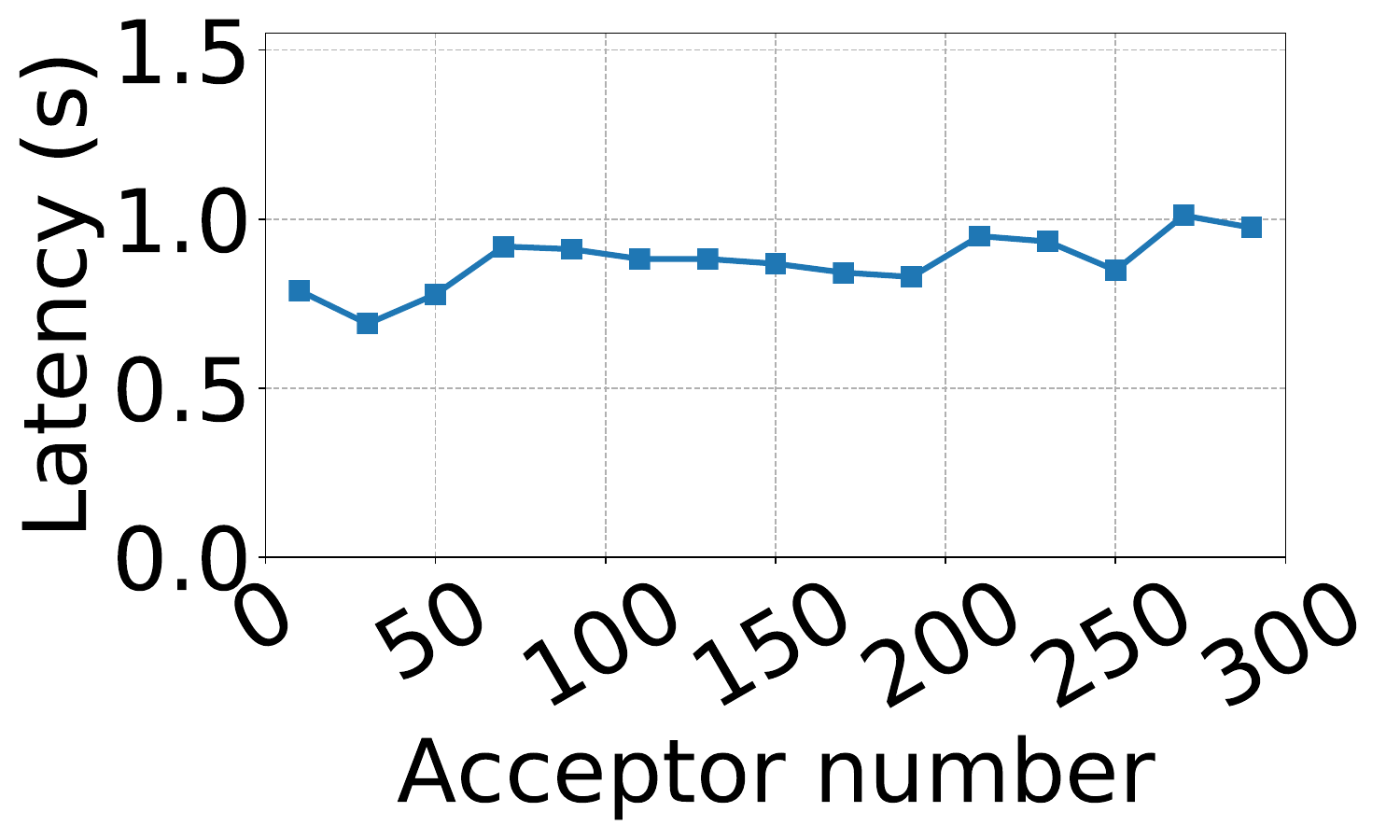}
		\label{fig:eval_valid}}
	\subfloat{
		\includegraphics[width=.45\columnwidth]{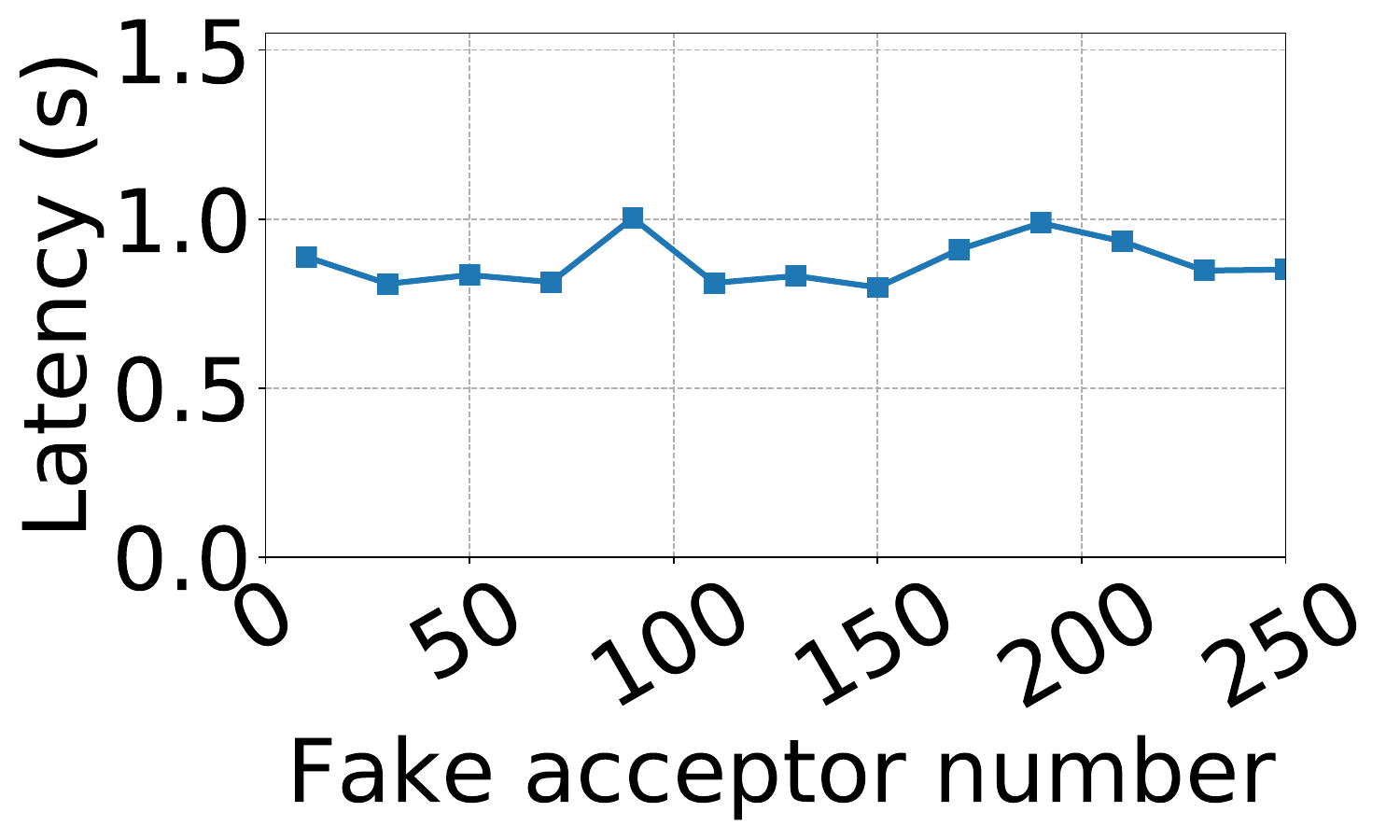}
		\label{fig:eval_committee_size}}
	
	\caption{\xxx's sensitivity on acceptor size and expected fake acceptor
	numbers (cluster setting).}\label{fig:eval_committee}
	
\end{figure}

\fig{fig:eval_blk_size} shows \xxx's performance sensitivity on block
size. When the block size was larger, \xxx's throughput did increase, but its
block confirm latency also increased. In our evaluation, we set \xxx's block size
to be 750KB, which is a near-optimal setting for both throughput and latency.

\subsection{Comparison to \smart and SBFT}\label{sec:eval:sbft}
\begin{figure}[tb]
	\centering
	\subfloat[No RTT delay (tc = 0)]{
		\centering
		\includegraphics[width=.45\columnwidth]{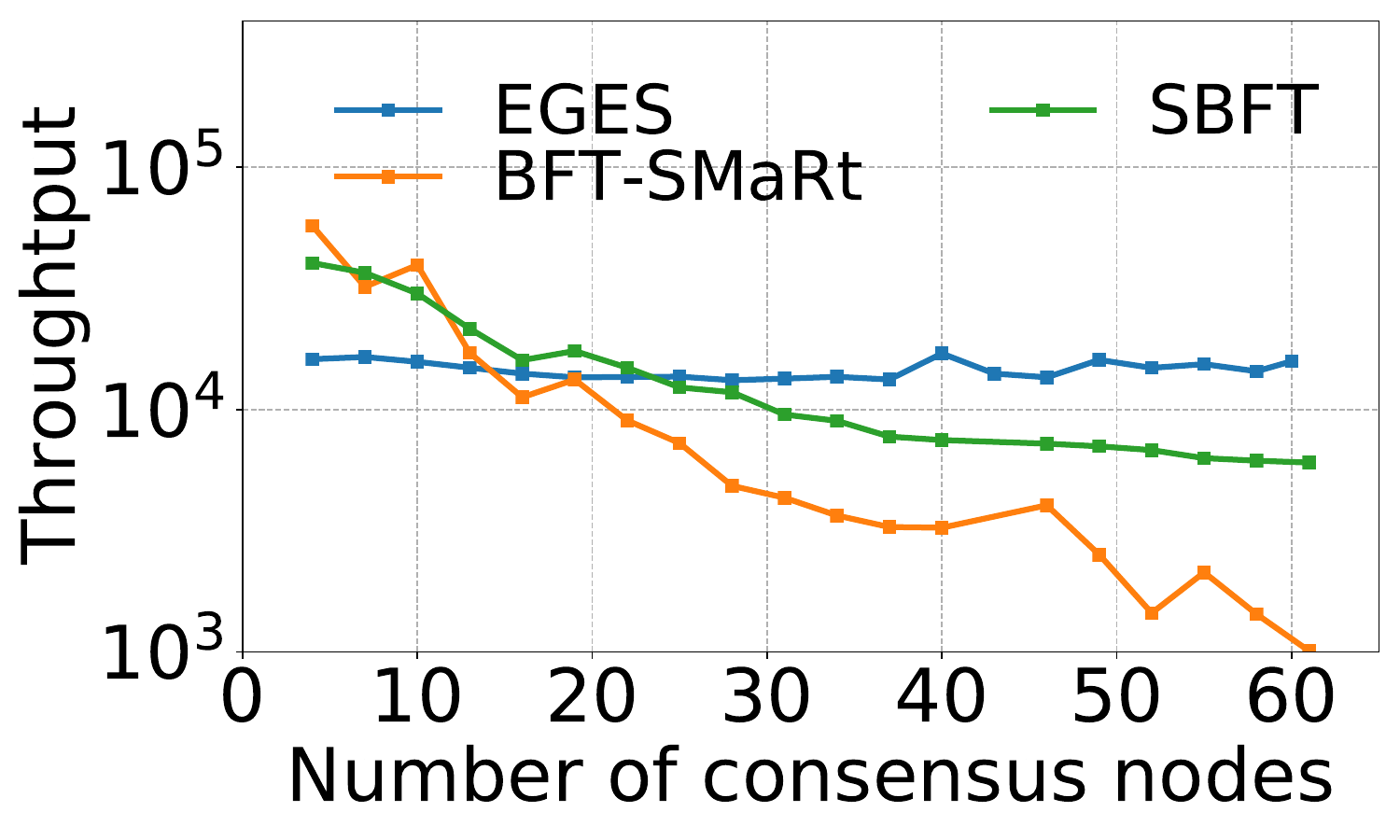}
		\label{fig:eval_bftsmart_0ms}
	}
	\subfloat[Geo-replicated mode.]{
		\centering
		\includegraphics[width=.45\columnwidth]{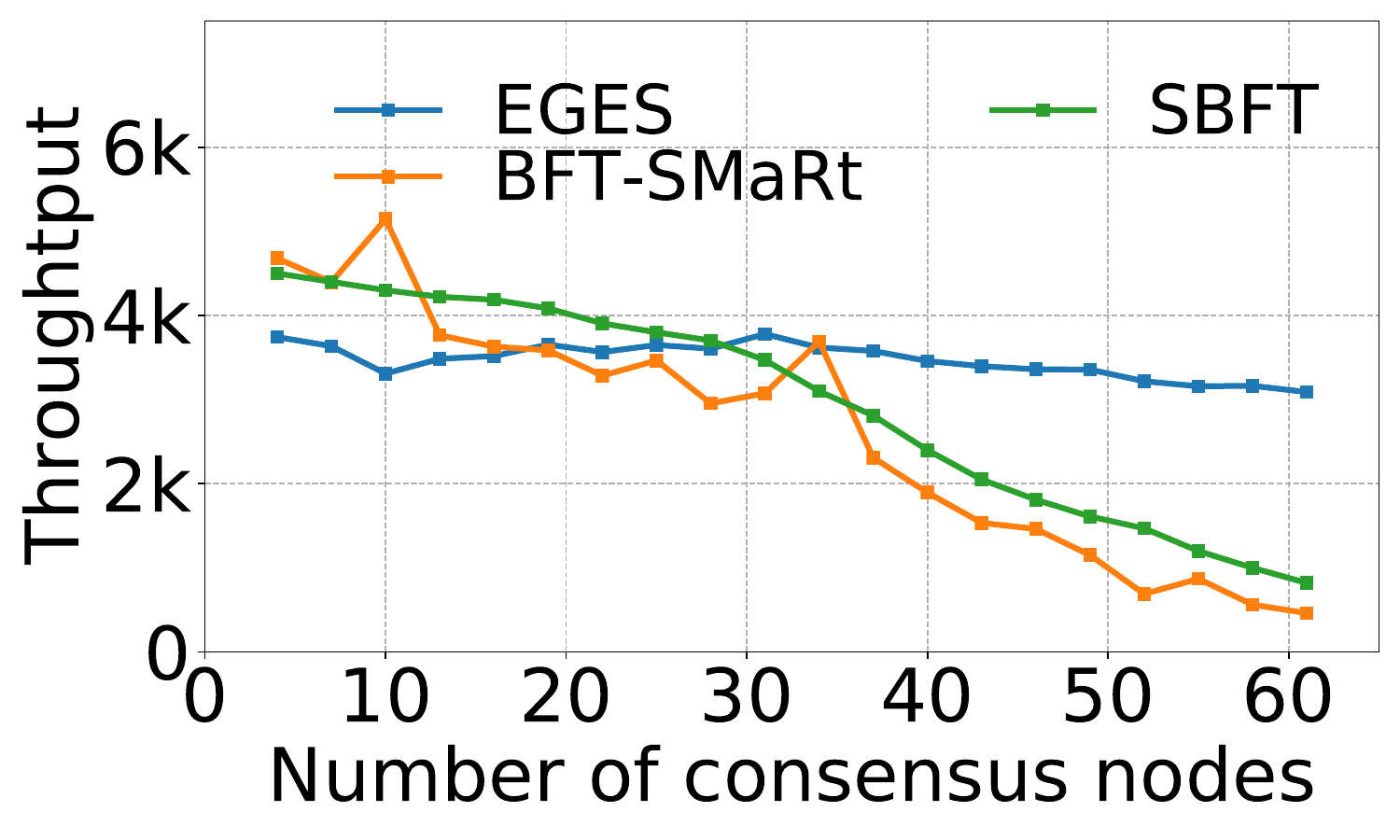}
		\label{fig:eval_bftsmart_200ms}}
	\\
	\subfloat[\smart with 10 nodes]{
		\centering
		\includegraphics[width=.45\columnwidth]{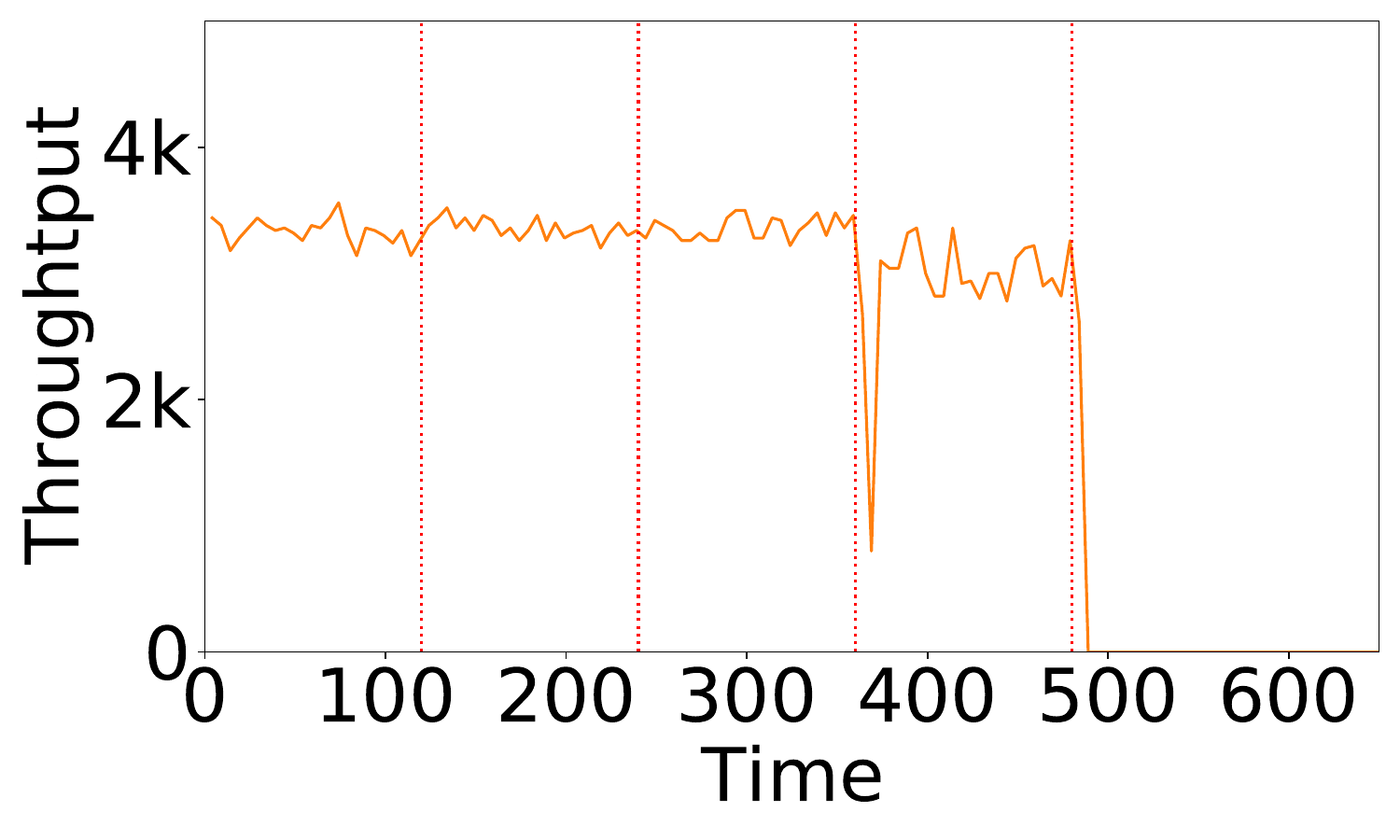}
		\label{fig:eval_bftsmart_failure}
	}
	\subfloat[SBFT with 62 nodes]{
		\centering
		\includegraphics[width=.45\columnwidth]{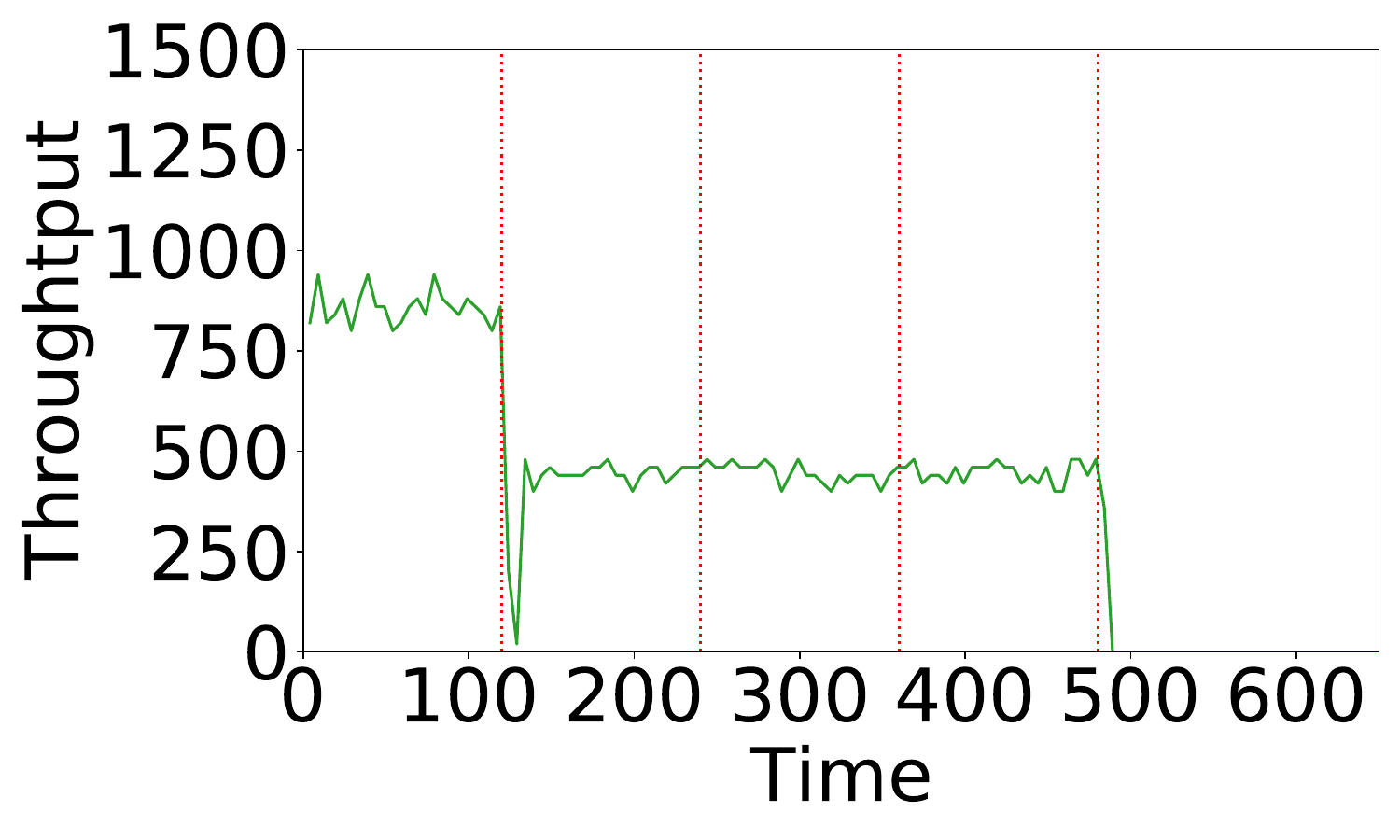}
		\label{fig:eval_sbft_failure}}
	\label{fig:eval_bftsmart}

	\caption{Comparing \xxx, SBFT, and \smart.}
\end{figure}

Since \smart with 10 (committee) nodes was faster than \xxx with 100 acceptors,
we evaluated both of them on a different number of nodes because more such nodes
can tolerate more faults and DoS attacks. \fig{fig:eval_bftsmart_0ms}
shows the results using the same setting for both systems (\eg, in our cluster,
TC disabled, and the same number of transactions in each batch). Overall, \xxx
throughput was stable because the number of acceptors affects little on the
latency in the seeking for quorum ACKs phase. \smart's throughput drops
dramatically because its protocol involves a quadratic number of messages on the
number of ordering nodes.
\looseness=-1

\fig{fig:eval_bftsmart_0ms} also shows SBFT's performance. 
% SBFT has better scalability than HLF-Smart because it
% greatly reduced the number of consensus messages (\chref{background}). However,
In the non-geo-replicated mode, when the number of nodes increased
from 4 to 62, SBFT's throughput dropped from 38.2K to 6.9K transactions/s. This
is because SBFT's collectors (\chref{sec:background:permissioned}) need to
collect more messages and to verify their signatures, so the time spent in
collectors increased from 2.5ms to 13.1ms.

\fig{fig:eval_bftsmart_200ms} shows the performance comparison of \xxx,
\smart, and SBFT in the geo-replicated setting. \xxx's throughput was at least
3.4X larger than both systems on 62 nodes. \smart's performance's
trend was similar to the no-delay setting because of its PBFT all-to-all
broadcasted messages. 
% \xxx's throughput is 3.4X timer than SBFT when there are
% 62 consensus nodes. 
SBFT's throughput also dropped dramatically because some nodes became
stragglers for the collectors due to the varied RTT. Since SBFT's fast path can
only tolerate a small number of straggler nodes (usually two
(\chref{sec:background:permissioned})), we observed that 87\% of the consensus
rounds in SBFT have reverted to the slow path (PBFT).

\looseness=-1 
More importantly, \xxx can safely switch its acceptor groups
across blocks, and it tolerated various failure scenarios, including DoS attacks
(\fig{fig:eval_down:1}). For comparison, we evaluated the performance of \smart
and SBFT on node failures (i.e., DoS attacks targeting consensus nodes).
\fig{fig:eval_bftsmart_failure} shows the result of \smart with its default
10-node setting. We randomly killed one node on each vertical line. The third
time we killed its leader coincidentally, so there was a noticeable performance
drop. \smart's throughput dropped to zero after we killed the fourth node. For
SBFT (\fig{fig:eval_sbft_failure}), we started with 62 nodes and killed 7 nodes
every time. Since SBFT's fast path can only tolerate two crashed or straggler
nodes, its throughput dropped significantly (reverted to PBFT) after the first
kill. 

\looseness=-1 Overall, \xxx is complementary to \smart and SBFT: \smart is the fastest
in a small scale; SBFT has better scalability, but its high performance
requires a synchronous network (stated in their paper). \xxx achieved reasonable
efficiency and DoS resiliency in a geo-replicated setting.

\subsection{Discussions}\label{sec:eval:discussion}

\xxx has two limitations. First, \xxx requires a joining node to have an SGX
device. We deem this requirement reasonable because SGX is available on
commodity hardware, and both academia and industry are actively improving the
security of SGX. Recent public blockchains~\cite{zhang2017rem} and permissioned
blockchains~\cite{poet,coco:ccf,veronese2013efficient} also use SGX. Second,
\xxx targets large-scale, geo-distributed permissioned blockchain systems (e.g.,
a global payment system~\cite{libra:smr}), while for small-scale deployments
(e.g., supply chain among a few small companies) in a single data
center, existing consensus protocols (e.g., \smart) are more suitable.

% Second,
% \xxx, as well as BitCoin and Algorand, enforces consistency with overwhelming
% probability, while other blockchains (\eg, \smart and ByzCoin) enforce
% deterministic consistency. All these systems are considered strongly
% consistent~\cite{bitcoinstrong} in practice. 

\looseness=-1
\xxx can incite the deployment of more Internet-scale applications that highly
desire the DoS resistance feature of \xxx, including e-voting, decentralized
auction, and payment systems. Moreover, the attested SGX enclaves on \xxx
nodes bring the potential to port existing centralized SGX-powered applications
on to \xxx, including SGX-protected distributed
database~\cite{bigmatrix:ccs17,enclavedb:sp18,shieldstore:eurosys19} and
SGX-powered anonymous networks~\cite{sgxtor:nsdi17}. For instance, one can
deploy an \xxx-ToR application by letting the \xxx enclave on each node do a
local attestation for an SGX-ToR enclave~\cite{sgxtor:nsdi17} and letting \xxx
nodes maintain the ToR directory service on the blockchain. By doing so,
\xxx-ToR removes SGX-ToR's centralized directory service that is vulnerable to
DoS attacks or service censoring.

\section{Conclusion}\label{sec:con}

We have presented \xxx, the first efficient permissioned blockchain consensus
protocol that can tolerate targeted DoS and partition attack. \xxx achieves
comparable performance to existing permissioned blockchain's consensus protocols
while achieving much stronger robustness. \xxx is carefully implemented with
two promising distributed applications, greatly improving the reliability and
security of their legacy, centralized versions. \xxx's source code is available
on \github.

% \xxx's source code and
% evaluation results are released on \github.

% Evaluation shows that
% \xxx can be an effective infrastructure to deploy blockchain applications. 

% % consensus protocol and its runtime system for public blockchains. 
% The stealth acceptor abstraction takes the first step to enable the strong 
% safety of Paxos to be integrated in \xxx's consensus protocol. 
% Extensive evaluation shows that \xxx{} is efficient, scalable, robust 
% and has the potential to support general applications, greatly improving 
% their security and reliability on Internet. \

\bibliographystyle{IEEEtran}
\bibliography{bib/biblio.bib,bib/blockchain.bib}

\vskip -3\baselineskip plus -1fil
\begin{IEEEbiography}
    [\vspace{-3.5em}{\includegraphics[width=0.8in, keepaspectratio]{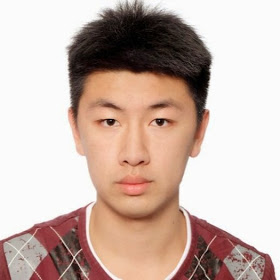}}]
    {Xusheng Chen} received his Bachelor degree in HKU. He is currently a PhD
    student in Computer Science of HKU (2017-now). He is under the supervision
    of Dr. Heming Cui. His research interests include distributed consensus
    protocols, distributed systems and system security.
\end{IEEEbiography}

\vskip -4\baselineskip plus -1fil
\begin{IEEEbiography}
    [\vspace{-3em}{\includegraphics[width=0.8in, keepaspectratio]{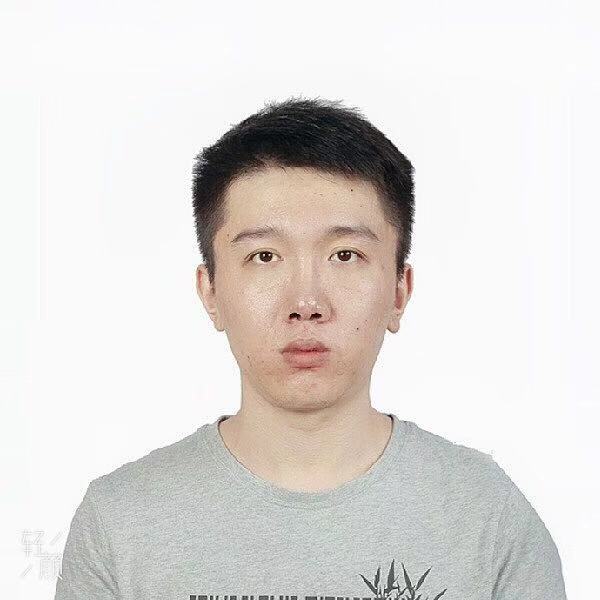}}]{Shixiong Zhao}
    received his Bachelor degree in HKU and his master degree in HKUST. 
    He is currently a PhD student in Computer Science of HKU (2017-now).
    He is under the supervision of Dr. Heming Cui. His research interests include distributed systems for high performance
    computing, distributed systems and system security.
\end{IEEEbiography}

\vskip -3\baselineskip plus -1fil
\begin{IEEEbiography}
    [\vspace{-3em}{\includegraphics[width=0.7in, keepaspectratio]{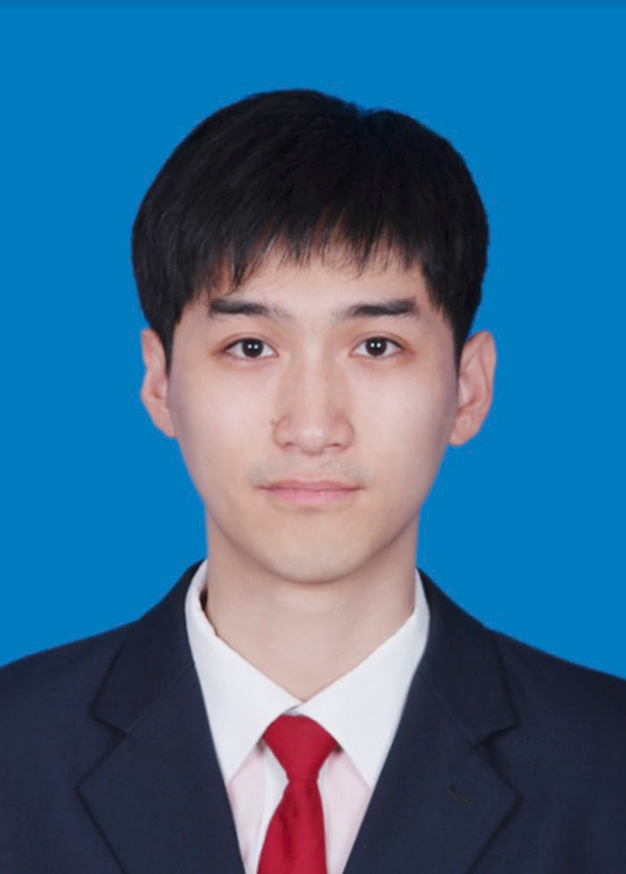}}]{Ji Qi}
   received his B.S (2015) degree from Beijing Institude of Technology, Beijing, China, and his M.S (2018)
   degree from Tsinghua University, Beijin, China. He is currently pursuing the PhD in computer science 
   at the University of Hong Kong under the supervision of Dr. Heming Cui.
   His interests include domain-specific modeling, distributed system and cloud computing. 
\end{IEEEbiography}

\vskip -3\baselineskip plus -1fil
\begin{IEEEbiography}
   [\vspace{-4em}{\includegraphics[width=0.75in, keepaspectratio]{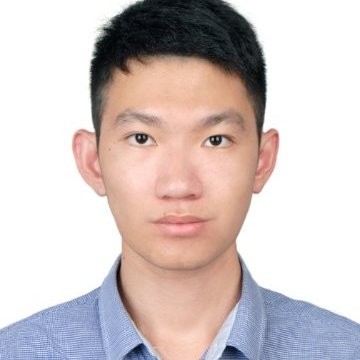}}]{Jianyu Jiang}
   is currently a third year PhD student in Computer Science Department at The University of Hong Kong. 
   He is working on topics in large scale computation platform under the supervision of Dr. Heming Cui.
   Jianyu received his Bachelor's Degree in Computer Science Department 
   at Xi'an Jiaotong University, under the supervision of Professor Qi Yong. 
\end{IEEEbiography}

\vskip -3\baselineskip plus -1fil
\begin{IEEEbiography}[\vspace{-3em}{\includegraphics[width=0.7in,
   keepaspectratio]{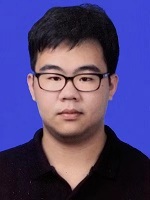}}] {Haoze Song} received the BS degree from
   Department of Computer Science, University of Science and Technology of
   China, in 2020. He is currently working towards the MPhil degree in Computer
   Science at HKU. His research interests mainly focus on distributed system and
   distributed computing. 
  % is an undergraduate student in computer science of USTC and a
  % research assistant of HKU. His research interests including distributed
  % systems, key-value storage. Contact him at  shz666@mail.ustc.edu.cn
\end{IEEEbiography}

\vskip -3\baselineskip plus -1fil
\begin{IEEEbiography}
    [\vspace{-4em}
    {\includegraphics[height=0.9in]{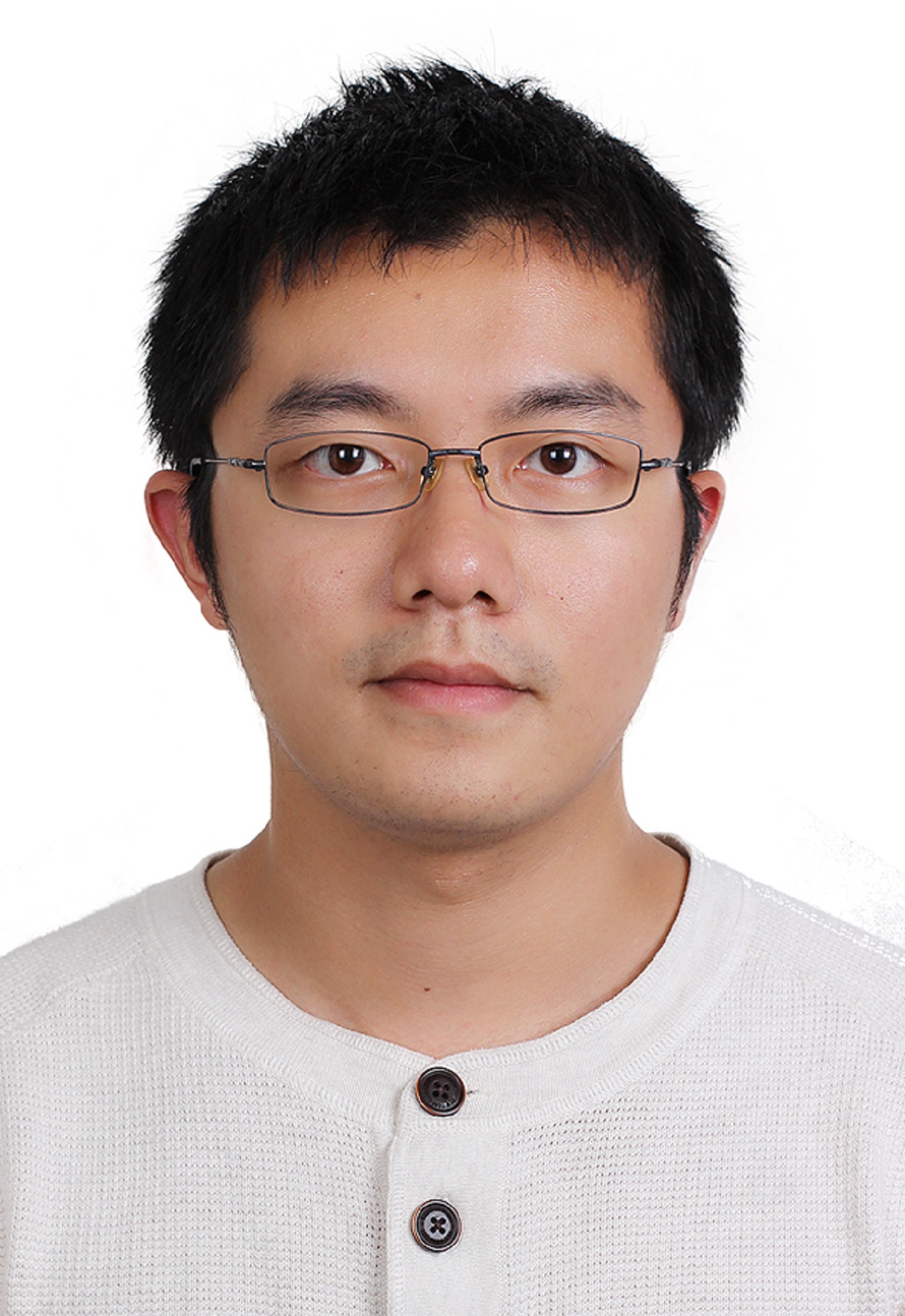}}] {Cheng
    Wang} received his PhD from the University of Hong Kong and B.Eng degree
    from Shanghai Tongji University. His research interests lie in distributed
    systems, with a particular focus on fault tolerance.
    He is currently working in Huawei Ltd.
\end{IEEEbiography}

\vskip -3\baselineskip plus -1fil
\begin{IEEEbiography}
    [\vspace{-3.5em}{\includegraphics[width=0.7in, clip, keepaspectratio]{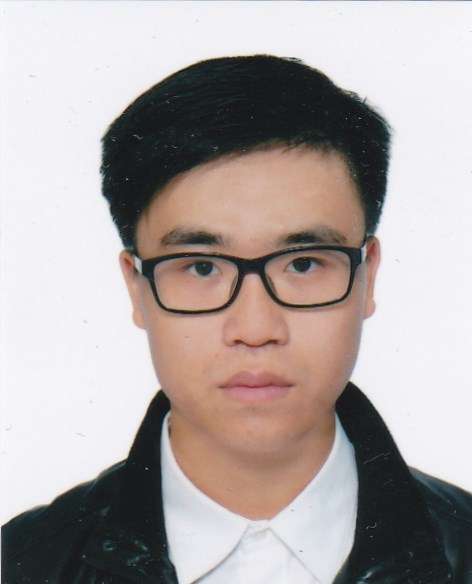}}]{Tsz On Li}
  received his Bachelor degree in HKU. He is currently an MPhil
  student in Computer Science of HKU (2019-now). He is under the supervision of
  Dr. Heming Cui. His research interests include differential privacy and big
  data systems.
\end{IEEEbiography}

\vskip -3\baselineskip plus -1fil
\begin{IEEEbiography}[{\includegraphics[trim={0.5cm 0cm 0cm 0cm}, width=0.8in, clip, keepaspectratio]{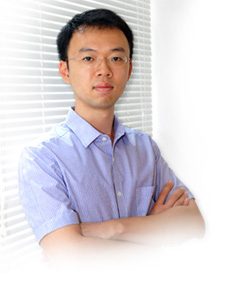}}]{T-H. Hubert Chan}
  T-H. Hubert Chan is an Associate Professor at the Department of
Computer Science at the University of Hong Kong. He completed his PhD
in Computer Science at Carnegie Mellon University in 2007.  His main
research interests are approximation algorithms, discrete metric
space, privacy and security inspired problems.
\end{IEEEbiography}
 
\newpage
\vskip -3\baselineskip plus -1fil
\begin{IEEEbiography}[{\includegraphics[width=0.8in,keepaspectratio]{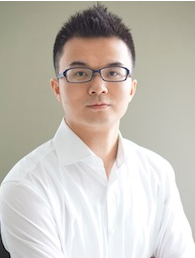}}]{\\Fengwei Zhang}
  is an Associate Professor at Department of Computer Science and Engineering at Southern University 
  of Science and Technology (SUSTech). 
  His primary research interests are in the areas of systems security, with a focus on trustworthy execution, 
  hardware-assisted security, debugging transparency, transportation security, and plausible deniability encryption.
  Before joining SUSTech, he spent four wonderful years as an Assistant Professor at Department of Computer Science 
  at Wayne State University. 
\end{IEEEbiography}

\vskip -3\baselineskip plus -1fil
\begin{IEEEbiography}[{\includegraphics[width=0.8in,keepaspectratio]{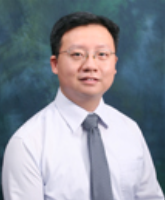}}]{\\Xiapu
  Luo} received his B.S. in Communication Engineering and M.S. in Communications
  and Information Systems from Wuhan University. He obtained his Ph.D. degree in
  Computer Science from the Hong Kong Polytechnic University, under the
  supervision of Prof. Rocky K.C. Chang. After that, he spent two years at the
  Georgia Institute of Technology as a post-doctoral research fellow advised by
  Prof. Wenke Lee. His current research interests include Network and System
  Security, Information Privacy, Internet Measurement, Cloud Computing, and
  Mobile Networks.
\end{IEEEbiography}

\vskip -3\baselineskip plus 1fil
\begin{IEEEbiography}[\vspace{-1em}{\includegraphics[width=0.8in,keepaspectratio]{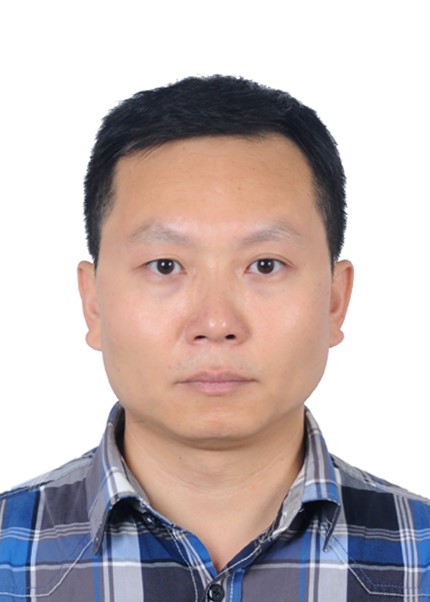}}]{\\Sen Wang}
  received the B.S. degree from the University of Science and Technology of
  China (USTC) in 2005, the M.S. degree from the Chinese Academy
  of Sciences (CAS) in 2008, and the Ph.D. degree from Tsinghua
  University in 2014, all in computer science. From 2014 to
  2019, he was a lecturer and then an associate professor at Chongqing
  University, China. Currently, he is a senior researcher at Huawei,
  Hongkong. His research interests include information-centric networking,
  Federated Learning and AI for System.
\end{IEEEbiography}

\vskip -3\baselineskip plus -1fil
\begin{IEEEbiography}[{\includegraphics[width=0.75in,keepaspectratio]{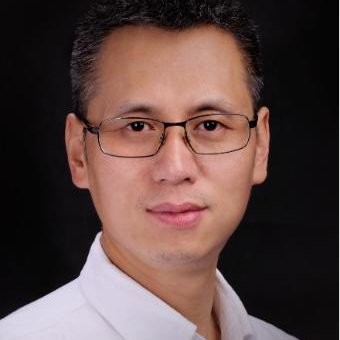}}]{\\Gong Zhang} 
  is a chief architect researcher
  scientist, director of the Huawei Future Network Theory
  Lab. His major research directions are network
  architecture and large-scale distributed systems. He has
  abundant experience on system architect in networks,
  distributed system and communication system for more than
  20 years. He has more than 90 global patents. 
\end{IEEEbiography}

\vskip -3\baselineskip plus -1fil
\begin{IEEEbiography}
   [\vspace{-3em}{\includegraphics[width=0.8in, clip, keepaspectratio]{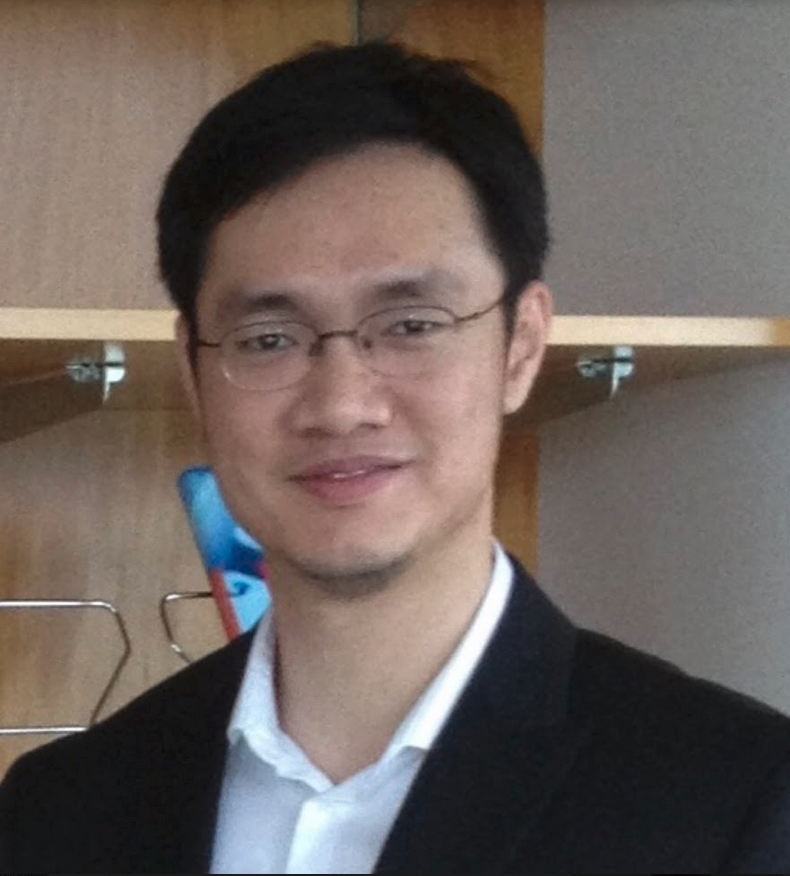}}]{Heming Cui}
   is an associate professor in computer science of HKU. His research interests include operating systems, programming 
   languages, distributed systems, and cloud computing, with a particular focus on building software infrastructures
   and tools to improve reliability and security of real-world software. Homepage: https://i.cs.hku.hk/~heming/.
   He is a member of IEEE.
\end{IEEEbiography}

\end{document}